# Domain Wall Motion in Magnetic Nanostrips

Oscar Alejos[1], Víctor Raposo[2] and Eduardo Martínez[2]*

[1] Dpto. Electricidad y Electrónica, Facultad de Ciencias, University of Valladolid, E-47011 Valladolid, Spain

[2] Dpto. Física Aplicada. University of Salamanca, Plaza de la Merced s/n, E-38008, Salamanca, Spain

* Corresponding author: edumartinez@usal.es

**Domain walls are the transition regions between two magnetic domains. These objects have been very relevant during the last decade, not only due to their intrinsic interest in the development of novel spintronics devices but also because of their fundamental interest. The study of domain wall has been linked to the research on novel spin-orbit coupling phenomena such as the Dzyaloshinskii-Moriya interaction and the spin Hall effect amount others. Domain walls can be nucleated in ferromagnetic nanostrips and can be driven by conventional magnetic fields and spin currents due to the injection of electrical pulses, which make them very promising for technological applications of recording and logic devices. In this review, based on full micromagnetic simulations supported by extended one-dimensional models, we describe the static and dynamic properties of domain walls in thin ferromagnetic and ferrimagnetic wires with perpendicular magnetic anisotropy. The present chapter aims to provide a fundamental theoretical description of the fundaments of domain walls, and the numerical tools and models which allow describing the DW dynamics in previous and future experimental setups.**

# 1. Introduction.

Domain walls (DWs) are the transition regions between two adjacent magnetic domains where the magnetization ($\vec{M}(\vec{r})$) adopts an almost uniform configuration [1,2]. The Domain Theory assumed that the DWs were abrupt regions without internal structure, and therefore, the magnetic state of a ferromagnetic system was a balance between the different contributions to the total energy of the system, which includes magnetostatic and anisotropy interactions. The domain pattern was traditional modified by external magnetic fields ($\vec{H}_{ext}$), which promote growth in size of the domains magnetized in the parallel direction to the external field at expenses of the other domains. Hysteresis loops under external fields are therefore a sequence of local equilibrium states.

More recently, with the advent of novel fabrication and observation techniques, it has been evident that the DWs indeed have an internal structure, which is essential to understand, not only the magnetic configuration of ferromagnetic samples at the nanoscale, but also because it plays a key role to further understand the dynamic properties of these samples. The Micromagnetic Model (mM) is the continuous theory of the magnetization [3,4,5], which in parallel with the development of numerical techniques and observation methods, has been essential to understand the interesting physics of these systems, not only from a fundamental point of view, but also to envisage novel technological applications where DWs are the main ingredients of recording and logic devices. This model assumes that the local magnetization $\vec{M}(\vec{r})$ is a continuous vector function, where the local magnitude is equal to the saturation magnetization of the sample ($M_s$), and its direction, which is given by a unit vector $\vec{m}(\vec{r},t)$, changes smoothly over a nanometer scale. Besides, of the mentioned interactions, the micromagnetic formalism takes into account the exchange interaction, which is the main responsible of the ferromagnetic order. Therefore, the magnetic state of a ferromagnetic system results from the balance between the exchange, magnetostatic and anisotropy interactions in the absence and the presence of external magnetic fields. Hysteresis loops, which are a sequence of equilibrium states under a series of external magnetic fields, are naturally obtained in the micromagnetic model by minimizing the energy of the system for each field. However, the micromagnetic model does not only allow to compute equilibrium states and hysteresis loops. It also allows us to study the magnetization dynamics at the nanoscale, where besides from domains and DWs, other magnetic patterns, such as magnetic vortices and skyrmions, appear. The temporal evolution of the local magnetization ($d\vec{m}(\vec{r},t)/dt$) is dictated by the Landau-Lifshitz-Gilbert equation [1-5], which is obtained from Newton's law. It includes precession and damping terms. In the presence of a magnetic field ($\vec{B} = \mu_0 \vec{H}$, with $\mu_0$ being the permeability of the free space), a magnetic moment ($\vec{\mu}$) experiences a torque ($\vec{\tau} = -\vec{\mu} \times \vec{B}$) which makes it precess around the field, that is, $\frac{d\vec{\mu}}{dt} = -\gamma\mu_0 \vec{\mu} \times \vec{H}$. Similarly, in the context of the micromagnetic formalism, the magnetization of an elementary volume cell $\vec{m}(\vec{r},t) = \vec{M}(\vec{r},t)/M_s$ processes around the local effective field ($\vec{H}_{eff} = \vec{H}_{eff}(\vec{r})$) which includes all the magnetic interactions, and it is derived from the energy density of the system ($\epsilon$) by means of a functional derivative, $\vec{H}_{eff}(\vec{r}) = -\frac{1}{\mu_0 M_s}\frac{\delta \epsilon}{\delta \vec{m}}$. The magnetization dynamics in the absence of damping read as $\frac{d\vec{m}(\vec{r},t)}{dt} = -\gamma_0 \vec{m} \times \vec{H}_{eff}$, where $\gamma_0 = |\gamma \mu_0| = 2.21 \times 10^5$ m A$^{-1}$s$^{-1}$ is the gyromagnetic factor. The damping term is included to account for the observation of the relaxation of the local magnetization towards the local effective field. There exist several mechanisms which introduce losses: eddy currents [6,7], macroscopic discontinuities (Barkhausen jumps), diffusion and reorientation of the lattice defects, or spin-scattering mechanism can all introduce losses [2]. The damping term is a phenomenological contribution accounting for the dissipation phenomena involved, which are described by a dimensionless Gilbert damping $\alpha$. The resulting Gilbert equation reads as [1,2]:

$$\frac{d\vec{m}}{dt} = -\gamma_0 \vec{m} \times \vec{H}_{eff} + \alpha \vec{m} \times \frac{d\vec{m}}{dt} \qquad (1)$$

Equation (1) describes the magnetization dynamics under the presence of magnetic fields which are applied externally. In particular, the field-driven DW dynamics can be studied by numerically solving this equation, where the effective field $\vec{H}_{eff}$ includes the exchange ($\vec{H}_{exc}$), the magnetostatic ($\vec{H}_{dmg}$), the magnetic anisotropy, the Zeeman ($\vec{H}_{dc}$) interactions, and other classical electromagnetic interactions. Further details of these contributions can be consulted elsewhere [1-4,8,9]. Indeed, the interactions included in the effective field determine the equilibrium state of ferromagnetic samples, and particularly, the DW configurations in strips. These equilibrium magnetization configurations will be reviewed in this chapter. Also, the field-driven DW dynamics will be reviewed in systems with perpendicular magnetic anisotropy (PMA). In particular, we will focus our attention on thin strips where $\ell$ is the length along the longitudinal $x$-axis, $w$ is the strip width along the transverse $y$-axis, and $t$ represents the thickness along the out-of-plane direction ($z$-axis) with $\ell \gg w \gg t$. In these PMA strips, a DW separates two domains magnetized along the easy axis ($\vec{u}_k = \vec{u}_z$), that is, either along the $+z$ (or *up* domain) or the $-z$ (or *down* domain) directions. When an out of plane field is applied, $\vec{B}_{ext} = \mu_0 \vec{H}_{ext} = B\vec{u}_z$, the size of the *up* domain increases, and consequently, the DW moves along the $x$-axis. The motion of DWs by external fields has been studied from both theoretical and experimental points of view during the last decades [10-15], and here we will review the main features of the field-driven DW motion. However, driving DWs by external field has both experimental and technological limitations. For instance, due to the reduction of the scale (the need of miniaturization) of the systems of interest, with dimensions at the nanoscale, it is difficult to generate local external field to excite the dynamics of a single DW in a real device without perturbing the magnetic state of other DWs. Besides, adjacent DWs, *up-down* and *down-up*, in a single ferromagnetic strip are driven in opposite directions under a static magnetic field, which could result in the annihilation of the DWs. Therefore, the manipulation of DWs by magnetic fields is not very promising for magnetic recording and logic applications, where series of DWs need to be driven back and forth without destroying the coded information within the domain between them.

It is possible to drive DWs by using spin polarized currents, which could be of different nature depending on the architecture. For instance, DWs can be driven by injecting an electrical current along a conductive ferromagnetic strip. The mechanism behind is the Spin Transfer Torque (STT), an effect which was first predicted theoretically by Berger [16] and Slonczewski [17]. See [18] for a review of most experimental observations and the progress in the understanding on the current-driven DW dynamics through 2011. STT corresponds to a torque on the magnetization which is due to a transfer of angular momentum between the conduction electrons and the local magnetization. An electrical current corresponds to a flux of electrons. Each electron carries a magnetic moment $\vec{\mu}$ that can interact with the local magnetization by exchange interaction. As a consequence, electrons' spins are aligned with the magnetization and the current becomes spin-polarized (namely, there is also a spin current associated with the electrical current since all the electron have the same spin alignment). On top of that, the spin current affects the local magnetization if it differs from the spin-polarization direction: as the spin polarized current follows the magnetic texture, it exchanges angular momentum with the local magnetization. Therefore, the STT can move DWs along nanostrips. Additional torques $\vec{\tau}_{STT}$ are included in the LLG Equation (1) to account for these STT mechanisms. The explicit form of these STTs will be introduced later in Sec. 2.2. The direction of DW motion only depends on the sign of the current and not on the alignment of the domains, as in the case of field-driven DW motion. As already commented, in this latter case, two DWs would move either towards or away from each other, depending on the applied field. On the contrary, neighboring DWs can be moved in the same direction under the STT. This feature has paved the way for promising applications based on DW motion such as racetrack memories [19] or DW logic devices [20], where series of DWs can be driven along ferromagnetic strips. Under the STT the DWs are displaced along the electron flow direction, that is, in the opposite direction to the electrical current.

More recently, several experiments on current-driven DW dynamics [21-24] have been demonstrated to be inconsistent with the STTs introduced above: DWs are driven along the current direction (against the electron flow). The evaluated systems consist of an ultrathin ferromagnetic (FM) strip sandwiched between a heavy metal (HM) and a non-magnetic layer, *e.g.* an oxide (Ox). In this case, the DWs are driven along the electric current or against it depending on the heavy metal and the spin orbit coupling [22]. In particular, the internal structure of the magnetic DW has been found to be essential to understand the DW dynamics in these systems. In a typical thin FM strip, the internal magnetization within a DW tends to be aligned along the transversal direction, *i.e.* $\vec{m}_{DW} \approx \pm \vec{u}_y$, and it rotates from the *up* to the *dow*n domain in the $yz$-plane. This is the so-called Bloch DW configuration, which results from the balance between the exchange and the magnetostatic interactions. On the other hand, it was recently shown than when a ultra-thin FM strip is sandwiched between a heavy metal and an oxide, the DW adopts a Néel configuration where its internal moment aligns along the longitudinal direction, that is, $\vec{m}_{DW} \approx \pm \vec{u}_x$ [22,23]. Therefore, the magnetization rotates from the *up* to the *down* domain in the $xz$-plane with a fixed chirality. The key new ingredient responsible of these chiral DW patterns is an interfacial effect, which results from the combination of spin-orbit interaction and structural broken inversion symmetry of the stacks. The interfacial effect is the Dzyaloshinskii-Moriya interaction (DMI) [25,26] and gives rise to an antisymmetric exchange interaction which needs to be accounted for in the energetic balance with the rest of micromagnetic interactions. Experimentally, several effects of the DMI have been observed in ultra-thin multilayers, such as chiral DW [22-24,27-34], asymmetric bubble expansion [35,36] or asymmetric spin wave propagation [37]. Furthermore, DMI can also favor new types of magnetization patterns such as skyrmions [38] or helices [39].

The current-driven DW motion in these multilayers is essentially due to the spin Hall effect (SHE), which describes the conversion of charge current into spin current due to spin-orbit coupling [40-44]. In ultrathin multilayers such effect can be exploited to inject a spin current into the FM layer, which affects the magnetization dynamics through spin orbit torques (SOTs) [45-47]. Due to its small thickness, ultrathin FM films (with thicknesses of ~1 nm or below) have a high electrical resistance and, therefore, it is possible to inject electrical current mainly passing through the neighboring high conductive metallic layers. If the layers have a high spin-orbit coupling, electrons with different spins are scattered in opposite directions, resulting in a spin current perpendicular to the injected charge current. The spin-polarization of the spin current ($\vec{\sigma}$) will also be perpendicular to the charge current ($\vec{J}_e$) and the spin current ($\vec{J}_s$). For instance, a spin current along the $z$ direction is polarized along the $y$ direction since the charge current is flowing along the $x$ direction. More details about the microscopic origin of the spin Hall effect (SHE) can be found in [40,41]. The spin Hall effect and other spin-orbit torques (SOTs) can be included in the magnetization dynamics by adding a new torque into the LLG Equation (1) ($\vec{\tau}_{SOT}$). As it will be discussed here, this SOT drives chiral DWs along FM strips with high efficiency, a feature which makes these multilayers very promising for DW-based applications.

The rest of this chapter is structured as follows. We firstly present in **Sec. 2** the details of the micromagnetic model (mM) which allows us to describe the static and dynamic properties of DWs in thin ferromagnetic strips with high Perpendicular Magnetic Anisotropy (PMA). In **Sec. 3** we present micromagnetic results of DWs and, in parallel, the bases of the one-dimensional model (1DM). The different approaches in the 1DM are described and the corresponding results are compared to full micromagnetic simulations. In particular, both mM and 1DM models will be used in the rest of the chapter to describe the different types of DWs and the different systems which have been the object of study by several groups during the last decade. We will review the field-driven case, and the current-driven DW dynamics in single ferromagnetic strips. We will study this dynamics in multilayers where the ferromagnetic strip is sandwiched between a heavy metal and an oxide. In the first systems, the current-driven DW dynamics is a consequence of the spin current flowing through the ferromagnetic strip which interacts with the DWs by a spin-transfer torque (STT) mechanism.

In multilayers, the spin current is mainly generated in the heavy metal due to the spin Hall effect (SHE). This spin current interacts with the DWs in the ferromagnetic strip, in particular, when the spin-orbit coupling imposes chiral configurations of the DWs. Also in Sec. 3, we will present results of the current-driven DW motion in systems with curved parts, a study which will be also essential for the development of novel DW based devices in the next years. In the last part of Sec. 3 we will analyze other platforms for current-driven DW dynamics, such as synthetic antiferromagnetic multilayers and ferrimagnetic stacks. Therefore, the topics covered in the present chapter aim to provide a full description of the current state of the art of DWs in systems with high PMA. The objective of the present chapter is to expose a detailed and compressive description of the numerical and theoretical models on current-driven DW dynamics along strips with high PMA, with the intention of being useful for researchers in the development of novel DW based devices in the next future. We state here, that is not our goal here to present an exhaustive enumeration of experimental results, and consequently, we refer to the interested reader on these experimental aspects to other works which are not cited here but can be easily reached from the references provided in this document.

## 2. Micromagnetic model (mM).

### *2.1. Total energy and effective field.*

Within the micromagnetic formalism, a ferromagnetic system is discretized in elementary volume cells, and the equilibrium magnetic state is the result of the balance of several interactions [1-5]. The dominant interaction is the exchange interaction, which is responsible for the ferromagnetic order. Ferromagnetic materials show spontaneous magnetization above the microscopic scale. This is due to the tendency of magnetic moments in such materials to align parallel to each other, giving rise to magnetic order inside the material. The origin is the exchange interaction, a quantum mechanical effect whose description and complete treatment can be found in Ref. [1,2]. It all boils down to a spin-dependent effective Hamiltonian, the Heisenberg Hamiltonian

$$\widehat{\mathcal{H}} = -2\sum_{\langle ij \rangle} J_{ij}\vec{S}_i \cdot \vec{S}_j = -2JS^2 \sum_{\langle ij \rangle} \cos\phi_{ij} \qquad (2)$$

where the sum is over first neighbors, $J_{ij}$ is the exchange integral between spins $i$ and $j$, and $\phi_{ij}$ is the angle between the two spins ($\vec{S}_i$ and $\vec{S}_j$). Such interaction is of short-range. When $J_{ij} > 0$, the energy of the system is minimized when spins lie parallel to one another. The formulation of this requirement in the continuum micromagnetic formalism considers that the angle between neighboring spins is small so that the scalar product in Equation (2) can be approximated as $\cos\phi_{ij} \approx 1 - \frac{1}{2}\phi_{ij}^2$ [3,4] and the exchange energy $E = cte +$ $2JS^2 \sum_{\langle ij \rangle} \phi_{ij}^2$. The passage to the continuous approximation results in the exchange energy density term

$$\epsilon_{exch} = A(\nabla \vec{m})^2 \tag{3}$$

where $A = n2JS^2/a$ is the exchange stiffness constant, which has units of J m$^{-1}$, and $a$ is the lattice constant, with $n$ depending on the type of lattice ($n = 1$ for simple cubic, $n = 2$ for $bcc$ and $n = 4$ for $fcc$). Exchange energy favors uniform magnetic patterns for which $\nabla \vec{m} = 0$.

Exchange interaction is isotropic, which means that in the ideal situation of a magnetic specimen whose spins are all aligned, the energy level will be the same regardless of the orientation of the magnetization $\vec{M}$. In the common case in which the specimen has a crystalline structure, the atoms, and hence the magnetic moments, are positioned at the vertices of the lattice. Due to spin-orbit interactions, orientations of magnetization along certain crystallographic axes are favored over others, breaking the symmetry of the isotropic exchange interaction. This effect is described by the so-called magnetocrystalline anisotropy. Hexagonal and tetragonal crystals show a uniaxial anisotropy. In the present work, we will focus our attention to strips with high uniaxial perpendicular magnetic anisotropy (PMA). The corresponding energy density is [1,2]

$$\epsilon_{ani} = K_u[1 - (\vec{m} \cdot \vec{u}_k)^2] \tag{4}$$

where $K_u$ is the anisotropy constant (in J m$^{-3}$) and $\vec{u}_k = \vec{u}_z$ the unit vector along the easy axis.

The energy of the ferromagnetic sample also includes the magnetostatic ($\epsilon_{dmg}$) and Zeeman ($\epsilon_{ext}$) interactions, which are given by

$$\epsilon_{dmg} = -\frac{1}{2}\mu_0 M_s \vec{H}_{dmg} \cdot \vec{m} \tag{5}$$

$$\epsilon_{ext} = -\mu_0 M_s \vec{H}_{ext} \cdot \vec{m} \tag{6}$$

where $\vec{H}_{dmg}$ is the demagnetizing field and $\vec{H}_{ext}$ is the external magnetic field. The magnetostatic field $\vec{H}_{dmg}$ is the classical magnetic field generated by the magnetization $\vec{M}$. It can be obtained from Maxwell's equations, which in the absence of electrical current are

$$\nabla \cdot \vec{H}_{dmg} = \rho_m \tag{7}$$

$$\nabla \times \vec{H}_{dmg} = 0 \tag{8}$$

where $\vec{H}_{dmg} = \vec{B}/\mu_0 - \vec{M}$ and $\rho_m = -\nabla \cdot \vec{M}$. Mathematically, Equations (7) and (8) have the same form of the electrostatic Maxwell's equations relating charges and electrical field (note that the analogy holds only in this particular case where there are no free electrical currents). In this case, $\rho_m = -\nabla \cdot \vec{M}$ plays the role of magnetic charge per unit volume. Thus, it is possible to write

$$\vec{H}_{dmg} = -\nabla \phi_m \tag{9}$$

where $\phi_m$ is the magnetic scalar potential, and $\nabla \phi_m$ represents its gradient. As $\vec{H}_{dmg}$ satisfies Equation (8), the $\phi_m$ satisfies the Laplace Equation $\nabla^2 \phi_m = -\rho_m$, which can be solved as

$$\phi_m = \frac{1}{4\pi} \int_{V'} \frac{\rho_m}{|\vec{r} - \vec{r}'|} dv' + \frac{1}{4\pi} \oint_{S'} \frac{\sigma_m}{|\vec{r} - \vec{r}'|} ds' \tag{10}$$

where $\rho_m = -\nabla \cdot \vec{M}$ and $\sigma_m = \vec{M} \cdot \vec{n}$ are the magnetic charge densities per unit volume and surface respectively. The first integral is over the volume of the magnetic system ($V'$) and the second one is over the closed surface $S'$ surrounding $V'$. $\vec{n}$ is the unit vector perpendicular to the surface pointing outwards from the volume $V'$. For a given magnetization distribution $\vec{M}(\vec{r}) = M_s \vec{m}(\vec{r})$, the magnetic potential $\phi_m$ can be computed from Equation (10), and from it, Equation (9) determines the magnetostatic field $\vec{H}_{dmg}$.

The total energy of a magnetic system is therefore,

$$E = \int_{V'} \left[ \epsilon_{exch} + \epsilon_{ani} + \epsilon_{dmg} + \epsilon_{ext} \right] dv' \tag{11}$$

The energy of the ferromagnetic system is a function of the magnetization over the sample, *i.e.*, $E = E[\vec{M}(\vec{r})]$, and the equilibrium state is the minimum of Equation (11), which must satisfy $\delta E = 0$, where $\delta E$ represents the variational of the energy, and which can be calculated explicitly with the energy terms that we have introduced above. Since for exchange interaction $\vec{M}(\vec{r}) = M_s \vec{m}(\vec{r})$, at $T \ll T_C$ ($T_C$ is the Curie temperature of the ferromagnetic material), the energy can be considered as a functional of the unitary vector $\vec{m}(\vec{r})$ and the minimization is performed with the constrain $|\vec{m}(\vec{r})| = 1$. The variation of the system energy is given by [1]

$$\delta E[\vec{m}(\vec{r})] = \int_{V'} \left[ 2A(\nabla \vec{m}) \cdot \nabla(\delta \vec{m}) - 2K(\vec{m} \cdot \vec{u}_k)\vec{m} \cdot \delta \vec{m} - \mu_0 M_s \vec{H}_{dmg} \cdot \delta \vec{m} \right.$$
$$\left. - \mu_0 M_s \vec{H}_{ext} \cdot \delta \vec{m} \right] dv' \tag{12}$$

where $(\nabla \vec{m}) \cdot \nabla(\delta \vec{m})$ stands for $(\nabla m_x)\nabla(\delta m_x) + (\nabla m_y)\nabla(\delta m_y) + (\nabla m_z)\nabla(\delta m_z)$, and we have used the reciprocity theorem for the variation of the magnetostatic energy. In fact, a variation of $\vec{m}$ generates a variation of $\vec{H}_{dmg}$ as well. By using the identity $\nabla \cdot (a\vec{b}) = a(\nabla \cdot \vec{b}) + \vec{b} \cdot (\nabla a)$ and the divergence theorem, Equation (12) can be expressed as

$$\delta E[\vec{m}(\vec{r})] = -\int_{V'} \left[ 2A\nabla \cdot (\nabla \vec{m}) + 2K(\vec{m} \cdot \vec{u}_k)\vec{m} + \mu_0 M_s \vec{H}_{dmg} + \mu_0 M_s \vec{H}_{ext} \right]$$
$$\cdot \delta \vec{m} \, dv' + 2\oint_{S'} A \left[ \frac{\partial \vec{m}}{\partial n} \cdot \delta \vec{m} \right] ds' \tag{13}$$
$$= -\int_{V'} \mu_0 M_s \vec{H}_{eff} \cdot \delta \vec{m} \, dv' + 2\oint_{S'} A \left[ \frac{\partial \vec{m}}{\partial n} \cdot \delta \vec{m} \right] ds'$$

where the first integral is over the body volume and the second over the body surface. $\frac{\partial \vec{m}}{\partial n}$ represents the derivative across the direction normal to the body surface and $\vec{H}_{eff}$ is the effective field defined as

$$\vec{H}_{eff} = -\frac{1}{\mu_0 M_s} \frac{\delta \epsilon}{\delta \vec{m}} = -\frac{1}{\mu_0 M_s} \left[ \frac{\partial \epsilon}{\partial \vec{m}} - \nabla \left( \frac{\partial \epsilon}{\partial (\nabla \vec{m})} \right) \right] \tag{14}$$

where $\epsilon = \epsilon(\vec{m})$ is the energy density per unit volume defined in the integrand of Equation (11). Since $\delta \vec{m}$ must obey the constrain $|\vec{m}(\vec{r})| = 1$, a generic variation $\delta \vec{m}$ can be expressed as $\delta \vec{m} = \vec{m} \times \delta \vec{\theta}$, ($\delta \vec{\theta}$ represents the variation of the magnetization orientation), and Equation (13) can be written as

$$\delta E[\vec{m}(\vec{r})] = -\int_{V'} \left[ \mu_0 M_s \vec{H}_{eff} \times \vec{m} \right] \cdot \delta \vec{\theta} \, dv' + 2\oint_{S'} A \left[ \frac{\partial \vec{m}}{\partial n} \times \vec{m} \right] \cdot \delta \vec{\theta} \, ds' \tag{15}$$

Since the minima of the functional must satisfy that $\delta E[\vec{m}(\vec{r})] = 0$ for any arbitrary variation $\delta \vec{\theta}$, Equation (15) implies that equilibrium states must satisfy the following conditions:

$$\vec{H}_{eff} \times \vec{m} = 0, \forall \vec{r} \in V' \tag{16}$$

$$\left[ \frac{\partial \vec{m}}{\partial n} \times \vec{m} \right] = 0, \forall \vec{r} \in S' \tag{17}$$

which state the mathematical conditions that an equilibrium configuration must satisfy. Equation (16) means that, at equilibrium, the magnetization $\vec{m}$ in each elementary volume within the ferromagnetic sample ($V'$) is aligned with the local effective field $\vec{H}_{eff}$. In other words, the torque between $\vec{m}$ and $\vec{H}_{eff}$ is null. Equation (17) indicates that the variation of the magnetization must be null at the surface ($S'$) of the ferromagnetic object.

Exchange, anisotropy, demagnetizing and Zeeman are the energy terms usually considered in standard micromagnetics. Recently, the fabrication of nanostructures with broken inversion symmetry led to the appearance of an additional interaction at the interfaces between magnetic materials and heavy metals, which turns out to have an important effect on the magnetization dynamics. Therefore, we present it in the following paragraphs. When an ultrathin ferromagnetic strip is sandwiched between a heavy metal and an oxide, or between two different heavy metals with strong orbit coupling, the energy of the ferromagnetic system must include the corresponding interfacial interaction. Exchange interaction has been presented with a fully symmetric Hamiltonian. However, its generalized form can be written as

$$\mathcal{H}_{ij} = \vec{S}_i \cdot (\mathcal{M}_{ij} \vec{S}_j) \tag{18}$$

where $\mathcal{M}_{ij}$ is a tensor representing the bilinear form of the energy of two spins. This can be decomposed in a symmetric and an antisymmetric part [39]. The symmetric part represents the usual exchange interaction that we presented as fully isotropic with a scalar product in Equation (2). The antisymmetric part can be rewritten as a cross product by a vector $\vec{D}_{ij}$:

$$\mathcal{H}_{ij} = \vec{S}_i \cdot (\mathcal{M}_{ij} \vec{S}_j) = J_{ij} \vec{S}_i \cdot \vec{S}_j - \vec{D}_{ij} \cdot (\vec{S}_i \times \vec{S}_j) \tag{19}$$

This antisymmetric energy term $(\vec{D}_{ij} \cdot (\vec{S}_i \times \vec{S}_j))$ is due to the presence of a spin-orbit interaction, which connects the lattice with the spin symmetry. The broken parity of the lattice gives rise to an additional interaction that breaks the inversion invariance of the Heisenberg Hamiltonian, which is the Dzyaloshinskii–Moriya interaction (DMI). In the systems studied in this chapter, DMI is an interfacial effect, arising from the interaction of a ferromagnetic layer with an adjacent lattice of heavy metal atoms with large spin-orbit coupling (see Figure 1f in [48]). The vector $\vec{D}$ is then determined by the interaction at the interface as $\propto \vec{r}_{ij} \times \vec{n}$, where $\vec{r}_{ij}$ is the vector connecting spins $\vec{S}_i$ and $\vec{S}_j$ and $\vec{n}$ is here the interface plane vector. Here we consider a broken inversion symmetry along the $\vec{u}_z$ direction, so that the DMI vector $\vec{D}$ is directed as $\vec{D} = D\vec{u}_z \times \vec{u}_{ip}$, with $D$ being the magnitude of $\vec{D}$, for any in-plane direction $\vec{u}_{ip}$. In order to minimize the DMI energy, $\vec{S}_i \times \vec{S}_j$ must be parallel or anti-parallel to $\vec{D}$, depending on the sign of $D$. Thus, the DMI competes with the symmetric exchange interaction and promotes spin spiral states and other non-uniform magnetization patterns.

In the context of micromagnetics, Thiaville et al. [26], following the same approach shown for the exchange interaction, introduced the DMI in a continuous approximation, which reads as

$$E_{DM} = \int_{V'} \epsilon_{DM}\, dv' = \int_{V'} D[m_z(\nabla \vec{m}) - (m \cdot \nabla)m_z]\, dv'$$
$$= \int_{V'} D\left[\left(m_z \frac{\partial m_x}{\partial x} - m_x \frac{\partial m_z}{\partial x}\right) - \left(m_z \frac{\partial m_y}{\partial y} - m_y \frac{\partial m_z}{\partial y}\right)\right] dv' \quad (20)$$

By calculating the variation $\delta E_{DM}$, it is possible to include the DMI into the equilibrium equations. After some algebra we get

$$\delta E_{DM}[\vec{m}(\vec{r})] = \int_{V'} \mu_0 M_s \vec{H}_{DM} \cdot \delta\vec{m}\, dv' + 2\oint_{S'} D[\vec{m} \times (\vec{n} \times \vec{u}_z)] \cdot \delta\vec{m}\, ds' \quad (21)$$

where $\vec{H}_{DM}$ is the additional DMI contribution to the effective field

$$\vec{H}_{DM} = \frac{2D}{\mu_0 M_s}[\nabla m_z - (\nabla \cdot \vec{m})\vec{u}_z]$$
$$= \frac{2D}{\mu_0 M_s}\left[\frac{\partial m_z}{\partial x}\vec{u}_x + \frac{\partial m_z}{\partial y}\vec{u}_y - \left(\frac{\partial m_x}{\partial x} + \frac{\partial m_y}{\partial y}\right)\vec{u}_z\right] \quad (22)$$

and the variational calculus shows that DMI also affects the system boundary conditions. In fact, the surface integral, including the exchange contribution, now reads like [26,49]

$$\oint_{S'} \left[2A\frac{\partial \vec{m}}{\partial n} + D[\vec{m} \times (\vec{n} \times \vec{u}_z)]\right] \cdot \delta\vec{m}\, ds' \quad (23)$$

which must be zero for any arbitrary variation of $\vec{m}$. Hence, the boundary conditions are given by

$$\frac{\partial \vec{m}}{\partial n} = \frac{D}{2A}[\vec{m} \times (\vec{n} \times \vec{u}_z)], \forall \vec{r} \in S' \quad (24)$$

By combining these results with Equation (13), it is possible to see that the effect of DMI is the addition of the $\vec{H}_{DM}$ field to the previously defined effective field $\vec{H}_{eff}$, and the modification of the boundary conditions according to Equation (17) and (24). In summary, the energy density of a ferromagnetic specimen is

$$\epsilon = \epsilon_{exch} + \epsilon_{ani} + \epsilon_{dmg} + \epsilon_{ext} + \epsilon_{DM}$$
$$= A(\nabla \vec{m})^2 + K_u[1 - (\vec{m} \cdot \vec{u}_k)^2] - \frac{1}{2}\mu_0 M_s \vec{H}_{dmg} \cdot \vec{m} - \mu_0 M_s \vec{H}_{ext} \quad (25)$$
$$\cdot \vec{m} + D[m_z(\nabla \vec{m}) - (m \cdot \nabla)m_z]$$

and the general form of the effective field is

$$\vec{H}_{eff} = \frac{2A}{\mu_0 M_s}\nabla^2 \vec{m} + \frac{2K}{\mu_0 M_s}(\vec{m} \cdot \vec{u}_k)\vec{m} + \vec{H}_{dmg} + \vec{H}_{ext}$$
$$+ \frac{2D}{\mu_0 M_s}[\nabla m_z - (\nabla \cdot \vec{m})\vec{u}_z] \quad (26)$$

Note that other classical magnetic fields, such as the Oersted field generated by an electrical current, can also be included in these expressions.

## *2.2. Magnetization dynamics.*

As it was anticipated, the magnetization dynamics of a ferromagnetic sample under an external field can be described by the Landau-Lifshitz-Gilbert Equation (1). This equation can be generalized to include the torques generated by injection of electrical currents along the ferromagnetic strip ($\vec{J}_{FM}$) and/or along a heavy metal in contact with it ($\vec{J}_{HM}$). The general equation now reads

$$\frac{d\vec{m}}{dt} = -\gamma_0 \vec{m} \times \vec{H}_{eff} + \alpha \vec{m} \times \frac{d\vec{m}}{dt} + \vec{\tau}_{STT} + \vec{\tau}_{SOT} \qquad (27)$$

where $\vec{\tau}_{STT}$ and $\vec{\tau}_{SOT}$ represents the spin-transfer torques (STTs) and the spin-orbit torques (SOTs) respectively. When an electric current is injected along a conducting ferromagnetic strip ($\vec{J}_{FM} = J_{FM}\vec{u}_J$, where $J_{FM}$ is the magnitude of the current density and $\vec{u}_J$ the unit vector along the current direction), the spin of the conduction electrons becomes spin-polarized along the local magnetization direction. When this spin polarized current crosses a DW or any non-uniform magnetic texture, spin angular momentum is transferred from the current to the magnetization, thereby inducing a torque which causes the DW to be pushed in the direction of the electron flow. This spin-transfer torque phenomenon, which was first predicted by Berger [**5**], has adiabatic (A) and non-adiabatic (NA) contributions [**50-52**]:

$$\vec{\tau}_{STT} = \vec{\tau}_A + \vec{\tau}_{NA} = -u(\vec{u}_J \cdot \nabla)\vec{m} + \beta u\, \vec{m} \times (\vec{u}_J \cdot \nabla)\vec{m} \qquad (28)$$

where $u \equiv -\frac{|g|\mu_B P J_{FM}}{2|e|M_s}$, with $\mu_B$ being the Bohr magneton, $P$ the polarization factor and $e$ the electric charge of the electron. $\beta$ is the dimensionless non-adiabatic parameter. The adiabatic STT ($\vec{\tau}_A = u(\vec{u}_J \cdot \nabla)\vec{m}$) represents the transfer of angular momentum between conduction electrons and the local magnetization, namely, it assumes that current polarization follows adiabatically the local magnetization, thus exchanging its total angular momentum. It is expected to be dominant in wide DWs. It acts as a hard-axis field perpendicular to the magnetization inside the DW and controls the initial DW velocity. The non-adiabatic STT ($\vec{\tau}_{NA} = \beta u \vec{m} \times (\vec{u}_J \cdot \nabla)\vec{m}$) represents the linear momentum transfer, which considers the transfer of linear momentum between conduction electrons and the local magnetization and it accounts for the fact that electron polarization might not entirely follow the local magnetization. The non-adiabatic STT is expected to be dominant in thin DWs. It mimics an easy axis magnetic field, and it is responsible for the terminal DW velocity as it will be discussed later.

Additionally, spin-orbit torques (SOTs) are a generic term for spin transport phenomena that can occur in systems with high spin-orbit coupling. In particular, these SOTs emerge as the electrical current is injected along a HM/FM stack. They can be classified in two types of torques: damping-like ($\vec{\tau}_{DL}$, also called Slonczewski-like) and field-like ($\vec{\tau}_{FL}$) SOTs [21-23,42-44], and they are essentially related to the spin Hall effect and/or to the Rashba effects.

The spin Hall effect (SHE) was first proposed by Dyakonov and Perel although its current name was introduced later by Hirsch [40]. It describes how the spins accumulate at the interface of certain materials, due to the spin-dependent deflection of the electrons in the transverse directions to the current flow direction. When a ferromagnetic strip is present on top of such a spin Hall material (a heavy metal, HM), the vertical current (polarized in the transverse $y$-direction) can be injected into it. The maximum perpendicular spin current $J_s$ induced by a longitudinal charge current density $J_{HM}$ is characterized by the spin Hall angle, $J_s = \theta_{SH} J_{HM}$. This spin current is injected into and absorbed by the ferromagnetic layer (FM), exerting a torque on the magnetization because of the angular momentum conservation. A similar process was firstly studied by Slonczewski [17], who used a perpendicular charge current polarized by a reference magnetic layer. Apart from the different generation of the spin current, the torques are identical, and therefore it is usually called Slonczewski-like or damping-like torque. Regarding the origin of the spin-dependent deflections that cause the SHE, several mechanisms exist, and different mechanisms dominate in different HMs. This fact manifests itself in a different sign of the spin Hall angle, as for example in Pt and Ta: $\theta_{SH}(Pt) > 0$ and $\theta_{SH}(Ta) < 0$.

Another SOT originates from the Rashba effect [53-56]. This effect can arise when there is an electric field gradient due to the symmetry breaking at an interface. Although first identified in semiconductors, the effect could also arise in metallic ferromagnets, particularly in a system consisting of a metal with large spin-orbit coupling, a ferromagnetic layer and an oxide layer. Due to a symmetry crystal field potential at the interface between two different materials, electric fields $\vec{E}$ can build up at the interfaces when the lateral current flows through the multilayer. In a classical picture, these electric fields transform to magnetic fields $\vec{H}_R$ in the rest frame of the electrons flowing at the interfaces. This effective field couples to the magnetic moment of the adjacent magnetic layers through the *s-d* exchange interaction. This can be modeled simply as an effective magnetic field in the $y$-direction, transverse to the current flow. Key ingredients to achieve a significant Rashba effect are a strong spin-orbit coupling (characterized by the Rashba parameter $\alpha_R$), a strong effective electric field at the interfaces, and a strong structural inversion asymmetry (SIA), *i.e.* asymmetry between the top and bottom interfaces. In addition to this effective field, a secondary SOT due to the Rashba effect was predicted to occur, perpendicular to the first. This non-adiabatic contribution to the Rashba effect can be caused by spin diffusion inside the ferromagnetic layer.

Having introduced the sources of spin orbit torques, we now describe how they enter in the LLG Equation that governs the magnetization dynamics. The SHE acts as a transverse-polarized spin current that is injected in the magnetic layer, and therefore, it has the form of a Slonczewski-like torque (or damping-like, DL),

$$\vec{\tau}_{SL} = -\gamma_0 H_{SL} \vec{m} \times (\vec{m} \times \vec{\sigma}) \quad (29)$$

where $H_{SL} = \frac{\hbar \theta_{SH} J_{HM}}{2|e|\mu_0 M_s t}$ with $\hbar$ being the Planck constant, and $\theta_{SH}$ the spin Hall angle, which determines the ratio between the electric current and the spin current ($J_s = \theta_{SH} J_{HM}$). $\vec{\sigma} = \vec{u}_J \times \vec{u}_z$ is the unit vector along the polarization direction of the spin current generated by the SHE in the HM. $\vec{\sigma}$ is orthogonal to both the direction of the electric current $\vec{u}_J$ and the vector $\vec{u}_z$ standing for the normal to the HM/FM interface. $t$ is the thickness of the ferromagnetic layer. For a longitudinal current along the HM ($\vec{J}_{HM} = J_{HM} \vec{u}_J$, $\vec{u}_J = \vec{u}_x$), the spin current is polarized along the transverse direction, $\vec{\sigma} = -\vec{u}_y$. Note that $\vec{u}_J$ does not include the sign of the current flow, which is taken into account in the sign of $J_{HM}$.

Differently, the Rashba effect acts as an effective field along the transverse $y$-direction, and although there could exist other physical origin for such a transverse field, the torque it exerts is often called the field-like (FL) torque, and it has the form [56]

$$\vec{\tau}_{FL} = -\gamma_0 H_{FL}(\vec{m} \times \vec{\sigma}) \quad (30)$$

where $H_{FL} = \frac{\alpha_R P J_{FM}}{\mu_B M_s}$, with $\alpha_R$ being the mentioned Rashba parameter and $P$ is the spin polarization in the ferromagnetic layer. As for the DL-SOT, $\vec{\sigma} = -\vec{u}_y$ for a longitudinal current.

Since the two effects take different form in the LLG Equation, it is possible in principle to separate the two contributions by measurements of the magnetization dynamics. However, several theoretical works have suggested that both the SHE and the Rashba effect could have a 'non-adiabatic' counterpart. This makes it less straightforward to distinguish both, since the non-adiabatic SHE torque has the same FL form as the direct Rashba torque, and the non-adiabatic Rashba torque takes the DL form as the SHE. However, nowadays most experiments suggest that the main driving force for DWs in HM/FM multilayers is due to the SHE, and that the Rashba effect plays a minor role. Consequently, in this chapter we will assume that the SOTs that enter in the LLG Equation have the following form

$$\vec{\tau}_{SOT} = \vec{\tau}_{DL} + \vec{\tau}_{FL} = -\gamma_0 H_{SL} \vec{m} \times (\vec{m} \times \vec{\sigma}) - \gamma_0 H_{FL}(\vec{m} \times \vec{\sigma}) \quad (31)$$

where $H_{SL}$ and $H_{FL}$ represent the magnitude of the DL-SOT and the FL-SOT respectively as defined above.

## 2.3. Numerical details.

Most of the micromagnetic results presented in the following sections were performed using Mumax3 [8]. This is an open-source micromagnetic software, which runs on Graphic Processing Units (GPUs). As we will study different systems, the corresponding magnetic parameters and dimensions will be given where corresponds in Sec. 3. The micromagnetic results presented in Sec. 3.8 were obtained by using a homemade micromagnetic solver [9].

The modeling is done by using computational cells well below than the characteristic length along which the magnetization changes spatially. Further details on numerical methods to find equilibrium states and solve the magnetization dynamics can be consulted in [8,9].

## 3. One-dimensional models (1DMs). Micromagnetic and 1DM results.

Once the micromagnetic equations in **Sec. 2** have been established, here we are going to introduce the one-dimensional model (1DM). In parallel, we present different micromagnetic results, and we compare them to the 1DM predictions. In the context of the DW motion, a useful tool is the Collective-Coordinates model (CC), which is also called the one-dimensional model (1DM). It rearranges the LLG Equation (27) in terms of the DW position $q$ and the internal DW angle $\psi$, and it implies an assumption on the magnetization profile, which is considered to follow a one-dimensional DW solution [**57-58**]. In developing the 1DM, it is convenient to write the LLG Equation (27) in spherical coordinates. The magnetization is written as $\vec{m} = (\sin\theta\cos\phi, \sin\theta\sin\phi, \cos\theta)$, where the angles $\theta$ and $\phi$ are defined in Figure 1(a). The dimensions of the single ferromagnetic (FM) strip and the FM strip on top of a heavy metal (HM) are also shown in Figure 1(b) and (c) respectively.

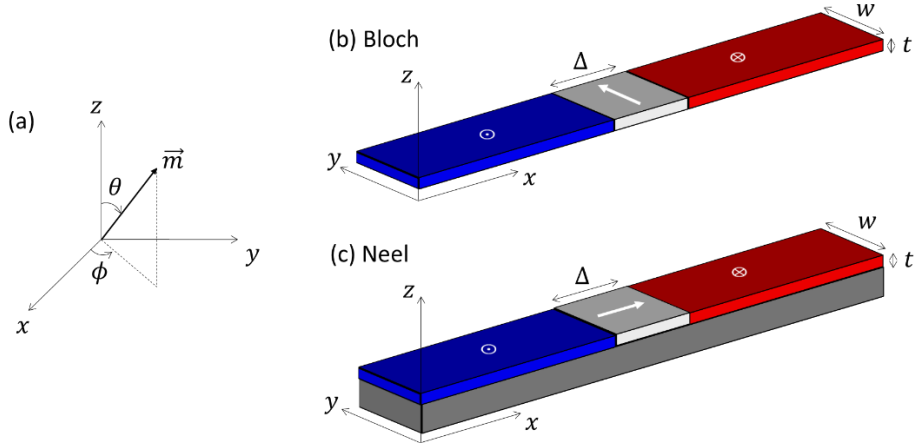

**Figure 1**. (a) Definition of the angles $\theta$ and $\phi$. (b) Scheme of a single FM layer with an *up-down* Bloch DW ($\vec{m}_{DW} = +\vec{u}_y$). (c) Scheme of a Neel DW ($\vec{m}_{DW} = +\vec{u}_x$) in a HM/FM bilayer, where the gray strip represents the HM. The cross section of the FM strips ($w \times t$) and the DW width ($\Delta$) are also defined.

### *3.1. Derivation of the 1DM equations*

In spherical coordinates ($d\vec{m} = d\theta \vec{u}_\theta + \sin\theta\, d\phi\, \vec{u}_\phi$, $\frac{d\vec{m}}{dt} = \frac{d\theta}{dt}\vec{u}_\theta + \sin\theta \frac{d\phi}{dt}\vec{u}_\phi = \dot{\theta}\vec{u}_\theta + \sin\theta\, \dot{\phi}\, \vec{u}_\phi$, $\frac{\partial \vec{m}}{\partial x} = \frac{\partial \theta}{\partial x}\vec{u}_\theta + \sin\theta \frac{\partial \phi}{\partial x}\vec{u}_\phi$), the LLG Equation (27) including the STTs and the SOTs reads as

$$\dot{\theta} = -\frac{\gamma_0}{\mu_0 M_s \sin\theta}\frac{\delta\epsilon}{\delta\phi} - \alpha \sin\theta\, \dot{\phi} - u\left(\frac{\partial \theta}{\partial x}\right) - \gamma_0 H_{FL}\cos\phi - \gamma_0 H_{SL}\cos\theta \sin\phi \quad (32)$$

$$\sin\theta\, \dot{\phi} = \frac{\gamma_0}{\mu_0 M_s}\frac{\delta\epsilon}{\delta\theta} + \alpha\dot{\theta} + \beta u\left(\frac{\partial \theta}{\partial x}\right) + \gamma_0 H_{FL}\cos\theta \sin\phi - \gamma_0 H_{SL}\cos\phi \quad (33)$$

where we have assumed that $\frac{\partial \phi}{\partial x} = 0$, as it is the case with the 1DM approach, $\epsilon$ is the energy density of the system (see Equation (25)), and $\vec{H}_{eff} = -\frac{1}{\mu_0 M_s}\frac{\delta\epsilon}{\delta\theta}\vec{u}_\theta - \frac{1}{\mu_0 M_s \sin\theta}\frac{\delta\epsilon}{\delta\phi}\vec{u}_\phi$ is the effective field. From (27), (28) and (30),

$$\frac{\delta\epsilon}{\delta\phi} = -\frac{\mu_0 M_s \sin\theta}{\gamma_0}\Big\{\dot{\theta} + \alpha \sin\theta\, \dot{\phi} + u\left(\frac{\partial \theta}{\partial x}\right) + \gamma_0 H_{FL}\cos\phi \\ + \gamma_0 H_{SL}\cos\theta \sin\phi\Big\} \quad (34)$$

$$\frac{\delta\epsilon}{\delta\theta} = \frac{\mu_0 M_s}{\gamma_0}\Big\{\sin\theta\, \dot{\phi} - \alpha\dot{\theta} - \beta u\left(\frac{\partial \theta}{\partial x}\right) - \gamma_0 H_{FL}\cos\theta \sin\phi \\ + \gamma_0 H_{SL}\cos\phi\Big\} \quad (35)$$

The different contributions to the energy density in polar coordinates are expressed as

$$\epsilon_{exch} = A(\nabla\vec{m})^2 = A[(\nabla\theta)^2 + \sin^2\theta\, (\nabla\phi)^2] \quad (36)$$

$$\epsilon_{ani} = K_u(1 - (\vec{m}\cdot\vec{u}_k)^2) = K_u \sin^2\theta \quad (37)$$

$$\begin{aligned}\epsilon_{dmg} &= -\frac{1}{2}\mu_0 M_s \vec{H}_{dmg}\cdot\vec{m} = -\frac{1}{2}\mu_0 M_s^2 \vec{m}\cdot\bar{\bar{N}}\vec{m} \\ &= -\frac{1}{2}\mu_0 M_s^2\left(N_x m_x^2 + N_y m_y^2 + N_z m_z^2\right) \\ &= -\frac{1}{2}\mu_0 M_s^2\left[N_z + (N_x - N_z)\sin^2\theta + (N_y - N_x)\sin^2\theta \sin^2\phi\right]\end{aligned} \quad (38)$$

$$\epsilon_{ext} = -\mu_0 M_s \vec{H}_{ext}\cdot\vec{m} = -\mu_0 M_s(H_x \sin\theta \cos\phi + H_y \sin\theta \sin\phi + H_z \cos\theta) \quad (39)$$

$$\begin{aligned}\epsilon_{DM} &= D[m_z(\nabla\vec{m}) - (m\cdot\nabla)m_z] \\ &= D\left[\cos\phi \frac{\partial \theta}{\partial x} + \sin\phi \frac{\partial \theta}{\partial y} + \sin\theta \cos\theta\left(\sin\phi \frac{\partial \phi}{\partial x} - \cos\phi \frac{\partial \phi}{\partial y}\right)\right]\end{aligned} \quad (40)$$

In calculating these terms, we have considered a demagnetizing field $\vec{H}_{dmg} = M_s \bar{\bar{N}} \vec{m}$ where $\bar{\bar{N}}$ is the demagnetizing tensor, which is assumed to be diagonal [59,60]. PMA and demagnetizing terms are usually combined in an effective anisotropy energy density, which reads as

$$\epsilon_{ani} + \epsilon_{dmg} = \left[K_u + \frac{1}{2}\mu_0 M_s^2 (N_x - N_z)\right] \sin^2 \theta \\
+ \left[\frac{1}{2}\mu_0 M_s^2 (N_y - N_x)\right] \sin^2 \theta \sin^2 \phi \\
= K_{eff} \sin^2 \theta + K_{sh} \sin^2 \theta \sin^2 \phi \quad (41)$$

where the constant $\frac{1}{2}\mu_0 M_s^2 N_z$ has been omitted, since it does not affect the system. The demagnetizing factors $(N_x, N_y, N_z)$ depend on the geometry of the system and can be calculated analytically as a function of the DW width ($\Delta$), the strip width ($w$) and the strip thickness ($t$) [57,58]. For the ultrathin films considered here $N_z \sim 1$ ($t \ll w$), and $N_z \gg N_x, N_y$. The effective out-of-plane anisotropy constant ($K_{eff} = K_u + \frac{1}{2}\mu_0 M_s^2 (N_x - N_z)$) reduces to $K_{eff} \approx K_u - \frac{1}{2}\mu_0 M_s^2 N_z$. Besides, the shape anisotropy constant $K_{sh} = \frac{1}{2}\mu_0 M_s^2 (N_y - N_x)$ determines the stable DW configuration in the absence of DMI, depending on the sign of $(N_y - N_x)$. If $N_y > N_x$, $K_{sh} > 0$ and the DW shape anisotropy favors Néel DWs. On the contrary, if $N_y < N_x$, $K_{sh} < 0$ and Bloch DWs are favored. For the samples studied here $w \gg \Delta$, and therefore, $N_x \gg N_y$. In this case, the shape anisotropy reduces to $K_{sh} \approx -\frac{1}{2}\mu_0 M_s^2 N_x$ which is negative, and therefore, favors the Bloch DW configuration in the absence of DMI. We will come back to the discussion of the equilibrium DW configurations later on.

The magnetization profile of a DW in a system with out-of-plane magnetization can be described by the following ansatz [57,58],

$$\theta(x,t) = 2 \arctan\left\{\exp\left[Q\left(\frac{x - q(t)}{\Delta}\right)\right]\right\} \quad (42)$$

$$\phi(x,t) = \psi(t) \quad (43)$$

where $q$ is the DW position along the longitudinal $x$-axis, $\psi$ is the internal DW angle with respect to the longitudinal $x$-axis, and $\Delta$ is the DW width. The parameter $Q$ selects if the domain configuration is *up-down* ($Q = +1$) or *down-up* ($Q = -1$) from left to right along the longitudinal $x$-axis (this convention will be used for all studies of DWs along straight strips). Expressions (42) and (43) only depend on the $x$ coordinate along the nanostrip. In this sense, the model is a 1D model since the variation of the magnetization along the transverse ($y$) and out-of-plane ($z$) directions are neglected. Figure 2 shows the vector representation of the Bloch (a) and Néel (b) DWs and the corresponding dependence of the

Cartesian components of the magnetization along the longitudinal axis (c)-(d), (Equation (42) and (43))

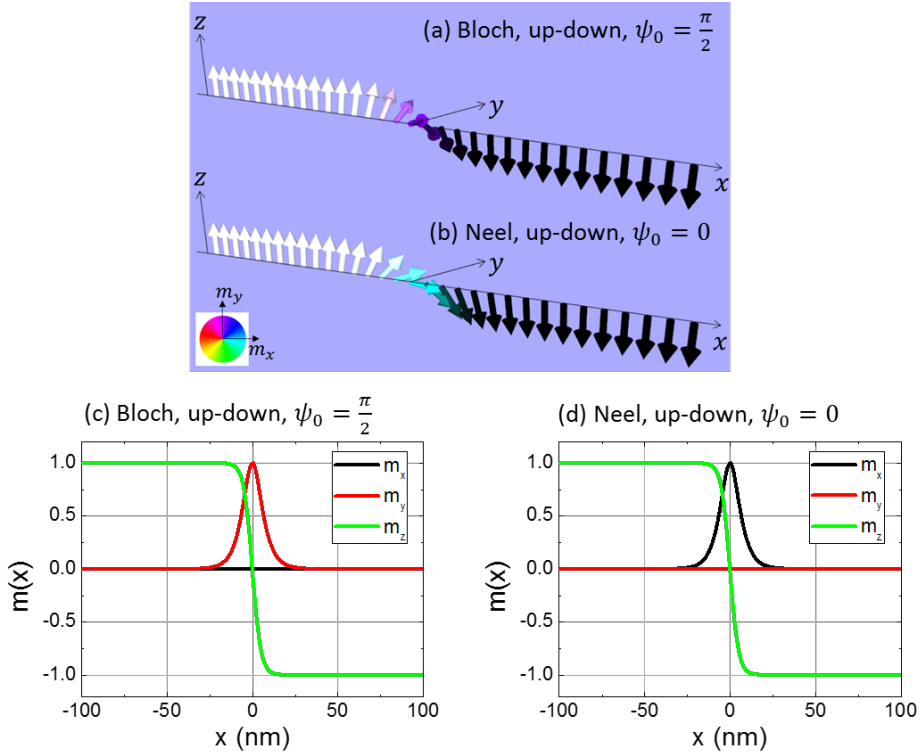

**Figure 2**. (a) and (b) show the vector representation of the Bloch ($\psi_0 = \pi/2$) and Neel ($\psi_0 = 0$) DW profiles along the $x$-axis. In (c) and (d) we plot the Cartesian components of the magnetization along the longitudinal $x$-axis as given by the Equation (42) and (43). The *up-down* DW configuration ($Q = +1$) is shown in all cases. The DW width parameter is $\Delta = 5$ nm.

By using this ansatz, we have that

$$\nabla\theta = \frac{\partial\theta}{\partial x} = +Q\frac{\sin\theta}{\Delta}, \quad \nabla\phi = \nabla\psi = 0 \tag{44}$$

$$\delta\theta = -Q\frac{\sin\theta}{\Delta}dq, \quad \delta\phi = \delta\psi = 0 \tag{45}$$

$$\dot\theta = -Q\frac{\sin\theta}{\Delta}\dot q, \quad \dot\phi = \dot\psi = 0 \tag{46}$$

and the energy density becomes

$$\epsilon = A\frac{\sin^2\theta}{\Delta^2} + K_{eff}\sin^2\theta + K_{sh}\sin^2\theta\sin^2\phi + QD\cos\phi\left(\frac{\sin\theta}{\Delta}\right) \\ - \mu_0 M_s(H_x\sin\theta\cos\phi + H_y\sin\theta\sin\phi + H_z\cos\theta) \tag{47}$$

where we have used the fact that $Q^2 = 1$. By integrating $\epsilon$ along the $x$ direction, it is possible to obtain the DW surface energy density $\sigma$ (J m$^{-2}$) in terms of the DW coordinates $(q, \psi)$, namely

$$\sigma = \int_{-\infty}^{+\infty} \epsilon dx = \frac{2A}{\Delta} + 2\Delta(K_{eff} + K_{sh} \sin^2 \psi) + \pi Q D \cos \psi \qquad (48)$$
$$- \mu_0 M_s \pi \Delta (H_x \cos \psi + H_y \sin \psi) - 2Q\mu_0 M_s q H_z$$

where we have used $\int_{-\infty}^{+\infty} \sin^2 \theta \, dx = 2\Delta$, $\int_{-\infty}^{+\infty} \cos \theta \, dx = 2Qq$. Hence, the variation of the DW energy density, in terms of the DW coordinates, reads as

$$\delta\sigma = \frac{\partial \sigma}{\partial q} \delta q + \frac{\partial \sigma}{\partial \psi} \delta \psi \qquad (49)$$

where

$$\frac{\partial \sigma}{\partial q} = -2Q\mu_0 M_s H_z \qquad (50)$$

$$\frac{\partial \sigma}{\partial \psi} = 4\Delta K_{sh} \sin \psi \cos \psi - \pi Q D \sin \psi - \mu_0 M_s \pi \Delta (H_y \cos \psi - H_x \sin \psi) \qquad (51)$$

At the same time, by using the LLG Equation (32) and (33), we have

$$\delta\epsilon = \frac{\partial \epsilon}{\partial \theta} \delta\theta + \frac{\partial \epsilon}{\partial \phi} \delta\phi$$
$$= \frac{\mu_0 M_s}{\gamma_0} \left\{ \left[ \sin\theta \, \dot\phi - \alpha\dot\theta - \beta u \left(\frac{\partial \theta}{\partial x}\right) - \gamma_0 H_{FL} \cos\theta \sin\phi \right. \right.$$
$$\left. + \gamma_0 H_{SL} \cos\phi \right] \delta\theta \right\} \qquad (52)$$
$$- \frac{\mu_0 M_s \sin\theta}{\gamma_0} \left\{ \left[ \dot\theta + \alpha \sin\theta \, \dot\phi + u \left(\frac{\partial \theta}{\partial x}\right) + \gamma_0 H_{FL} \cos\phi \right. \right.$$
$$\left. + \gamma_0 H_{SL} \cos\theta \sin\phi \right] \delta\phi \right\}$$

and integrating along the $x$-axis

$$d\sigma = \int_{-\infty}^{+\infty} \delta\epsilon \, dx$$
$$= 2\frac{\mu_0 M_s}{\gamma_0} \left\{ \left[ -Q\dot\psi - \alpha \frac{\dot q}{\Delta} + \frac{\beta u}{\Delta} - \gamma_0 \frac{\pi}{2} Q H_{SL} \cos\psi \right] dq \right. \qquad (53)$$
$$\left. + \left[ Q\dot q - \alpha \Delta \dot\psi - Qu - \gamma_0 \frac{\pi}{2} \Delta H_{FL} \cos\psi \right] d\psi \right\}$$

where we have used $\int_{-\infty}^{+\infty} \sin^2 \theta \, dx = 2\Delta$, $\int_{-\infty}^{+\infty} \sin \theta \, dx = \pi\Delta$, and $\int_{-\infty}^{+\infty} \sin \theta \cos \theta \, dx = 0$. From (53) and (50) and (51), we finally obtain

$$\left(\alpha\frac{\dot{q}}{\Delta}+Q\dot{\psi}\right)=\gamma_0 Q\left[H_z-\frac{\pi}{2}H_{SL}\cos\psi\right]+\beta\frac{u}{\Delta} \tag{54}$$

$$\left(-Q\frac{\dot{q}}{\Delta}+\alpha\dot{\psi}\right)=\gamma_0\left[-\frac{H_k}{2}\sin(2\psi)+\frac{\pi}{2}QH_D\sin\psi+\frac{\pi}{2}(H_y\cos\psi-H_x\sin\psi)\right.$$
$$\left.-\frac{\pi}{2}H_{FL}\cos\psi\right]-Q\frac{u}{\Delta} \tag{55}$$

where

$$H_D=\frac{D}{\mu_0 M_s \Delta} \tag{56}$$

$$H_k=\frac{2K_{sh}}{\mu_0 M_s}=M_s(N_y-N_x) \tag{57}$$

$$H_{SL}=\frac{\hbar\theta_{SH}J_{HM}}{2|e|\mu_0 M_s t} \tag{58}$$

$$u=-\frac{g\mu_B P J_{FM}}{2|e|M_s} \tag{59}$$

represent the magnitude of the DMI ($H_D$), the shape anisotropy ($H_k$), and the SOT ($H_{SL}$, $H_{FL}$) fields respectively, and $u$ is the constant related to the STT with units of m s$^{-1}$. Note that here we are interested in strips where $w \gg \Delta$. Therefore, $K_{sh}=\frac{1}{2}\mu_0 M_s^2(N_y-N_x)\approx -\frac{1}{2}\mu_0 M_s^2 N_x$ and $\frac{2K_{sh}}{\mu_0 M_s}\approx -N_x M_s$. In this context, we can write $\frac{2K_{sh}}{\mu_0 M_s}\approx -H_k$ where $H_k=N_x M_s$. Equations (53) and (54) describe the DW dynamics under an applied field $\vec{H}_{ext}=(H_x, H_y, H_z)$ and STTs and/or SOTs in the framework of the 1DM. They can be easily written in a different manner to separate $\dot{q}$ and $\dot{\psi}$. By defining

$$\Omega_A=\gamma_0 Q\left[H_z-\frac{\pi}{2}H_{SL}\cos\psi\right] \tag{60}$$

$$\Omega_B=\gamma_0\left[-\frac{H_k}{2}\sin(2\psi)+\frac{\pi}{2}QH_D\sin\psi+\frac{\pi}{2}(H_y\cos\psi-H_x\sin\psi)\right.$$
$$\left.-\frac{\pi}{2}H_{FL}\cos\psi\right] \tag{61}$$

we have

$$(1+\alpha^2)\frac{\dot{q}}{\Delta}=\alpha\Omega_A-Q\Omega_B+(1+\alpha\beta)\frac{u}{\Delta} \tag{62}$$

$$(1+\alpha^2)\dot{\psi}=Q\Omega_A+\alpha\Omega_B+Q(\beta-\alpha)\frac{u}{\Delta} \tag{63}$$

Before describing the field-driven DW dynamics and the current-driven DW dynamics, we here firstly review the DW static configurations.

## 3.2. DW static configurations

The DW surface energy density was derived in Equation (48). In the absence of external fields $\vec{H}_{ext} = (H_x, H_y, H_z) = 0$ and currents $\vec{J}_{FM} = \vec{J}_{HM} = 0$, it reduces to

$$\sigma = \frac{2A}{\Delta} + 2\Delta(K_{eff} + K_{sh} \sin^2 \psi) + \pi QD \cos \psi \tag{64}$$

The static DW configuration with $q_{eq} = 0$ and $\Delta_{eq} \equiv \Delta_0$, and $\psi_{eq} \equiv \psi_0$ must satisfy $\frac{\partial \sigma}{\partial \Delta} = 0$ and $\frac{\partial \sigma}{\partial \psi} = 0$, which yield respectively to

$$\Delta_0 = \sqrt{\frac{A}{K_{eff} + K_{sh} \sin^2 \psi_{eq}}} \tag{65}$$

$$\psi_0 = \begin{cases} 0 & |\pi D| > |4K_{sh}\Delta|, QD < 0 \\ \pi & |\pi D| > |4K_{sh}\Delta|, QD > 0 \\ \arccos\left(\frac{\pi QD}{4\Delta K_{sh}}\right) & |\pi D| < |4K_{sh}\Delta| \end{cases} \tag{66}$$

For ultrathin films with high PMA, $K_{eff} \gg K_{sh}$ and the DW width at rest is $\Delta_0 \approx \sqrt{\frac{A}{K_{eff}}}$. Regarding the equilibrium DW angle $\psi_0$ with respect to the longitudinal $x$-axis, we can distinguish two cases:

i) In the absence of DMI ($D = 0$), the condition $\frac{\partial \sigma}{\partial \psi} = 0$ means $4K_{sh}\Delta \cos \psi_0 \sin \psi_0 = 0$, and therefore, $\psi_0: 0, \pi$ (Néel) or $\psi_0: \frac{\pi}{2}, \frac{3\pi}{2}$ (Bloch) depending on the sign of $K_{sh} = \frac{1}{2}\mu_0 M_s^2 (N_y - N_x)$. Typically, for ultrathin films, $N_y < N_x$ and $K_{sh} < 0$, which favors Bloch DWs.

ii) On the other hand, the DMI ($D \neq 0$) favors chiral Néel DWs if $|\pi D| > |4K_{sh}\Delta|$. If $|\pi D| < |4K_{sh}\Delta|$, the equilibrium DW angle is between Bloch and Néel configurations. Note that the sign of $D$ and the DW configuration ($Q$) fix the chirality, as it was anticipated. The in-plane internal magnetic moments and equilibrium DW angle $\psi_0$ are plotted in Figure 3(a) and (b) respectively as a function of the DMI parameter for an *up-down* DW. A good agreement with micromagnetic results is achieved. See [61] for an alternative derivation of Equation (66) for thicker FM samples.

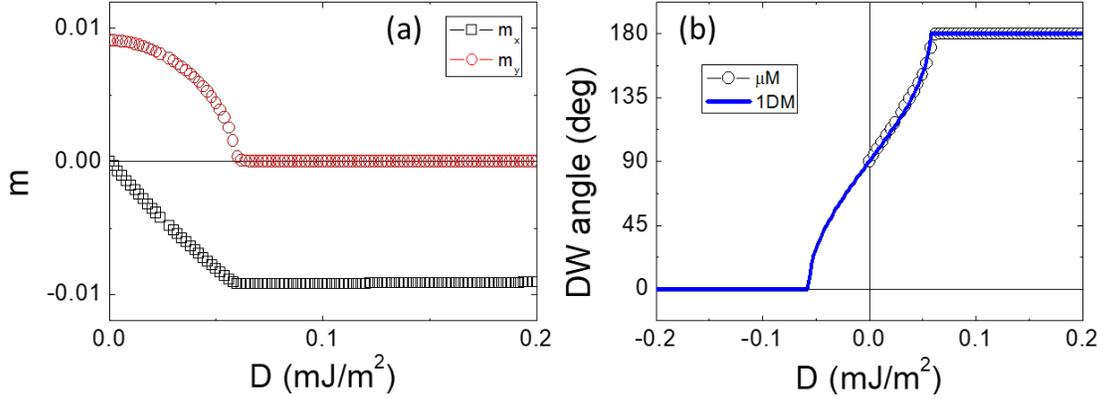

**Figure 3**. (a) Micromagnetically computed in-plane components of the internal DW moment ($m_x^{DW}$ and $m_y^{DW}$) at rest as a function of the DMI parameter ($D > 0$). (b) Internal DW angle ($\psi_{eq}$) at rest as a function of the DMI parameter. The following inputs were adopted: $A = 16$ pJ/m, $M_s = 0.8 \times 10^6$ A/m, $K_u = 0.8 \times 10^6$ J/m³, $K_{sh} = -7.0 \times 10^6$ J/m³, $\Delta_0 \approx \sqrt{\frac{A}{K_{eff}}}$ and $Q = +1$.

Figure 4 presents the micromagnetic equilibrium DW configurations for the left-handed ($D > 0$) and right-handed ($D < 0$) cases, when $|\pi D| > |4K_{sh}\Delta|$. In the left-handed case, the *up-down* DW has its internal magnetic moment aligned along the $-x$-axis ($\vec{m}_{DW} = -\vec{u}_x, \psi = \pi$), whereas in the *down-up* DW the internal moments points along the $+x$-axis ($\vec{m}_{DW} = +\vec{u}_x, \psi = 0$). The opposite occurs for the right-handed case. This naming can be understood looking to the palm of the corresponding hand: for instance, if the DW is left-handed, looking to the palm of the left hand with the four fingers pointing along the magnetization of the domain at the left side of the DW, the thumb will indicate the direction of the internal DW moment ($\vec{m}_{DW}$). Same criterium is valid for all *up-down* and *down-up* combinations. Obviously, for right-handed DWs, we must look to the palm of the right hand.

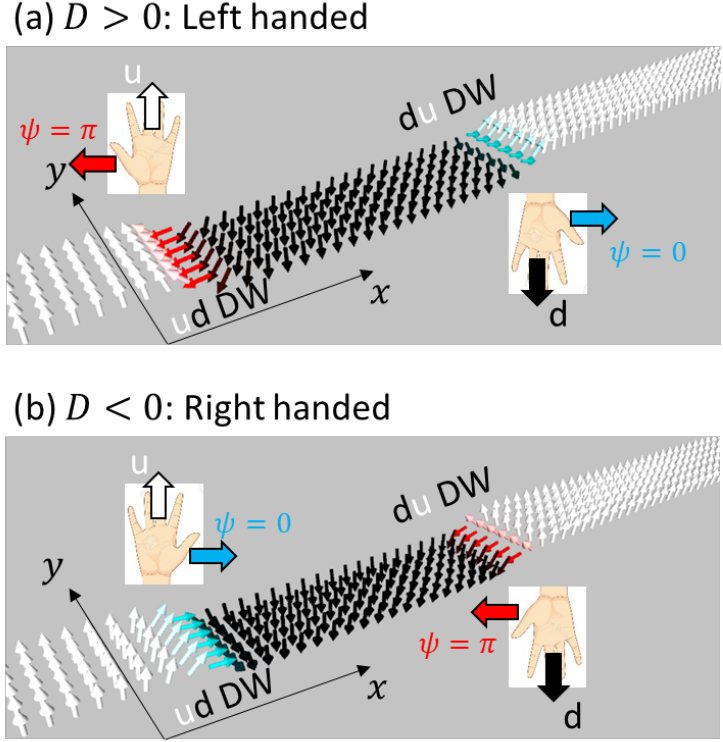

**Figure 4**. Typical micromagnetic snapshots of the DW states at rest for the left-handed ($D > 0$, (a)) and right-handed ($D < 0$ (b)) cases when $|\pi D| > |4K_{sh}\Delta|$.

## 3.3. Field-driven DW dynamics

Once the static DW configurations have been described, here we analyze the predictions of the 1DM for the field-driven DW dynamics, firstly in the absence of currents ($\vec{J}_{FM} = \vec{J}_{HM} = 0$). We consider an *up-down* DW ($Q = +1$). Under an out-of-plane field $\vec{H}_{ext} = H_z \vec{u}_z$ the domain aligned with the field will tend to expand while the domain opposite to the field will shrink, and DWs will move accordingly. The magnetization dynamics is driven by two kind of torques: the precessional torque, which leads to the spin precession around the effective field, and the dissipative torque (or damping torque), which leads to the spin alignment with the effective field. Consider, for instance, an out-of-plane strip with a Néel DW, whose internal magnetization points in the positive $x$ direction (see Figure 2(b) and (d)), and an external field along the positive $z$ direction. The precession torque $-\gamma_0 \vec{m}_{DW} \times H_z \vec{u}_z$ points in the positive $y$ direction leading to a DW rotation in the plane of the strip ($xy$). On the other hand, the dissipative torque $(-\alpha\gamma_0 \vec{m}_{DW} \times (\vec{m}_{DW} \times H_z \vec{u}_z)$ or $(\alpha \vec{m}_{DW} \times \left(\frac{d\vec{m}_{DW}}{dt}\right))$ points in the positive $z$ direction, leading to DW motion and expansion of the *up* domain. Note indeed that it is the damping torque that does not conserve the system energy and drives the system towards the energy minimum. The DW rotation further triggers the internal DW field, which includes the shape ($H_k$) and the DMI ($H_D$) fields, which both will try to restore the DW equilibrium configuration. Hence, the DW motion is characterized by two types of dynamics: DW precession and DW translation.

All these effects are enclosed in the 1DM Equations (54) and (55). We can look for a stationary solution for the DW angle ($\psi_s$). By imposing $\dot{\psi} = 0$, and $H_x = H_y = 0$ in Equations (54) and (55), we obtain

$$H_z = \alpha \left[\frac{H_k}{2}\sin(2\psi_s) - \frac{\pi}{2}QH_D \sin\psi_s\right] \tag{67}$$

In the absence of DMI ($D = 0$),

$$\sin(2\psi_s) = \frac{2H_z}{\alpha H_k} \tag{68}$$

where $\psi_s$ represents the dynamical equilibrium angle (or steady state), which is the result of compensation between $H_z$ and the shape anisotropy field ($H_k$). As a consequence, the DW moves at constant velocity, given by

$$v_s \equiv \dot{q} = \frac{\Delta \gamma_0}{\alpha} H_z \tag{69}$$

where the ratio $v_s/H_z$ defines the DW mobility, $\mu = \frac{\Delta \gamma_0}{\alpha}$. However, there is a limit for this rigid DW motion since $\sin(2\psi_s) < 1$, which implies that

$$|H_z| \leq \left|\frac{\alpha}{2}H_k\right| \tag{70}$$

This threshold field is called Walker breakdown field ($H_W = \left|\frac{\alpha}{2}H_k\right|$). For fields larger than the Walker breakdown, $|H_z| > \left|\frac{\alpha}{2}H_k\right|$, the precessional torque cannot be compensated by the internal DW field, and the internal DW angle $\psi$ steadily precesses during the DW motion. Due to this precessional motion, the DW velocity drops abruptly just above the Walker breakdown $H_W$. For driving fields high above Walker breakdown ($H_z \gg H_W$), the DW velocity scales again linearly with $H_z$, but the mobility is smaller than in the rigid regime.

In the presence of a strong DMI, that is, in the limit of $H_D \gg H_k$, the equilibrium angle is given by

$$\sin(2\psi_s) = \frac{2H_z}{\alpha \pi Q H_D} \tag{71}$$

In this case, the rigid DW velocity is still given by (69). However, the Walker field is now given by

$$H_W = \left|\frac{\alpha}{2}\pi Q H_D\right| \tag{72}$$

Since $H_D \gg H_k$, the DMI can lead to considerably faster DW motion by increasing the Walker field. In the presence of strong DMI the threshold internal DW angle before Walker breakdown is $\psi_W = \pi/2$, whereas in the absence of DMI, the limiting angle is $\psi_W = \pi/4$.

Figure 5 and 6 show the micromagnetic (mM) snapshots of the magnetization evolution under two different fields: $B_z = 1.5$ mT $< B_W$ and $B_z = 5$ mT $> B_W$ respectively. In both cases material parameters are: $A = 10$ pJ/m; $M_s = 0.3 \times 10^6$ A/m; $K_u = 0.2 \times 10^6$ J/m$^3$; $\alpha = 0.1$; and $\theta_{SH} = 0$. The strip has a $w \times t = 120 \times 3$ nm$^2$ cross section. For $B_z = 1.5$ mT $< B_W$ the DW moves rigidly, whereas for $B_z = 5$ mT $> B_W$ the internal DW moment precesses as the DW is displaced along the longitudinal strip axis. The temporal evolutions of the DW position predicted by the 1DM are presented in Figure 7(a) for the same fields. Figure 7(b) shows the average DW velocity as a function of the driving field $B_z$ predicted by the 1DM (lines) and the mM results (dots). A good agreement between both models is observed.

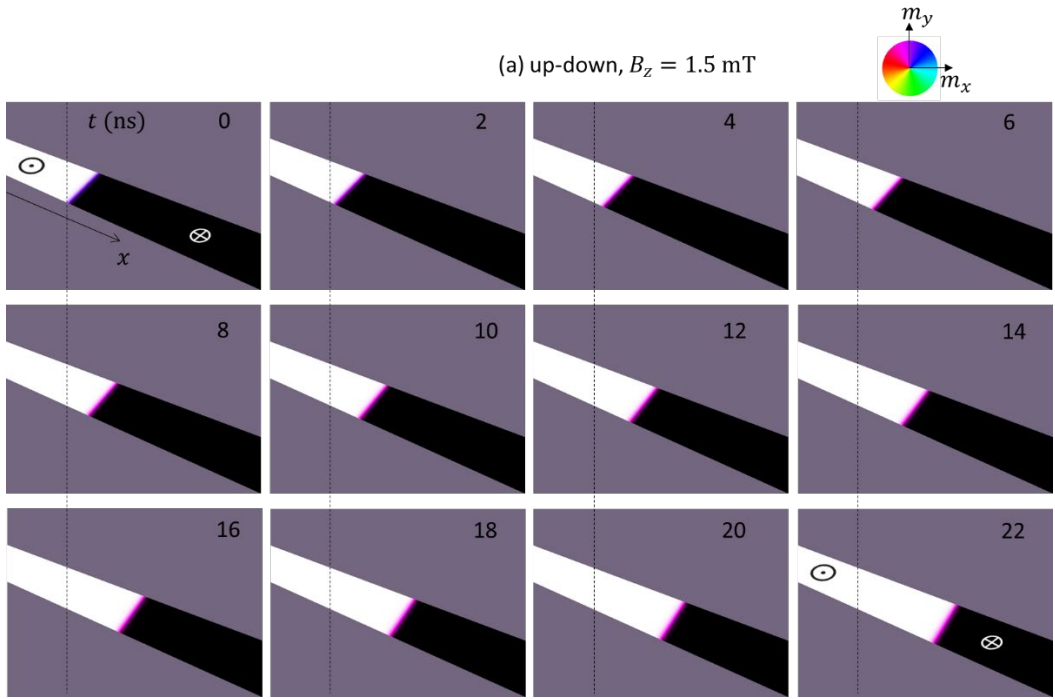

**Figure 5**. Micromagnetic snapshots showing the field-driven DW dynamics along a ferromagnetic strip with $D = 0$ for a static field smaller than the Walker field ($B_z = 1.5$ mT). The vertical dashed line indicates the initial DW position. Notice that the color representing the internal DW moment holds during the whole dynamics.

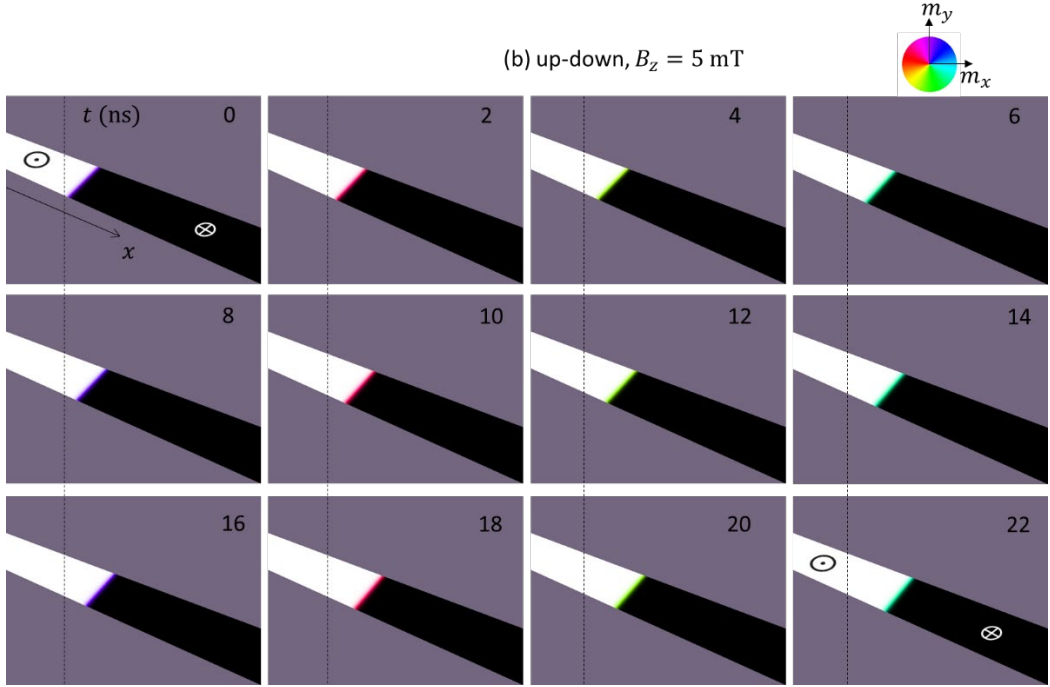

**Figure 6**. Same as Figure 5 but for a static field larger than the Walker field ($B_z = 5$ mT).

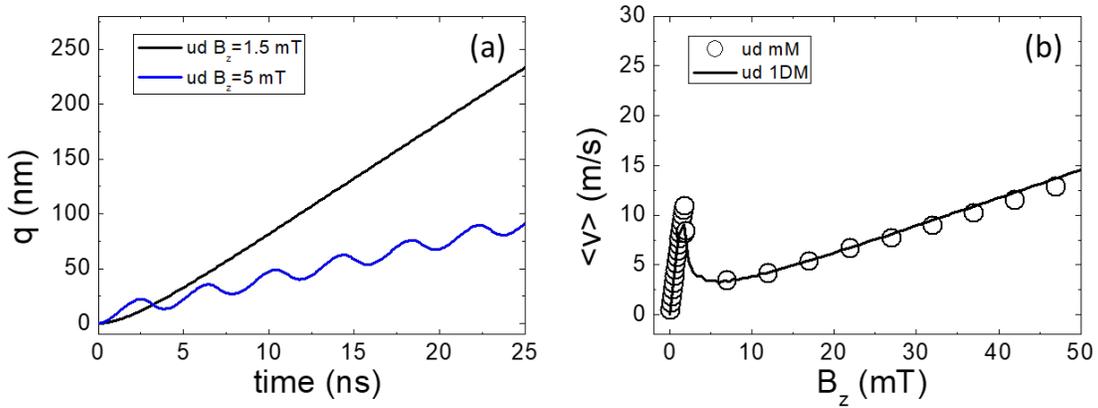

**Figure 7**. (a) Temporal evolution of the DW position ($q$) as predicted by the 1DM for two different driving fields below and above the Walker breakdown: $B_z = 1.5$ mT and $B_z = 5$ mT. (b) DW velocity as a function of the driving field $B_z$ predicted by both the 1DM (lines) and the compared to micromagnetic results (mM, dots). Parameters and dimensions for the mM results are the same as for Figure 5 and 6. For 1DM results, $H_k = 1.25 \times 10^4$ A/m and $\Delta = 8.5$ nm.

Figure 8(a) shows "$v_{DW}$ vs $B_z$" in a HM/FM stack with DMI ($D = 1$ mJ/m$^2$). Again, a good agreement is observed between the 1DM and the mM results. Figure 8(b) and (c) show the 1DM predictions for "$v_{DW}$ vs $B_z$" for different combinations of $D$ and $\alpha$. As predicted by the analytical calculations (Equation (72)), the Walker breakdown field increases with $D$ and $\alpha$.

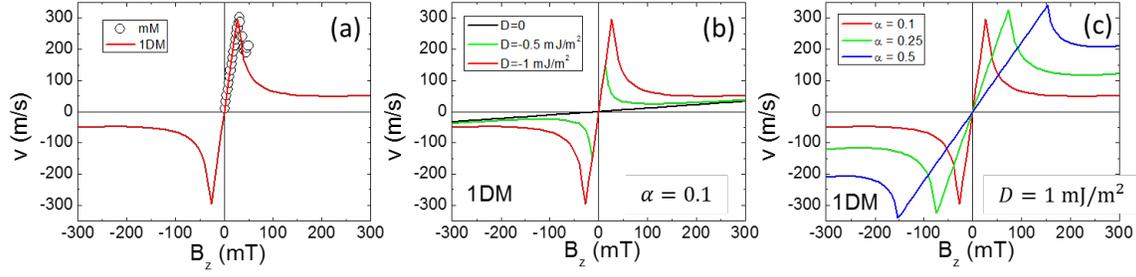

**Figure 8**. Field-driven DW dynamics in a HM/FM stack with DMI ($D = 1 \text{ mJ/m}^2$). (a) $v$ vs $B_z$. Here, the material parameters are $A = 16 \text{ pJ/m}$; $M_s = 0.8 \times 10^6 \text{A/m}$; $K_u = 0.8 \times 10^6 \text{ J/m}^3$; $\alpha = 0.1$; $P = \theta_{SH} = 0$, and the cross section $w \times t = 120 \times 0.6 \text{ nm}^2$. (b) 1DM results for different values of the DMI parameter and fixed damping ($\alpha = 0.1$). (c) 1DM results for different values of the damping ($\alpha$) and fixed DMI ($D = 1 \text{ mJ/m}^2$). The shape anisotropy field used in the 1DM is $H_k = -M_s N_x$ where $N_x = t \log 2 /(\pi \Delta)$ [60] and $\Delta = \sqrt{A/K_{eff}}$.

### *3.4. Current-driven DW dynamics under STTs*

As it was already mentioned in Sec. 2 (see Equation (27)-(31)), DWs can also be driven by electrical currents. When the electric current is injected along a conducting FM layer ($\vec{J}_{FM}$), it is usually called current-driven DW motion by STTs, which are related to the exchange interaction between the spin of the conduction electrons and the local magnetic moments of the FM strip. The main advantage over the field-driven DW motion is that neighboring DWs within a FM strip can be displaced in the same direction, which is along the electron flow (or against the current). Again, the DW dynamics is given by Equation (54) and (55). For the moment, we assume a single FM layer or no current injected along the HM, so $\vec{J}_{HM} = 0$ and only $\vec{J}_{FM} \neq 0$. In the perfect adiabatic case, $\beta = 0$, the DW does not move until a critical current density is reached, which can be calculated analogously to the Walker breakdown field by imposing $\dot{\psi} = 0$. For $\beta = \alpha$, the DW moves rigidly, independently of the current (infinite Walker breakdown). In any other case, that is, if $\beta \neq 0$ and $\beta \neq \alpha$, the current-driven DW motion presents rigid and turbulent regimes below and above a threshold Walker current density $J_{FM}^W$ respectively. In these cases, analogously to the field-driven dynamics, the dynamical equilibrium internal DW angle increases as $J_{FM}$ approaches to $J_W$. Note that for positive currents ($J_{FM} > 0$) the DW velocities are always negative, *i.e.* the DW moves along the electron flow. Both *up-down* and *down-up* DWs move with the same velocity for a given $J_{FM}$.

An example of a typical "$v_{DW}$ vs $J_{FM}$" curve is shown in Figure 9 for different values of the non-adiabatic parameter $\beta$. These are 1DM results, but full mM simulations (not shown here) are also in good quantitative agreement (see Ref. [62-64]). In the perfect adiabatic case ($\beta = 0$, black line in Figure 9), there is a threshold density current ($J_{FM}^W(\beta = 0)$) below which the DW stops after a short transient, and therefore no monotonous DW motion is achieved. Above this threshold, the DW moves by precessing similar to the field-driven case for fields larger than the Walker breakdown. In the non-adiabatic case ($\beta \neq 0$), the DW moves for any finite current, and it does it without changing its initial structure if the non-adiabatic parameter matches the damping ($\beta = \alpha$, green line in Figure 9). For any other case, there is a Walker breakdown density current $J_{FM}^W(\beta)$ above which the internal DW moment rotates periodically around the z-axis. Analytical expressions for the Walker current density ($J_{FM}^W(\beta)$) and for the DW velocity as a function of the injected current can be directly obtained from the 1DM. These expressions can be consulted, for instance, in [64,65].

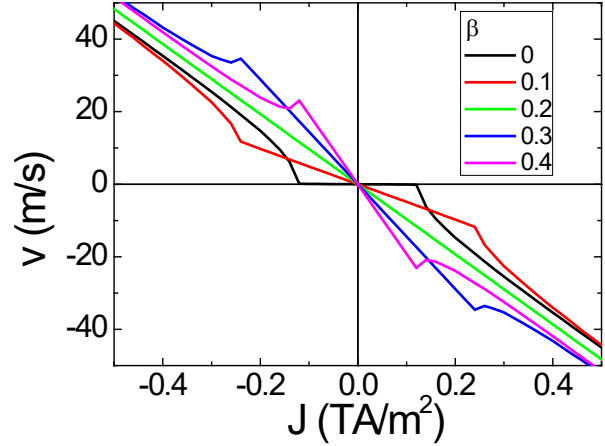

**Figure 9**. Current-driven DW dynamics in a single FM strip in the absence of DMI under STTs. Here the injected current flows through the FM layer and it is supposed to generate adiabatic and non-adiabatic STTs. The presented results correspond to 1DM predictions. The material parameters are $A = 10$ pJ/m; $M_s = 0.3 \times 10^6$ A/m; $K_u = 0.2 \times 10^6$ J/m$^3$; $\alpha = 0.2$, $P = 0.5$ and $\theta_{SH} = 0 = 0$. The strip has a $w \times t = 120 \times 3$ nm$^2$ cross section. The shape anisotropy field is $H_k = 1.25 \times 10^4$ A/m and $\Delta = 8.5$ nm.

### 3.5. Current-driven DW dynamics under SOTs

DWs can also be driven by SOTs. The dominant driving contribution due to a current along a HM ($\vec{J}_{HM}$) under the FM layer is the Slonczewski-like SOT (SL-SOT, or the damping-like DL-SOT). This SOT is typically due to the spin Hall effect (SHE) which takes place in a HM under the FM layer [62], as it was already discussed. Again, in this case, Equations (54) and (55) provide an explanation of this DW dynamics. Besides, the FL-SOT, represented by the $\vec{H}_{FL} = H_{FL}\vec{u}_y$, could also influence the DW dynamics. However, this FL-SOT, which enters in the 1DM Equation similarly to a transverse magnetic field $H_y$, is not the main driving force and it cannot drive DWs by itself [62].

In the absence of applied fields ($\vec{H}_{ext} = 0$), STTs ($P = J_{FM} = 0$) and FL-SOT ($k = 0$), the SL-SOT has no effect if the internal DW angle is already aligned along the spin-polarization ($\vec{\sigma} = -\vec{u}_y$). Therefore, Bloch DWs ($\psi = \pm\frac{\pi}{2}$) are not driven by the SL-SOT, because the DW needs to have a finite longitudinal component of the magnetization ($m_x \neq 0$) to be driven by the SL-SOT. Indeed, Néel DWs are efficiently driven by the SL-SOT. Depending on the current density along the HM, $J_{HM}$, the DW reaches a new equilibrium dynamical regime where the internal DW angle and the DW velocity are given by

$$\psi_s = \arctan\left(\frac{QH_{SL}}{\alpha H_D}\right) \tag{73}$$

$$v_{DW} \equiv \dot{q} = -\frac{\Delta}{\alpha}\gamma_0 Q \frac{\pi}{2} H_{SL} \cos \psi_s \tag{74}$$

The direction of the DW velocity depends on both the sign of the spin Hall angle $\theta_{SH}$ and the chirality of the DW imposed by the DMI parameter $D$. For a left-handed DW ($D > 0$) with $|\pi D| > |4K_{sh}\Delta|$, $\vec{m}_{DW} = \pm \vec{u}_x$ for $Q = \mp 1$, and therefore, $v_{DW} > 0$ for $\theta_{SH} > 0$ and $J_{HM} > 0$, so DWs move along the current direction ($\vec{u}_J$). As the current increases, the equilibrium angle $\psi_s$ approaches to $-\frac{\pi}{2}$ (which is also the direction of the spin polarization, $\vec{\sigma} = \vec{u}_J \times \vec{u}_z = -\vec{u}_y$), and the efficiency of the SHE is reduced. Eventually, the DW approaches asymptotically to a limiting velocity (see Figure 13(b)) which only depends on the DMI field, $H_D$, namely

$$v_{DW} \equiv \dot{q} = \gamma_0 \frac{\pi}{2} H_D \tag{75}$$

Before presenting the 1DM predictions, we show in Figure 10 the micromagnetic snapshots of the current-driven DW velocity in a HM/FM stack with $|D| = 1$ mJ/m². The rest of material parameters are $A = 16$ pJ/m; $M_s = 0.8 \times 10^6$ A/m; $K_u = 0.8 \times 10^6$ J/m³; $\alpha = 0.1$; $P = 0$, and $|\theta_{SH}| = 0.1$. The cross section $w \times t = 120 \times 0.6$ nm². Different combinations of the signs of the DMI parameter ($D$) and the spin Hall angle ($\theta_{SH}$) are shown: (a) $D > 0$ and $\theta_{SH} > 0$; (b) $D > 0$ and $\theta_{SH} < 0$; (c) $D < 0$ and $\theta_{SH} > 0$; and (d) $D < 0$ and $\theta_{SH} < 0$. $D > 0$ corresponds to left-handed DWs, whereas $D < 0$ corresponds to right-handed DWs [22,23]. Top graphs show the static equilibrium configuration at rest ($J_{HM} = 0$, $t = 0$) for *up-down* and *down-up* DWs simultaneously. The corresponding bottom graphs depict the micromagnetic snapshots under a current of $J_{HM} = 1$ TA/m² at $t = 0.5$ ns. Left-handed DWs ($D > 0$) move along the current flow if the spin Hall angle is positive ($\theta_{SH} > 0$, Figure 10(a)). On the contrary, right-handed DWs move with negative velocity ($v_{DW} < 0$) for the same current (Figure 10(c)). Note that the internal DW angle rotates towards the direction of the spin polarization, *i.e.* $\vec{\sigma} = -\vec{u}_y$. If the spin Hall angle is negative $\theta_{SH} < 0$, the opposite behavior is observed: the left-handed DWs move with $v_{DW} < 0$ (Figure 10(b)) whereas right handed DWs move with $v_{DW} > 0$ (Figure 10(d)). In this case ($\theta_{SH} < 0$), the rotation of the internal DW angle is towards the positive transverse direction, which is the direction of the spin accumulation, $-\vec{\sigma} = +\vec{u}_y$.

This convention allows us to describe the experimental observation that a left-handed DW in a FM strip on top of a Pt heavy metal ($\theta_{SH} > 0$) is driven along the current direction ($\vec{u}_J$). Although the choice of the sign of DMI parameter ($D$) is arbitrary, here we will adopt the convention that $D > 0$ corresponds to the Left-handed chirality. In such a case, and if $D$ is sufficiently high, an *up-down* DW has its internal magnetization pointing along the $-x$-axis, *i.e.* ⊙ ← ⊗, with the central arrow indicating the internal DW moment ($\psi = \pi$) and the circles the *up* (⊙) and *down* (⊗) symbols represent domains at the left and the right side of the DW. The corresponding left-handed *down-up* configuration will be therefore, ⊗ → ⊙ ($\psi = 0$). The right-handed chirality ($D < 0$) results in the opposite behavior: ⊙ → ⊗ and ⊗ ← ⊙, which will be driven along the electron flow (against the current, $-\vec{u}_J$) for $\theta_{SH} > 0$.

Apart from these observations, which are naturally explained by the presented 1DM model (see Equation (73), (74) and (75)), the micromagnetic snapshots also indicate that the DW plane, or the normal of the DW plane ($\vec{n}_{DW}$), also tilts (rotates) as DWs are driven by the current [66,67]. This DW tilting is a degree of freedom which is not taken into account in the already presented 1DM (Equation (54) and (55)), and therefore, it is convenient to extend the 1DM to account for this DW tilting. This is done in the following section.

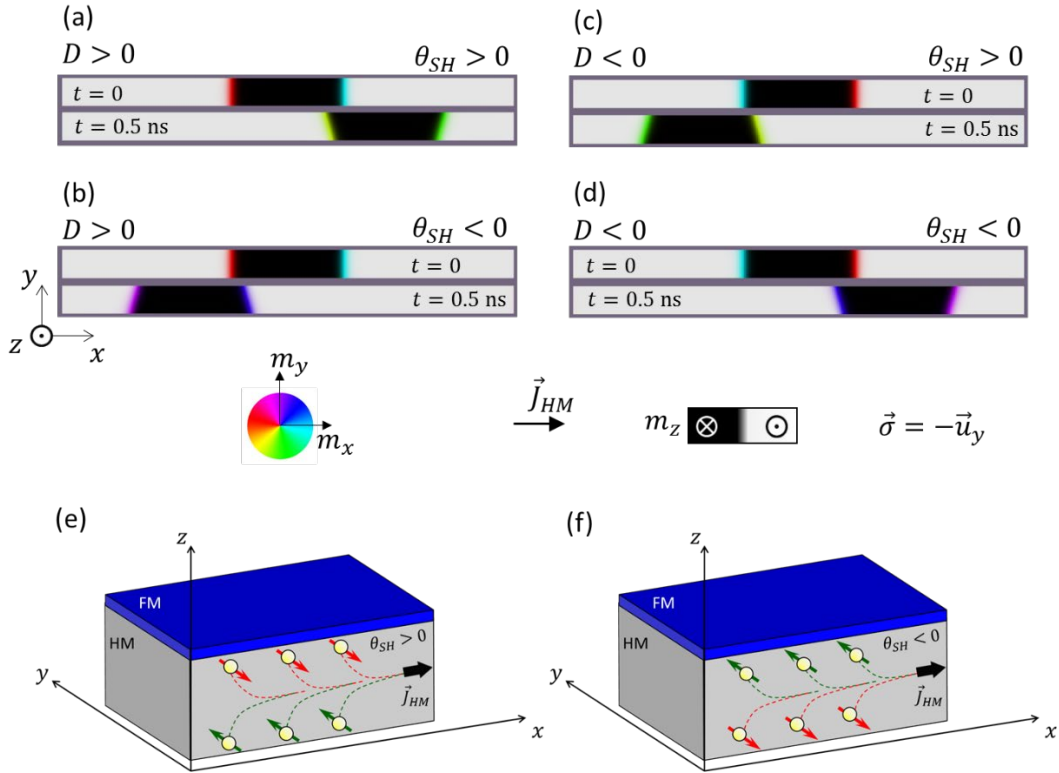

**Figure 10**. (a)-(d) Micromagnetic results of the current-driven DW dynamics in a HM/FM stack for different combinations of the DMI and the spin Hall angle parameters. The material parameters and the value of the injected current along the HM are given in the text. (e) and (f) show a schematic representation of the spin Hall effect in HM with $\theta_{SH} > 0$ and $\theta_{SH} < 0$ respectively.

### *3.6. Lagrangian formalism and 1DM equations including DW tilting.*

Here we add a new degree of freedom to the DW dynamics in the 1DM Equations, which is the rotation of the DW normal plane. To do it, here we also illustrate the derivation of the 1DM based on the Lagrangian formalism. Note that this formalism could alternatively be followed to obtain the same 1DM Equations (54) and (55) that we already presented in previous sections.

Our micromagnetic simulations of Figure 10 clearly shown that in asymmetric multilayers consisting of a FM layer sandwiched between a HM and an oxide with strong interfacial DMI, the DW normal ($\vec{n}_{DW}$) can rotate along with the internal DW angle ($\psi$) under external fields and/or currents. To consider this observation, we can extend the 1DM by including an additional tilting angle, $\chi$, which represents the orientation of the DW normal with respect to the longitudinal $x$-axis. The definition of the DW plane and the DW angles are given in the Figure 11.

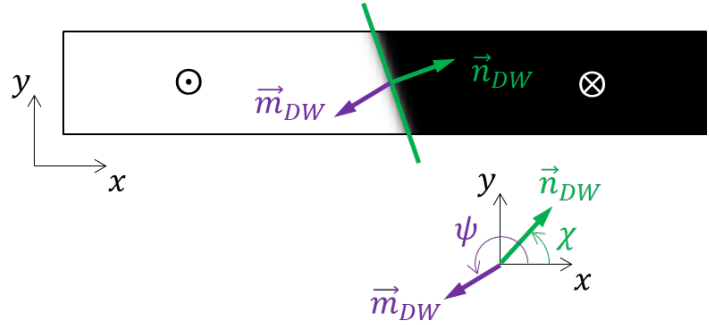

**Figure 11**. Definitions of the internal DW angle $\psi$ and the DW tilting angle $\chi$.

Including the DW tilting angle ($\chi$), the magnetization profile of a DW can be described by the following ansatz [**66,67**],

$$\theta(x,t) = 2\arctan\left\{\exp\left[Q\left(\frac{(x-q(t))\cos\chi + y\sin\chi}{\Delta}\right)\right]\right\} \quad (76)$$

$$\phi(x,t) = \psi(t) \quad (77)$$

We could also proceed anomalously as we did for the rigid 1DM case (no tilting) to derive the new 1DM equations from the LLG Equations (27) augmented with the STTs and SOTs. However, here we derive the 1DM Equations including the DW tilting by using the Lagrangian formalism. Indeed, the LLG Equation (27) in spherical coordinates can be derived from the following Lagrangian ($\mathcal{L}$, in [J/m³]) and Rayleigh dissipative ($\mathcal{F}$, in [J/(m³s)]) functions:

$$\mathcal{L} = \mathcal{K} + \epsilon \quad (78)$$

$$\mathcal{F} = \frac{\alpha\mu_0 M_s}{2\gamma_0}\left\{\frac{d\vec{m}}{dt} - \frac{\beta}{\alpha}u(\vec{u}_J \cdot \nabla)\vec{m} - \frac{\gamma_0}{\alpha}H_{SL}\vec{m}\times\vec{\sigma}\right\}^2 \quad (79)$$

where $\epsilon$ is again the energy density of the system and $\mathcal{K}$ the kinetic energy density (in [J m$^{-3}$]), which is given by

$$\mathcal{K} = \left[\frac{\mu_0 M_s}{\gamma_0} \varphi \sin\theta \, \dot{\theta} - \frac{\mu_0 M_s}{\gamma_0} \varphi \sin\theta \, u \, (\vec{u}_J \cdot \vec{\nabla})\theta\right] \tag{80}$$

The 1DM Equations can be deduced from the Euler-Lagrange-Rayleigh Equations:

$$\frac{\partial L}{\partial X} - \frac{d}{dt}\left(\frac{\partial L}{\partial \dot{X}}\right) + \frac{\partial F}{\partial \dot{X}} = 0 \tag{81}$$

where $X = \{q, \psi, \chi\}$. By using the ansatz (76) and (77) and integrating along the x-axis, we can obtain the Lagrangian ($L = \int \mathcal{L} dx$) and the dissipation function ($F = \int \mathcal{F} dx$) in terms of the DW coordinates $\{q, \psi, \chi\}$. The resulting Lagrangian ($L = \frac{1}{wt}\int \mathcal{L} dv$) and the dissipation function ($F = \frac{1}{wt}\int \mathcal{F} dv$) are expressed in terms of the DW coordinates $\{q, \psi, \chi\}$ as follows:

$$\begin{aligned} L = \frac{1}{wt}\int \mathcal{L} dv = &-2Q\frac{\mu_0 M_s}{\gamma_0}(\dot{q} + u)\psi \\ &+ \frac{1}{\cos\chi}\Big\{4\sqrt{AK_{eff} + 2\Delta K_{sh}\sin^2(\psi-\chi)} + Q\pi D\cos(\psi-\chi) \\ &- \mu_0 M_s \pi \Delta(H_x \cos\psi + H_y \sin\psi) + \mu_0 M_s \pi \Delta H_{FL}\sin\psi\Big\} \\ &- 2Q\mu_0 M_s q H_z \end{aligned} \tag{82}$$

$$\begin{aligned} F = \frac{1}{wt}\int \mathcal{F} dv &= \frac{1}{wt}\int \frac{\alpha \mu_0 M_s}{2\gamma_0}\left\{\frac{d\vec{m}}{dt} - \frac{\beta}{\alpha}u(\vec{u}_J \cdot \nabla)\vec{m} - \frac{\gamma_0}{\alpha}H_{SL}\vec{m}\times\vec{\sigma}\right\}^2 dv \\ &= \frac{\alpha\mu_0 M_s}{\gamma_0}\frac{\Delta}{\cos\chi}\left[\frac{\dot{q}^2}{\Delta^2}\cos^2\chi + \frac{\dot{\chi}^2}{12}\left(\pi^2\tan^2\chi + \frac{w^2}{\Delta^2}\frac{1}{\cos^2\chi}\right) + \dot{\psi}^2\right] \\ &+ \frac{\mu_0 M_s}{\gamma_0}2\beta u\frac{\dot{q}}{\Delta}\cos\chi + Q\mu_0 M_s \pi H_{SL}\dot{q}\cos\psi + \cdots \end{aligned} \tag{83}$$

where in the last equality of (83) we only write down the relevant terms, i.e., the ones depending on $\dot{q}$, $\dot{\psi}$ or $\dot{\chi}$. In the derivation of these expressions we have used the following identities: $\frac{\partial \theta}{\partial x} = Q\frac{\sin\theta}{\Delta}\cos\chi$, $\frac{\partial \theta}{\partial y} = Q\frac{\sin\theta}{\Delta}\sin\chi$, $\dot{\theta} = -Q\frac{\sin\theta}{\Delta}\dot{q}$, $\int_{-\infty}^{+\infty} d\theta = Q\pi$, $\int_{-\infty}^{+\infty}\sin^2\theta \, dx = \frac{2\Delta}{\cos\chi}$, $\int_{-\infty}^{+\infty}\cos\theta \, dx = 2Qq$, $\int_{-\infty}^{+\infty}\sin\theta \, dx = \frac{\pi\Delta}{\cos\chi}$, $\int_{-\infty}^{+\infty}\dot{\theta}\sin\theta \, dx = -2Q\dot{q}$, $\int_{-\infty}^{+\infty}\sin\theta\cos\theta \, dx = 0$, and $\int_{-\infty}^{+\infty}\dot{\theta}^2 dx = \frac{\Delta}{\cos\chi}\left[\frac{\dot{q}^2}{\Delta^2}\cos^2\chi + \frac{\dot{\chi}^2}{12}\left(\pi^2\tan^2\chi + \frac{w^2}{\Delta^2}\frac{1}{\cos^2\chi}\right) + \dot{\psi}^2\right]$.

The equations describing the DW dynamics in the framework of the 1DM can be directly deduced from the Euler-Lagrange-Rayleigh Equations (81). Including the tilting [67] the resulting equations are:

$$\left(\alpha \frac{\dot{q}}{\Delta} \cos \chi + Q \dot{\psi}\right) = \gamma_0 Q \left(H_z - \frac{\pi}{2} H_{SH} \cos \psi\right) - \beta \frac{u}{\Delta} \cos \chi \tag{84}$$

$$\begin{aligned}\left(-Q \frac{\dot{q}}{\Delta} \cos \chi + \alpha \dot{\psi}\right) \\ = \gamma_0 \Big[ -\frac{H_k}{2} \sin[2(\psi - \chi)] + \frac{\pi}{2} Q H_D \sin(\psi - \chi) \\ + \frac{\pi}{2}(H_y \cos \psi - H_x \sin \psi) - \frac{\pi}{2} H_{FL} \cos \psi \Big] + Q \frac{u}{\Delta} \cos \chi\end{aligned} \tag{85}$$

$$\begin{aligned}\alpha \frac{\pi^2}{12} \dot{\chi} \left(\tan^2 \chi + \frac{w^2}{\pi^2 \Delta^2} \frac{1}{\cos^2 \chi}\right) \\ = -\gamma_0 \frac{\sigma \sin \chi}{2\mu_0 M_s \Delta} + \gamma_0 \frac{H_k}{2} \sin[2(\psi - \chi)] - \gamma_0 \frac{\pi}{2} Q H_D \sin(\psi - \chi) \\ - \gamma_0 \frac{\pi}{2} H_{FL} \cos \psi \tan \chi\end{aligned} \tag{86}$$

where

$$\begin{aligned}\sigma = \frac{1}{\cos \chi} \Big\{ 4\sqrt{AK_{eff} + 2\Delta K_{sh} \sin^2(\psi - \chi)} + Q\pi D \cos(\psi - \chi) \\ - \mu_0 M_s \pi \Delta (H_x \cos \psi + H_y \sin \psi) \Big\}\end{aligned} \tag{87}$$

Note that here the STT parameter is defined as $u = g\mu_B P J_{FM}/(2|e|M_s)$, whereas in the derivation of the rigid 1DM Equations we used the notation $u = -g\mu_B P J_{FM}/(2|e|M_s)$. The rest of parameters are the same (see Equations (56)-(59)). As it can be verified, these Equations (84)-(87) reduce to the rigid 1DM Equations (54)-(55) in the absence of DW tilting ($\chi = 0$).

With the inclusion of the DW tilting in the 1DM, we can now provide a more accurate description of the micromagnetic results. To show it we fix the DW chirality to the left-handed case ($D > 0$) and consider $\theta_{SH} > 0$. The dependence of terminal values of the DW velocity ($v_{DW}$) and DW angles ($\psi, \chi$) on the current along the HM ($J_{HM}$) is shown in Figure 12 for both *up-down* ($Q = +1$) and *down-up* ($Q = -1$) DW configurations. The predictions of the 1DMs (red line: rigid 1DM; blue line: 1DM including tilting) are compared to the micromagnetic results (dots) for "$v_{DW}$ vs $J_{HM}$". Although the rigid 1DM already provides a good description, the 1DM including the tilting provides a more accurate quantitative description of the micromagnetic results.

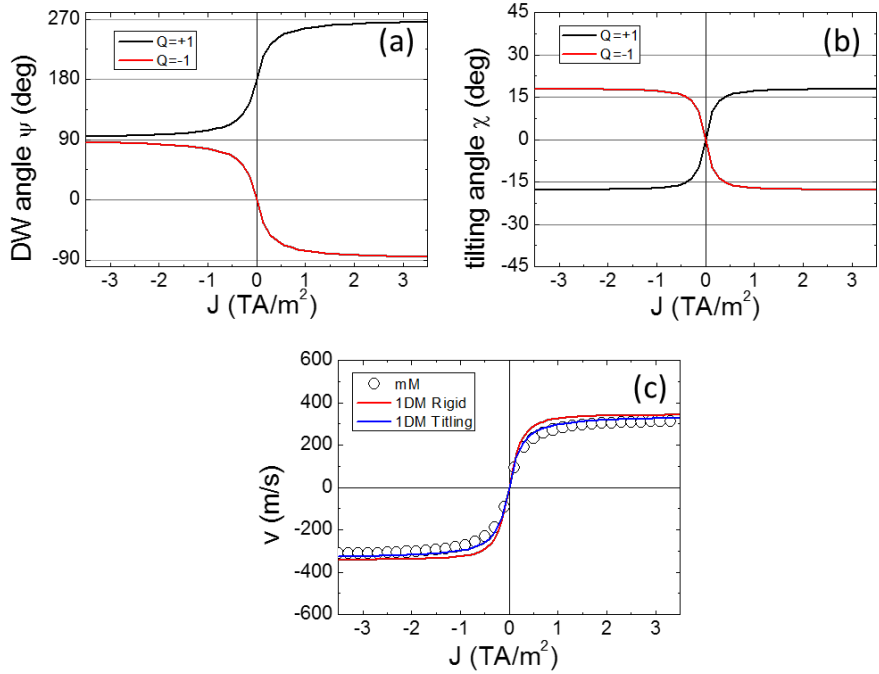

**Figure 12**. Current-driven DW dynamics in a HM/FM stack with DMI ($D = 1 \text{ mJ/m}^2$). Here, the material parameters are $A = 16 \text{ pJ/m}$; $M_s = 0.8 \times 10^6 \text{A/m}$; $K_u = 0.8 \times 10^6 \text{ J/m}^3$; $\alpha = 0.1$; $P = 0$, and $\theta_{SH} = 0.1$. The FM layer cross section is $w \times t = 120 \times 0.6 \text{ nm}^2$. (a) and (b) correspond to the predictions of the 1DM including the DW tilting for the terminal DW angles $\psi$ and $\chi$ as a function of $J_{HM}$ for *up-down* ($Q = +1$) and *down-up* ($Q = -1$) configurations. (c) shows the comparison of "$v_{DW}$ vs $J_{HM}$" between the micromagnetic model (dots) and the 1DMs, without (red line) and with (blue line) tilting. The shape anisotropy field used in the 1DMs is $H_k = -M_s N_x$ where $N_x = t \log 2 / (\pi \Delta)$ and $\Delta = \sqrt{A/K_{eff}}$.

The 1DM predictions of the velocity *vs* $J_{HM}$ for different combinations of the material parameters $\theta_{SH}$, $\alpha$, and $D$ are presented in Figure 13. For high currents, the velocity increases with $D$ (Figure 13(a)). In the linear regime, the mobility increases with $\theta_{SH}$ (Figure 13(b)) and decreases with $\alpha$ (Figure 13(c)). These results were already explained by the analytical expressions (72)-(75). Extensions of these expressions including the DW tilting can be consulted in [67].

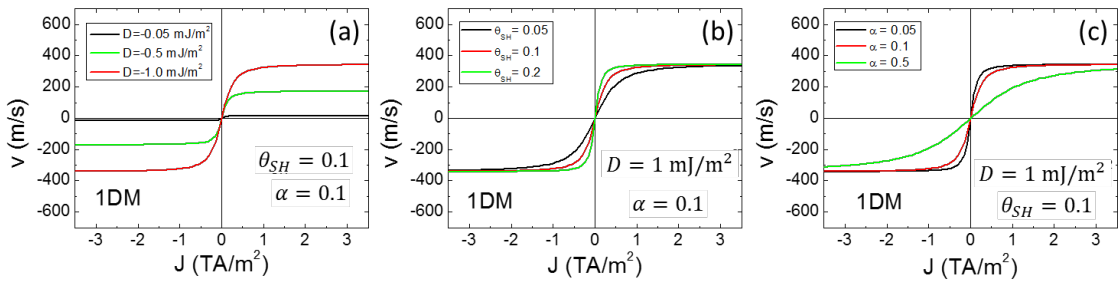

**Figure 13**. 1DM predictions for "$v_{DW}$ vs $J_{HM}$" for different combinations of the material parameters: (a), (b) and (c). Except indicated otherwise, material parameters and dimensions are the same as in Figure **12**.

It is also interesting to evaluate the current driven DW dynamics in the presence of longitudinal fields ($B_x = \mu_0 H_x$), which directly support or act against the DMI effective field. In Figure 14 we compare the micromagnetic results (dots) to the predictions of the 1DM (lines) for different combinations of the DW configuration (*up-down* $Q = +1$, and *down-up* $Q = -1$) and the direction of the injected current along the HM ($J_{HM} > 0$ and $J_{HM} < 0$). Lines in Figure 14(a) correspond to the rigid 1DM, whereas in Figure 14(b) the 1DM includes the DW tilting. Again, a better agreement with the micromagnetic results is obtained with the 1DM which includes the DW tilting. See also [67] for the analytical expression describing these results.

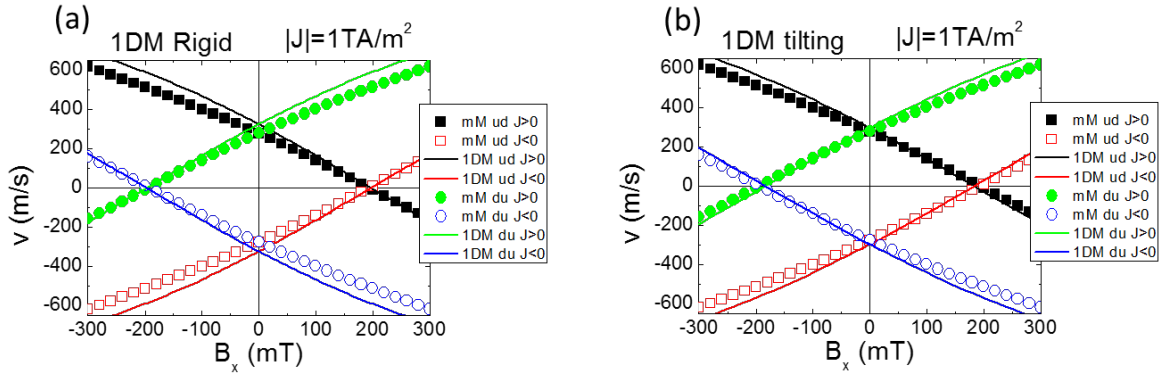

**Figure 14**. Current-driven DW dynamics in a HM/FM stack with DMI ($D = 1 \text{ mJ/m}^2$) under longitudinal fields. The magnitude of the applied current is $|J_{HM}| = 1 \text{ TA/m}^2$. The material parameters and the cross section of the FM layer are the same as in previous Figure **12**. Micromagnetic (mM) results are represented by dots, whereas lines correspond to the 1DM predictions: (a) 1DM-rigid and (b) 1DM-tilting. "ud" indicates an up/down DW: u: up; d: down.

### 3.7. Curved samples. Micromagnetic results and 1DM equations for curved strips.

Up to here, we have studied the DW dynamics along straight strips or straight HM/FM stacks. However, it is also interesting to evaluate the DW dynamics along systems with curved parts, both for the development of DW-based devices and from a theoretical point of view also. We consider a HM/FM stack with the geometry presented in Figures 15(a)-(b), where all the geometrical parameters are defined. The criteria for positive and negative currents are also defined in Figure 15(b). The FM layer has inner and outer radius given by $r_i$ and $r_o$ respectively. Its thickness is again $t$, and the cross section is $w \times t$ with $w = r_o - r_i$. The FM layer is on top of a HM with the same $r_i$ and $r_o$. Here we will consider the case where the electrical current is mainly flowing along the HM in the azimuthal direction ($\vec{J}_{HM} = J_{HM}(\rho)\vec{u}_\varphi$). Due to the curvature, the current density $J_{HM}(\rho)$ is no longer homogeneous over the HM cross section, but its magnitude depends on the polar radial coordinate ($\rho$). A simple calculation allows us to express $J_{HM}(\rho)$ in terms of the injected current along a straight strip ($J_0$) with the same cross section, where the current is homogeneous, as follows

$$J_{HM}(\rho) = \frac{J_0 w}{\rho \log\left(1 + \frac{w}{r_i}\right)} \tag{88}$$

The local value of $J_{HM}(\rho)$ decreases as $1/\rho$ from $\rho = r_i$ to $\rho = r_o$ (see Figure 15(c)). For a given $w$, the difference $J_{HM}(r_i) - J_{HM}(r_o)$ decreases as $r_i$ increases, and it tends to 0 as $r_i \to \infty$, which corresponds to the straight strip. On the contrary, $J_{HM}(r_i) - J_{HM}(r_o)$ increases with the curvature, that is, as $r_i \to 0$. As it can be easily understood, due to the non-homogenous character of $J_{HM}(\rho)$ in the HM, the resulting SOT in the FM will be also non-homogenous, and therefore, it will play a role for the DW dynamics in curved strips. In the figures we use the notation $J_{HM} \equiv J$ for simplicity.

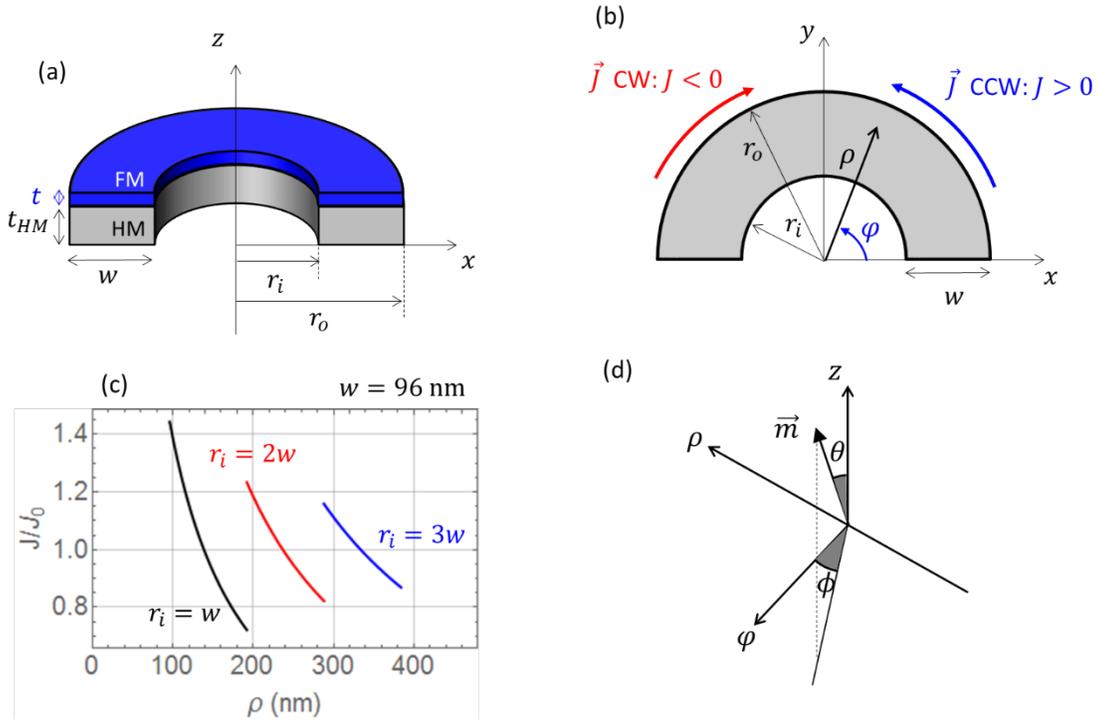

**Figure 15**. (a)-(b) Geometry of a curved HM/FM strip with the definition of the geometrical parameters. The convention for clockwise (CW, $J_{HM} < 0$) and counterclockwise (CCW, $J_{HM} > 0$) direction for the current is also showed. (c) $J_{HM}(\rho)/J_0$ vs $\rho$ for three different cases with $w = 96$ nm and $r_i = w$, $r_i = 2w$ and $r_i = 3w$. (d) Shows the definition of the coordinates adopted for the 1DM along curves.

The temporal micromagnetic snapshots of the evolution of different DW configurations and directions of the injected current are presented in Figure 16 and 17 for curved strips. In both cases, $w = 96$ nm, $r_i = 2w$, and $t = 0.6$ nm. The same material parameters are considered: $A = 16$ pJ/m, $M_s = 0.8 \times 10^6$ A/m, $D = 1$ mJ/m$^2$, $K_u = 0.8 \times 10^6$ J/m$^3$; $\alpha = 0.5$; $P = 0$, and $\theta_{SH} = 0.1$. The local current density is given by Equation (88) with $J_0 = 3$ TA/m$^2$. Here we adopt the following convention: the polarization of the spin current is fixed to $\vec{\sigma} = +\vec{u}_\rho$ for positive spin Hall angles $\theta_{SH} > 0$, and therefore the CCW (CW) current is $J > 0$ ($J < 0$).

When dealing with straight strips, we assumed the naming *up-down* or *down-up* as going from left to right along the $x$-axis. For curved strips we consider that we will cross through the DW always in the CCW direction, and therefore $Q = +1$ ($Q = -1$) indicates a transition from *up* to *down* (from *down* to *up*) across the DW. These snapshots of Figure 16 (inverted "U" shape) and Figure 17 ("U shape"), clearly show that DW dynamics depends on the DW configuration and on the direction of the current. For example, the total DW displacement is the same for the cases shown in Fig. 16(a) and (b), and Fig. 17(c) and (d). In short, in all cases included in $Q = +1$, the DWs move faster than for $Q = -1$ cases. Note, that this convention is independent of the direction of the current. A general feature of the presented results in Figure 16 and 17 is that DWs with smaller tilting (or smaller rotation of its normal $\vec{n}_{DW}$) are the fastest, whereas the DW velocity is reduced as the tilting increases. This is also shown in Figure 18 where two DWs are forced to propagate along an S-shape FM strip with straight and curved parts.

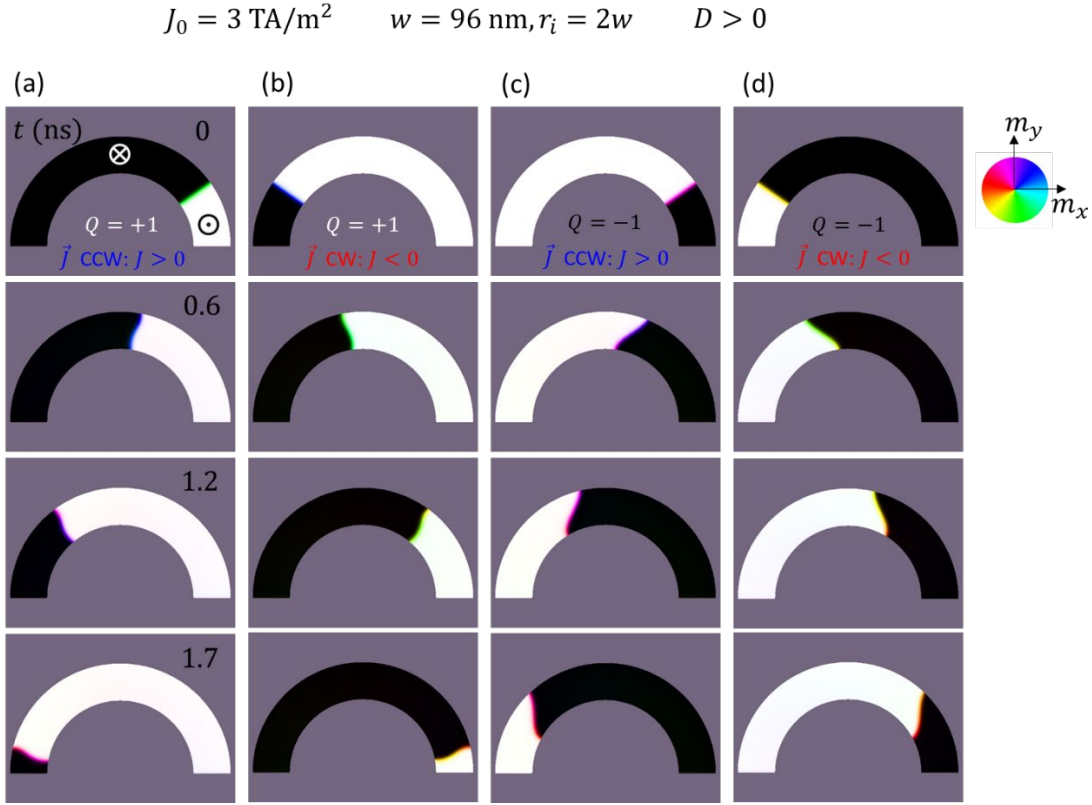

**Figure 16**. Micromagnetic snapshots showing the temporal evolution of different DW configuration along a curved FM layer with **inverted "U" shape** for different directions of the injected current along the HM. All parameters are given in the text.

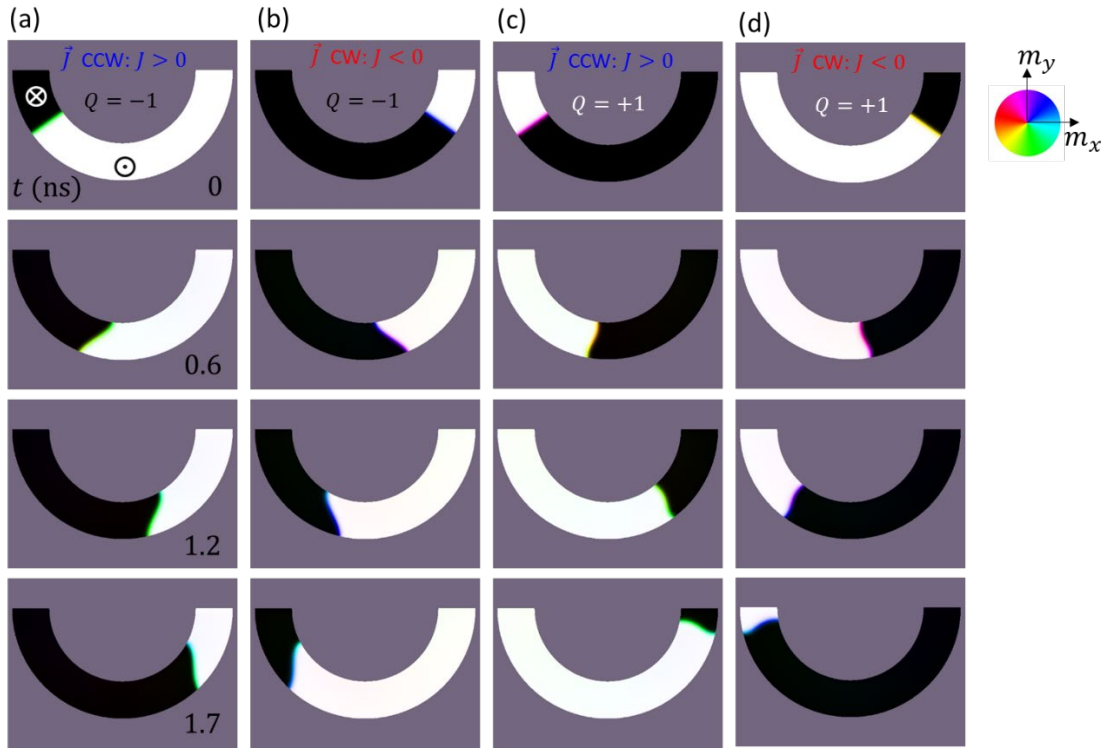

**Figure 17**. Micromagnetic snapshots showing the temporal evolution of different DW configuration along a curved FM layer with **"U" shape** for different directions of the injected current along the HM.

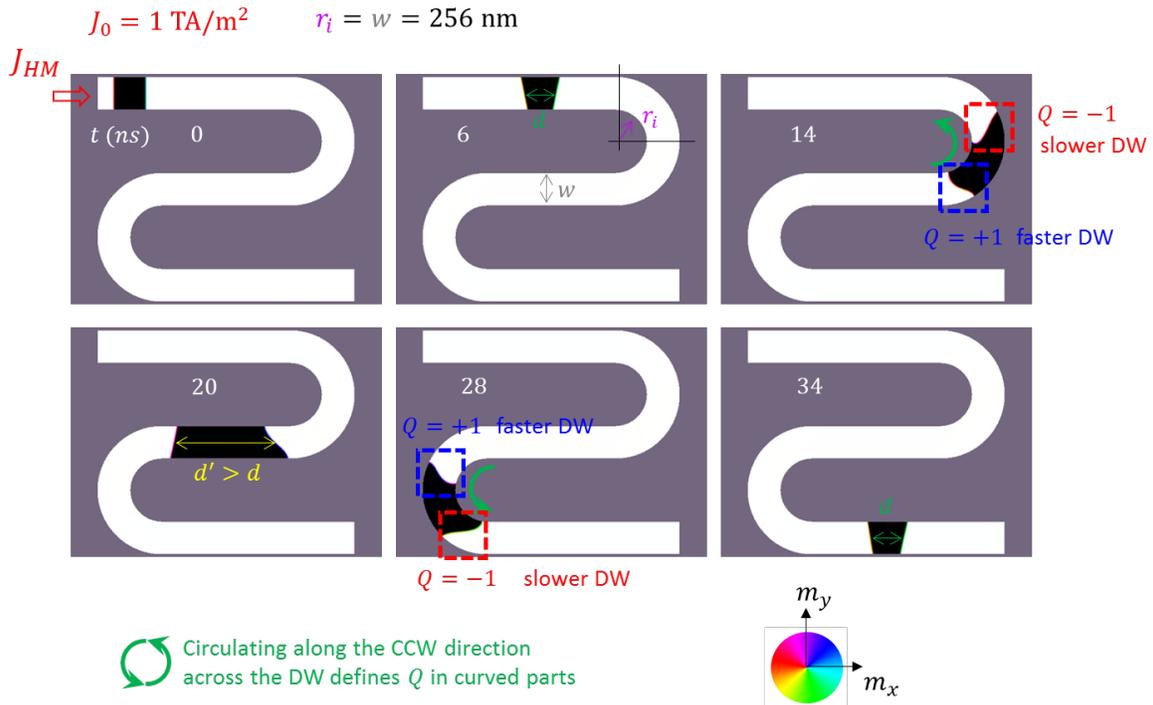

**Figure 18**. Micromagnetic snapshots showing the temporal evolution of two DW along a S-shape FM strip with straight and curved parts. In the straight parts, we assumed the naming *up-down* or *down-up* as going from left to right along the $x$-axis. For curved strips we consider that we will cross through the DW always in the CCW direction, and therefore $Q = +1$ ($Q = -1$) indicates a transition from *up* to *down* (from *down* to *up*) across the DW. This convention is defined at the bottom. All parameters are given in the text. The spatial distribution of ($\vec{J}_{HM} = \vec{J}_{HM}(\vec{r})$) in the HM is taken into account (not shown).

In Figure 19, we present the micromagnetic results for the dependence of the DW velocity as a function of the current for three different cases with $w = 96$ nm: (a) $r_i = w$, $r_i = 2w$ and $r_i = 3w$. The results for curved HM/FM stacks ($Q = \pm 1$) are compared to the straight case with the same cross section ($w \times t$). In all cases, and for sufficient high current, the DWs with $Q = +1$ move faster than in a straight sample, whereas the ones with $Q = -1$ move with smaller velocity than along the straight counterpart. The different increase with the curvature, that is, $|v_{curved} - v_{straigth}|$, increases as $r_i$ decreases for a given $w$. In order to elucidate which is the role of the inhomogeneous current distribution in curved HM/FM stacks ($Q = \pm 1$), we also present the micromagnetic results obtained by assuming that the current is homogenous over the HM cross section. Although the non-homogenous current distribution is also due to the curved geometry, the deviation from the homogenous results with respect to the inhomogeneous current distribution indicates that the different velocities for $Q = +1$ and $Q = -1$ cases is also ascribed to the curved geometry, and the corresponding different DW tilting. In short, the smaller the DW tilting the larger the DW velocity (the tilting is larger for the $Q = -1$ cases). And, the larger the curvature (or the smaller the internal radius $r_i$), the larger the deviation of the DW velocity with respect to the straight case. Moreover, as the DW tilting increases with the strip width ($w$) [66,67], the difference in the velocity for the $Q = +1$ and $Q = -1$ cases will increase for wider FM strips in these curved stacks.

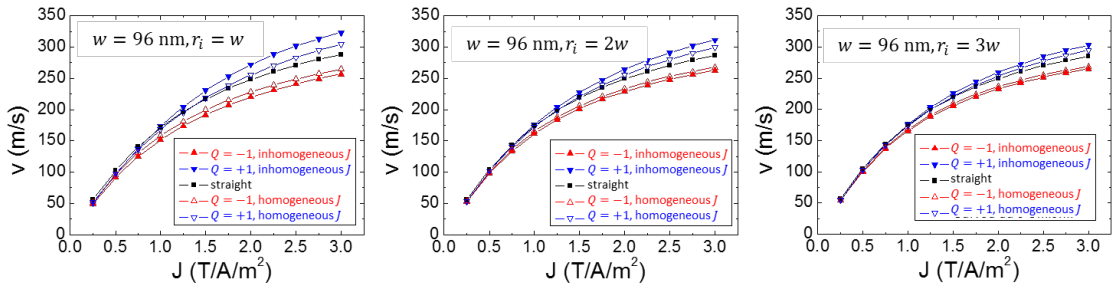

**Figure 19**. Micromagnetic results for the current-driven DW dynamics along curved strips for three different cases with fixed width $w = 96$ nm: (a) $r_i = w$, $r_i = 2w$ and $r_i = 3w$. The results for curved HM/FM stacks ($Q = \pm 1$) are compared to the straight case with the same cross section ($w \times t$).

In order to describe these micromagnetic results, we can also develop a 1DM for curved strips. Strips are considered to have cross-sectional dimensions $w \times t$. Strip width $w = r_o -$ $r_i$, is calculated as the difference between outer and inner radii $r_o$ and $r_i$, respectively. Based on the micromagnetic results the DW profile is proposed as given by the ansatz:

$$\theta(\rho,\varphi,t) = 2\arctan\exp\left\{Q\frac{R}{\Delta}\left[(\varphi-\varphi_0)\cos\chi - \ln\frac{\rho}{R}\sin\chi\right]\right\} \tag{89}$$

$$\phi(\varphi,t) = \psi(t) \tag{90}$$

where $R = \sqrt{r_i r_o}$ is the geometrical mean radius. Curvature is defined through the dimensionless parameter $\kappa = \sqrt{\frac{r_o}{r_i}}$, ranging from 1 for straight strips to infinity for the most extreme curvatures. See Figure 15(d) for the definitions of the coordinates adopted for the 1DM along curves. In order to adequately compare results for straight strips with those for curved ones, the instantaneous DW position must be defined as $q = \frac{r_i+r_o}{2}\varphi_0 = \frac{1+\kappa^2}{2\kappa}R\varphi_0$, while $R$ and $w$ are linked by the relationship $R = w\frac{\kappa}{\kappa^2-1}$. Finally, the same criteria of $Q = \pm 1$ *up-down / down-up* is considered when crossing the DW following the counterclockwise direction as previously stated in the micromagnetic part.

In the absence of STTs ($P=0$) and in-plane fields ($H_x = H_y = 0$), the LLG Equation in spherical coordinates can again be derived from the following Lagrangian ($\mathcal{L}$, in [J/m$^3$]) and Rayleigh dissipative ($\mathcal{F}$, in [J/(m$^3$s)]) functions:

$$\mathcal{L} = \mathcal{K} + \epsilon \tag{91}$$

$$\mathcal{F} = \frac{\alpha\mu_0 M_s}{2\gamma_0}\left\{\frac{d\vec{m}}{dt} - \frac{\gamma_0}{\alpha}H_{SL}\vec{m}\times\vec{\sigma}\right\}^2 \tag{92}$$

where $\epsilon$ is again the energy density of the system and $\mathcal{K}$ the kinetic energy density, which is given by

$$\mathcal{K} = \frac{\mu_0 M_s}{\gamma_0}\varphi\sin\theta\,\dot\theta \tag{93}$$

By using the ansatz (89) and (90) and integrating along the $x$-axis, we can obtain the Lagrangian ($L = \int \mathcal{L}dx$) and the dissipation function ($F = \int \mathcal{F}dx$) in terms of the DW coordinates $\{q,\psi,\chi\}$. The resulting Lagrangian ($L = \frac{1}{wt}\int \mathcal{L}dv$) and the dissipation function ($F = \frac{1}{wt}\int \mathcal{F}dv$) are expressed in terms of the DW coordinates $\{q,\psi,\chi\}$ as follows:

$$L = -2Q\frac{\mu_0 M_s}{\gamma_0}\left[\dot{q} + \frac{1}{2}\frac{\kappa^2+1}{\kappa^2-1}\left(\frac{\kappa^4+1}{\kappa^4-1}\ln\kappa - \frac{1}{2}\right)w\frac{\dot{\chi}}{\cos^2\chi}\right]\psi + \frac{2A}{\Delta\cos\chi}\frac{2\kappa\ln\kappa}{\kappa^2-1}$$
$$+ \frac{2K_u\Delta}{\cos\chi}\frac{\kappa^2+1}{2\kappa}$$
$$+ \frac{\mu_0 M_s^2 \Delta}{\cos\chi}\frac{\kappa^2+1}{2\kappa}\left[(N_\rho - N_\varphi)\sin^2(\psi-\chi) - (N_z - N_\varphi)\right]$$
$$+ Q\frac{\cos(\psi-\chi)}{\cos\chi}\pi D \tag{94}$$
$$- 2Q\mu_0 M_s H_z\left[q + \frac{1}{2}\frac{\kappa^2+1}{\kappa^2-1}\left(\frac{\kappa^4+1}{\kappa^4-1}\ln\kappa - \frac{1}{2}\right)w\tan\chi\right]$$
$$+ \mu_0 M_s \frac{\pi\Delta}{\cos\chi}H_{FL}^0\frac{\kappa^2-1}{2\kappa\ln\kappa}\sin\psi$$

$$F = \frac{1}{wt}\int\frac{\alpha\mu_0 M_s}{2\gamma_0}\left(\frac{d\vec{m}}{dt} - \frac{\gamma_0}{\alpha}H_{SL}^0 \vec{m}\times\vec{\sigma}\right)^2 dv$$
$$= \frac{\alpha\mu_0 M_s \Delta}{\gamma_0\cos\chi}\left\{\frac{2\kappa}{\kappa^2+1}\frac{\dot{q}^2}{\Delta^2}\cos^2\chi\right.$$
$$+ 2\frac{\dot{q}}{\Delta}\cos\chi\frac{\dot{\chi}}{\cos\chi}\frac{w}{\Delta}\frac{\kappa\left(\frac{\kappa^4+1}{\kappa^4-1}\ln\kappa - \frac{1}{2}\right)}{\kappa^2-1}$$
$$+ \frac{\kappa^2+1}{2\kappa}\frac{\dot{\chi}^2}{\cos^2\chi}\frac{\pi^2}{12}\left[\sin^2\chi\right.$$
$$+ \frac{w^2}{\pi^2\Delta^2}\frac{12\kappa^2\left(\ln^2\kappa - \frac{\kappa^4+1}{\kappa^4-1}\ln\kappa + \frac{1}{2}\right)}{(\kappa^2-1)^2}\right] + \frac{\kappa^2+1}{2\kappa}\dot{\psi}^2\right\} \tag{95}$$
$$+ Q\frac{\mu_0 M_s \pi\Delta}{\cos\chi}H_{SL}^0\frac{\kappa^2-1}{2\kappa\ln\kappa}\cos\varphi\left[\frac{2\kappa}{\kappa^2+1}\frac{\dot{q}}{\Delta}\cos\chi\right.$$
$$+ \frac{w}{\Delta}\frac{\dot{\chi}}{\cos\chi}\frac{\kappa\left(\frac{\kappa^2+1}{\kappa^2-1}\ln\kappa - 1\right)}{\kappa^2-1}\right] + \cdots$$

In previous equations, it must be noted that the demagnetizing terms $N_\varphi$ and $N_\rho$ play the role equivalent to those played by $N_x$ and $N_y$, respectively, in straight strips. The derivation of the dynamical equations requires the following intermediate results: $\frac{\partial\theta}{\partial\varphi} = Q\sin\theta\frac{R}{\Delta}\cos\chi$, $\frac{\partial \theta}{\partial \rho} = -Q \sin\theta \frac{R \sin\chi}{\Delta} \frac{1}{\rho}$, $\quad\quad \frac{\partial \theta}{\partial \chi} = -Q \sin\theta \frac{R}{\Delta} \Big[(\varphi - \varphi_0) \sin\chi + \ln\frac{\rho}{R} \cos\chi\Big]$, $\quad\quad \dot\theta = -\frac{R}{\Delta} Q \sin\theta \cos\chi \, \dot\varphi_0 - Q \sin\theta \frac{R}{\Delta}\Big[(\varphi - \varphi_0)\sin\chi + \ln\frac{\rho}{R}\cos\chi\Big]\dot\chi$, $\quad \int_{\varphi=-\infty}^{\varphi=+\infty} d\theta = Q\pi$, $\int_{\varphi=-\infty}^{\varphi=+\infty} \sin\theta \, d\varphi = \frac{\pi}{\cos\chi}\frac{\Delta}{R}$, $\int_{\varphi=-\infty}^{\varphi=+\infty} \sin^2\theta \, d\varphi = \frac{2}{\cos\chi}\frac{\Delta}{R}$, $\int_{\varphi=-\infty}^{\varphi=+\infty} \cos\theta \, d\varphi = -2Q\Big(\frac{q}{R}\frac{2\kappa}{\kappa^2+1} - \tan\chi \ln\frac{\rho}{R}\Big)$, $\int_{\varphi=-\infty}^{\varphi=+\infty}\sin\theta\cos\theta \, d\varphi = 0$ and $\int_{\varphi=-\infty}^{\varphi=+\infty} \dot\theta \sin\theta \, d\varphi = -2Q\Big(\frac{\dot q}{R}\frac{2\kappa}{\kappa^2+1} - \frac{1}{\cos^2\chi}\ln\frac{\rho}{R}\dot\chi\Big)$. In Equations (94) and (95) we have used the notation $H_{SL}^0 = \hbar\theta_{SH}J_{HM}/(2|e|\mu_0 M_s t)$ and $H_{FL}^0$, which both correspond to the definitions for straight strips (see Equation (58)). Note that the non-homogenous character of $J_{HM}(\rho)$ in the HM (Equation (88)) is taken into account in this derivation. Minimization of the energy density $\epsilon$ with respect to the DW width $\Delta$ results in a dependence of such parameter with curvature, in the form:

$$\Delta = 2\kappa \sqrt{\frac{\ln\kappa}{\kappa^4 - 1}} \Delta_0 \tag{96}$$

with $\Delta_0$ being the DW width for straight strips as obtained in (65). $\Delta$ decreases as $\kappa$ is increased, that is, for higher curvatures. Note also that here $\vec\sigma = \vec u_J \times \vec u_z = \vec u_\rho$, for $\vec u_J = \vec u_\varphi$ (counterclockwise current is considered as positive).

The 1DM Equations for curved strips can be deduced from the Euler-Lagrange-Rayleigh equations (81) where $X = \{\theta, \psi, \chi\}$. The final equations are

$$\alpha \frac{2\kappa}{\kappa^2 + 1}\frac{\dot q}{\Delta}\cos\chi + Q\dot\psi + \alpha \frac{w}{\Delta}\frac{\kappa\Big(\frac{\kappa^4+1}{\kappa^4-1}\ln\kappa - \frac{1}{2}\Big)}{\kappa^2 - 1}\frac{\dot\chi}{\cos\chi} = Q\gamma_0\Big(H_z - \frac{\pi}{2}H_{SL}\cos\varphi\Big) \tag{97}$$

$$-Q\frac{2\kappa}{\kappa^2+1}\frac{\dot q}{\Delta}\cos\chi + \alpha\dot\psi - Q\frac{w}{\Delta}\frac{\kappa\Big(\frac{\kappa^4+1}{\kappa^4-1}\ln\kappa - \frac{1}{2}\Big)}{\kappa^2-1}\frac{\dot\chi}{\cos\chi} = -\gamma_0\frac{1}{2}H_k\sin 2(\psi - \chi) + Q\gamma_0\frac{\pi}{2}H_D\sin(\psi - \chi) - \gamma_0\frac{\pi}{2}H_{FL}\cos\psi \tag{98}$$

$$\left(\alpha \frac{2\kappa}{\kappa^2+1}\frac{\dot{q}}{\Delta}\cos\chi + Q\dot{\psi}\right)\frac{\frac{w}{\Delta}}{\cos\chi}\frac{\kappa\left(\frac{\kappa^4+1}{\kappa^4-1}\ln\kappa - \frac{1}{2}\right)}{\kappa^2-1}$$

$$+ \alpha\dot{\chi}\frac{\pi^2}{12}\left[\tan^2\chi + \frac{\frac{w^2}{\pi^2\Delta^2}}{\cos^2\chi}\frac{12\kappa^2\left(\ln^2\kappa - \frac{\kappa^4+1}{\kappa^4-1}\ln\kappa + \frac{1}{2}\right)}{(\kappa^2-1)^2}\right]$$

$$= -\gamma_0\frac{\sigma_0\sin\chi}{2\mu_0 M_s\Delta}\frac{2\kappa}{\kappa^2+1} + \gamma_0\frac{1}{2}H_k\sin 2(\psi-\chi) \quad (99)$$

$$- Q\gamma_0\frac{\pi}{2}H_D\sin(\psi-\chi) + Q\gamma_0 H_z\frac{\frac{w}{\Delta}}{\cos\chi}\frac{\kappa\left(\frac{\kappa^4+1}{\kappa^4-1}\ln\kappa - \frac{1}{2}\right)}{\kappa^2-1}$$

$$- Q\gamma_0\frac{\pi}{2}H_{SL}\cos\psi\frac{\frac{w}{\Delta}}{\cos\chi}\frac{\kappa\left(\frac{\kappa^2+1}{\kappa^2-1}\ln\kappa - 1\right)}{\kappa^2-1}$$

$$- \gamma_0\frac{\pi}{2}H_{FL}\cos\psi\tan\chi$$

where

$$\sigma_0 = \frac{2A}{\Delta\cos\chi}\frac{2\kappa\ln\kappa}{\kappa^2-1} + \frac{2\Delta\left[K_u - \frac{1}{2}(N_z - N_\varphi)\mu_0 M_s^2\right]}{\cos\chi}\frac{\kappa^2+1}{2\kappa} \quad (100)$$

$$+ \frac{\mu_0 M_s^2\Delta}{\cos\chi}(N_\rho - N_\varphi)\frac{\kappa^2+1}{2\kappa}\sin^2(\psi-\chi) + Q\frac{\cos(\psi-\chi)}{\cos\chi}\pi D$$

and $H_k = (N_\rho - N_\varphi)M_s$, $H_D = \frac{2\kappa}{\kappa^2+1}H_D^0$, $H_{FL} = H_{FL}^0\frac{1}{\ln\kappa}\frac{\kappa^2-1}{\kappa^2+1}$, and $H_{SL} = H_{SL}^0\frac{1}{\ln\kappa}\frac{\kappa^2-1}{\kappa^2+1}$. Again, in these definitions $H_D^0$, $H_{FL}^0$ and $H_{SL}^0$ correspond to the definitions of $H_D$, $H_{FL}$ and $H_{SL}$ for straight strips as given in Equations (56)-(58). Note that previous Equations (97)-(100) converge to those of straight strips when $\kappa \to 1$, with $N_\rho \to N_y$ and $N_\varphi \to N_x$. Equations (97)-(100) were derived by us. An alternative collective coordinate model was obtained by Garg et al. [68].

The 1DM results for curved strips are shown in Figure 20 (a), (b) and (c) for the same material parameters and the same geometries as previously considered in the full micromagnetic simulations of Figure 19. Although the 1DM results do not show a marked difference between the velocities for curved and straight strips, the general trends of the micromagnetic results are reproduced by our 1DM model. Indeed, the discrepancy was expected, as the DW profile observed in the micromagnetic simulations could deviate from the assumed DW ansatz (84)-(85). Note that this 1DM ansatz assumes that the DW tilting $\chi$ is homogeneous for all points within the DW. However, the DW tilting in the micromagnetic simulations is not homogeneous and the DW adopts a S-shape configuration (see snapshots in Figure 16, 17 and 18), which is not described by the 1DM. Therefore, although the 1DM for curved strips is in qualitative agreement with full micromagnetic results, the study of DW motion should be done by means of full micromagnetic simulations, which nowadays are accessible with reasonable computational effort using GPU-based solvers for reduced strip widths ($w$, as $w$ approaches $w \sim 1$ μm the computational effort of full micromagnetic simulation becomes time prohibitive for systematic analysis). See [68] and [69] for others experimental and numerical studies of the current-driven DW dynamics along curved stacks.

Taking into account these issues, the 1DM still provides very relevant information to understand the DW dynamics in curved HM/FM stacks. In particular, the 1DM indeed indicates that higher velocity is achieved in the cases where the absolute value of the DW tilting ($|\chi|$) is smaller (See Figs. 20, 21 and 22).

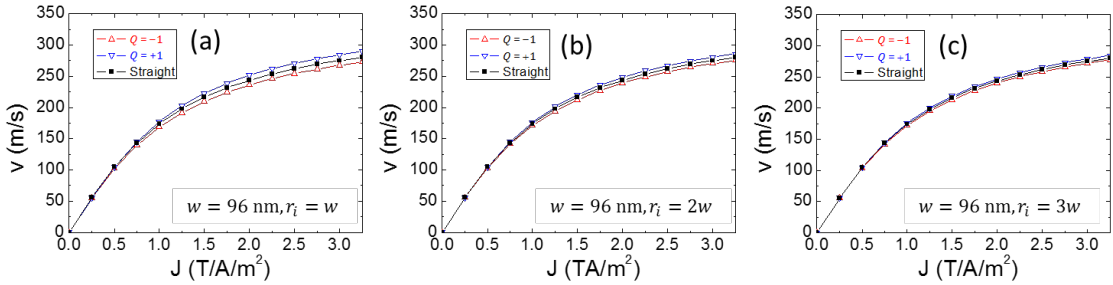

**Figure 20**. 1DM results for the current-driven DW dynamics along curved strips for three different cases with fixed width $w = 96$ nm: (a) $r_i = w$, $r_i = 2w$ and $r_i = 3w$. The results for curved HM/FM stacks ($Q = \pm 1$) are compared to the straight case with the same cross section ($w \times t$). Same material parameters as for the micromagnetic results of Figure 19 are considered.

In Figure 21(a) we plot the dependence of the internal DW angle $\psi$ as a function of $J_{HM}$ for straight and curved strips with the same cross section. It indicates that the internal DW angle is identical for straight and curved strips. However, the remarkable difference is related to the tilting angle ($\chi$). For a given current, the same $|\chi|$ is obtained for both $Q = +1$ and $Q = -1$ DW configuration in straight strips. However, the situation is different for curved strips: for a given current $|\chi|$ is larger for $Q = -1$ than for $Q = +1$, which explains the larger velocity of the $Q = +1$ configuration as the DW velocity scales with $1/\cos \chi$.

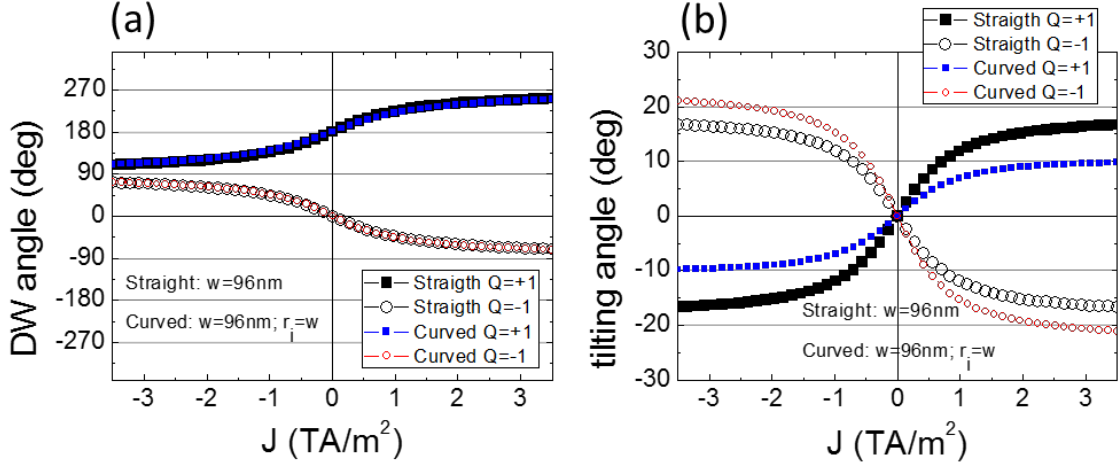

**Figure 21**. 1DM results for the current-driven DW dynamics along straight and curved strips with $w = 96$ nm and $r_i = w$. (a) DW angle $\psi$ as a function of $J_{HM}$. (b) Tilting angle $\chi$ as a function of $J_{HM}$. The results for curved HM/FM stacks ($Q = \pm 1$, blue and red symbols) are compared to the straight case (black filled symbols: $Q = +1$; and open symbols $Q = -1$) with the same cross section ($w \times t$). Same material parameters as for the micromagnetic results of Figure 19 are considered.

The difference in the DW velocity between the $Q = +1$ and $Q = -1$ cases for curved strips also increases with the strip width $w$. This is exemplified in Figure 22 for three different cases where now the inner radius is fixed to $r_i = w$, but the strip width is varied from $w = 96\ nm$ to $w = 500$ nm.

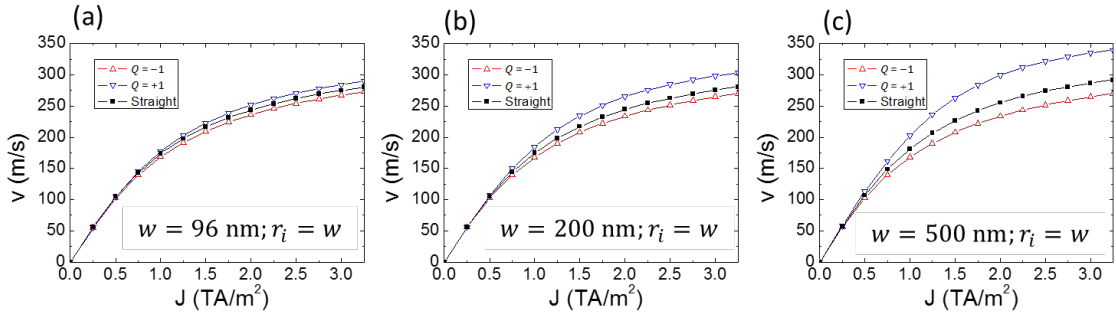

**Figure 22**. 1DM results for the current-driven DW dynamics along curved strips with $r_i = w$ and three different values of the strip width: (a) $w = 96$ nm, (b) $w = 200$ nm, and (c) $w = 500$ nm Same material parameters as for the micromagnetic results of Figure 19 are considered. The strip thickness is also the same $t = 0.6$ nm.

### 3.8. Current-driven DW motion in Synthetic Anti-Ferromagnets and Ferrimagnetic systems.

Previous analyses were carried out in a single FM layer, or in a single FM layer sandwiched between a heavy metal (HM) and an oxide (not shown), or between two dissimilar HMs. In what follows we are going to evaluate the current-driven DW dynamics in other systems which are of relevance for fundamental and technological reasons. In particular, we have seen that in HM/FM systems the DWs in the FM layer are efficiently driven along the strip axis by injection of electrical current through the HM ($\vec{J}_{HM}$). The main driving force is due to the spin Hall effect, which generates a spin current on the FM, and drives series of homochiral DWs along the FM layers. We have shown that such spin current generates an out-of-plane effective field, and also exerts a torque on the internal DW magnetization. In a straight FM stack (HM/FM), the DW velocity scales linearly with $J_{HM}$ for low currents. As $J_{HM}$ increases, the DW velocity tends to saturate as due to the rotation of the internal DW moment along the transverse $y$-axis, which is the direction of the spin current polarization (Sec. 3.5 and 3.6; Figure 12; and Equations (74) and (75)). Besides, in curved HM/FM stacks, consecutive DWs are driven with different velocities as due to the different DW tilting (Sec. 3.7). Therefore, these two issues constitute limiting factors for the performance of DW-based devices, where series of DWs should be driven along straight and curved parts of a track. Recent experimental and theoretical studies have pointed out that these limiting factors for the current-driven DW motion can be overcome in synthetic antiferromagnets stacks (SAF) [68-70] and in ferrimagnetic layers [71-74] or bilayers [75]. The field driven DW motion in the vicinity of the angular momentum compensation temperature of ferrimagnets have been also recently addressed [76]. The current-driven DW motion in these stacks is described hereafter.

### 3.8.A. Synthetic Anti-Ferromagnets (SAF)

A schematic representation of a SAF stack is depicted in Figure 23. It consist of two FM layers (LFM: Lower FM layer; UFM: Upper FM layer) separated by a spacer (S) which generates an antiferromagnetic exchange coupling between the magnetization of the LFM and UFM layers [77-81]. In the general case, the LFM is on top of a heavy metal (LHM: Lower HM) and the UFM layer is under another different HM (UHM: Upper HM). These HMs introduce an interfacial DMI at the LHM/LFM and at the UFM/UHM interfaces, and also generate a spin current in the adjacent FM layer that drives DWs in the FM layers. The length ($\ell$) and the width ($w$) of the stack, and also the thickness ($t$) of each layer are defined in Figure 23(a) and (b).

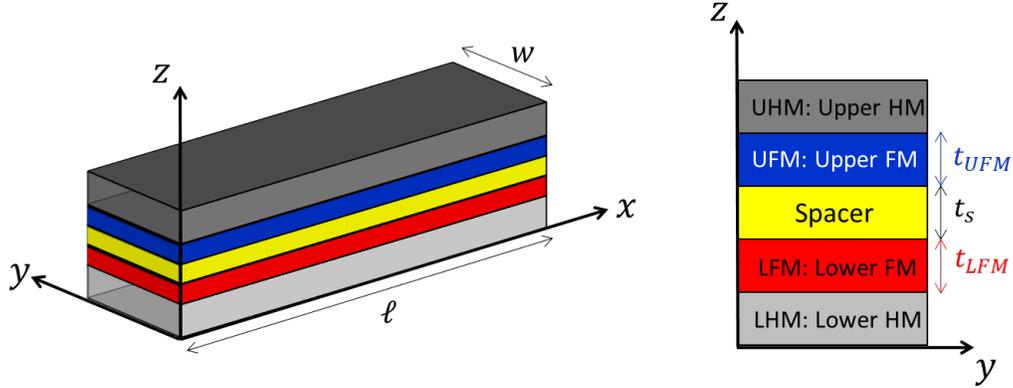

**Figure 23**. Schematic representation of two FM layers (Lower and Upper FM layers: LFM and UFM respectively) separated by a spacer ($S$) in a SAF stack. In the general case, the bottom layer is a heavy metal (Lower HM: LHM) and also the UFM is covered by another heavy metal (Upper HM: UHM). Both can generate an interfacial DMI in the corresponding FM layer, and also serve to inject an electrical current which generate a spin current in the adjacent FM layer.

In the framework of the micromagnetic model, the magnetization dynamics of a single ferromagnetic strip was described by the Landau-Lifshitz-Gilbert Equation (27). This equation can be generalized to study the magnetization dynamics in the SAF [69]. The general LLG Equation for each FM layer is

$$\frac{d\vec{m}_i}{dt} = -\gamma_i \vec{m}_i \times \vec{H}_{eff,i} + \alpha_i \vec{m}_i \times \frac{d\vec{m}_i}{dt} + \vec{\tau}_{STT,i} + \vec{\tau}_{SOT,i} \qquad (101)$$

where the sub-index $i$ stands for $i$: L (lower), U (upper) FM layers respectively. $\gamma_i = g_i \mu_B/\hbar$ and $\alpha_i$ are the gyromagnetic ratios and the Gilbert damping constants, respectively. $g_i$ is the Landé factor of each layer, and $\vec{m}_i(\vec{r},t) = \vec{M}_i/M_{s,i}$ is the normalized local magnetization to its saturation value ($M_{s,i}$), defined differently for each FM layer: $M_{s,i}$ ($i$: L, U). $\vec{H}_{eff,i}$ is the effective field, which includes not only the intralayer exchange and the uniaxial anisotropy, but also the interlayer exchange and the magnetostatic interactions adequately weighed to account for the different saturation magnetizations. Also, the DMI is included in the effective field. The material parameters of these interactions are denoted as: $A_i$ (intralayer exchange parameter), $K_{u,i}$ (PMA constant), and $D_i$ (DMI parameter), where, again, $i$: L, U for the LFM and the UFM layers. The interlayer exchange contribution $\vec{H}_{AF,i}$ to the effective field $\vec{H}_{eff,i}$, acting on each FM layer is computed from the corresponding energy density, $\omega_{AF,i} = -B_{ij} \vec{m}_i \cdot \vec{m}_j$, where $B_{ij}$ (in [J m$^{-3}$]) is a parameter describing the interlayer exchange coupling between the Lower and the Upper FM layers (here, we used the notation $i$: L and $j$: U). For the SAF case, the interlayer exchange coupling parameter is defined as $B_{ij} = J^{ex}/t_S$, where $J^{ex}$ represents the magnitude of the interlayer exchange coupling (in [J m$^{-2}$]) and $t_S$ is the thickness of the spacer separating the FM layers [69,70]. The resulting interlayer exchange contribution $\vec{H}_{AF,i}$ to the effective field reads as

$$\vec{H}_{AF,i} = -\frac{1}{\mu_0 M_{s,i}} \frac{\delta \omega_{AF,i}}{\delta \vec{m}_i} = \frac{J^{ex}}{\mu_0 M_{s,i} t_S} \vec{m}_j \tag{102}$$

where $i$ : L, U for the LFM and the UFM layers. Ferromagnetic (FM) and antiferromagnetic (AF) coupling cases are evaluated by a positive $J^{ex}$, and by a negative $J^{ex}$, respectively. Here we study the most relevant case of AF coupling. In Equation (101), $\vec{\tau}_{STT,i}$ and $\vec{\tau}_{SOT,i}$ are the STTs and the SOTs acting on each FM layer. In the general case, $\vec{\tau}_{STT,i}$ is due to the electrical current along the layer $i$ ($\vec{J}_{FM,i}$), whereas $\vec{\tau}_{SOT,i}$ are related to the electrical current along the HMs ($\vec{J}_{HM,i}$). The $\vec{\tau}_{STT,i}$ is parameterized by the polarization factor ($P_i$) and the non-adiabatic parameter ($\beta_i$) of each layer. Besides, the spin-orbit torques ($\vec{\tau}_{SOT,i}$) are proportional to the spin Hall angles $\theta_{SH,i}$: $\theta_{SH,L}$ and $\theta_{SH,U}$ represent the spin Hall angles of the Lower and the Upper HMs respectively. The expressions of these torques are similar to the ones for the single-FM stack expressions (see Equations (28) and (31)) with the corresponding parameters for each FM layer.

In order to illustrate the current-driven DW dynamics in the SAF, typical values of the parameters above have been chosen in our simulations. Except where the contrary is indicated, some common material parameters for the UFM and the LFM layers have been adopted: $M_{s,i} = 0.6 \times 10^6$ A/m, $K_{u,i} = 0.8 \times 10^6$ J/m³, $A_{ex} = 16$ pJ/m, $\alpha_i = 0.1$, and $\gamma_i = \gamma_0$ ($g_i \approx 2$) for both Lower and Upper FM layers ($i$: L, U). To illustrate the key ingredients for the current-driven DW along a SAF, we assume that STT is negligible ($P_i = 0$) in the evaluated samples. The spin Hall angle and the DMI parameter are only different from zero in the LFM: $\theta_{SH,L} = 0.1$ and $D_L = 1$ mJ/m². The interlayer exchange coupling is fixed to $J^{ex} = -0.5$ mJ/m². Here, the thickness of the FM layers (LFM and UFM) and the spacer (S) are $t_{LFM} = t_S = t_{UFM} = 0.6$ nm. See Ref. [69] for a detailed analysis of other combinations of material parameters and thicknesses.

We can present the main advantages of the SAF in a single micromagnetic simulation of a S-shape in-plane geometry, where two DWs are driven along each of the two FM layers of the SAF by a current density along the Lower HM with a magnitude of $J_{LHM} = J_0 = 1$ TA/m² in the straight parts, where it is uniform. Note that this current is non-uniform along the curved parts, as it was already described in Figure 15. The spatial distribution of the current was computed as described in Ref. [69] by using a commercial software, and taken into account for the micromagnetic calculations. The considered SAF and the micromagnetic snapshots showing the temporal evolution of the DWs in the LFM and the UFM layers are shown in Figure 24(b). The corresponding single FM case is shown in Figure 24(a) for comparison. When the interlayer exchange coupling ($|J^{ex}|$) is sufficiently strong, the DWs in the LFM and in the UFM move coupled to each other, and they are mirror to each other: an *up-down* in the LFM corresponds to a *down-up* in the UFM and vice versa. Note that the spin Hall effect is only considered in the LHM, and that the DMI is only different from zero at the LHM/LFM interface. However, due to the strong AF coupling all DWs in the system are left-handed ($D_L > 0$), as it can be seen by examining their internal moments. Moreover, DWs are driven along the SAF without tilting (Figure 24(b)), whereas the tilting is evident for the single FM case (Figure 24(a)). It is also verified that the internal magnetic moment of each DW points along the normal to the DW plane for the SAF, which is also linked to the lack of tilting. Therefore, the AF coupling between the FM layers mitigates the rotation of the internal DW moments towards the direction of the spin polarization, and consequently, higher velocities (and larger driving currents) than in the single-FM stack are achieved (Ref. [68-70]). Indeed, as discussed in Sec. 3.5 for the single-FM stack, the terminal DW velocity scales linearly with the current, since $v_{DW} \propto H_{SL} \cos \psi_s$, (Equation (74)) and the terminal DW angle ($\psi_s$) remains equal to its value at rest for sufficiently high interfacial DMI, that is, if $\psi_s \approx 0, \pi$ for *down-up* and *up-down* DW configurations. As it was deduced in [69] (see also following 1DM discussion, and Equation (117) below), the analytical expression for the DW velocity in a SAF also indicates the DW velocity increases as the sum of the saturation magnetization of the two FM layers ($M_{S,L} + M_{S,U}$) decreases. On the other hand, it is also observed in Figure 24(b) that adjacent DWs move with the same velocity along the curved parts of the SAF: the relative distance between these DWs remains the same after passing the first and the second turns of the S-shape. This is also different from the single FM case, where the distance between adjacent DWs is different after crossing the first turn (see the corresponding snapshot in Figure 24(a)). For these reasons, these SAF systems are promising to develop efficient DW-based devices in-planar geometries: DWs do not depict tilting and the distance between adjacent does not change after passing through curved parts. The velocity as a function of the current is discussed below, after presenting an extended 1DM for SAF systems.

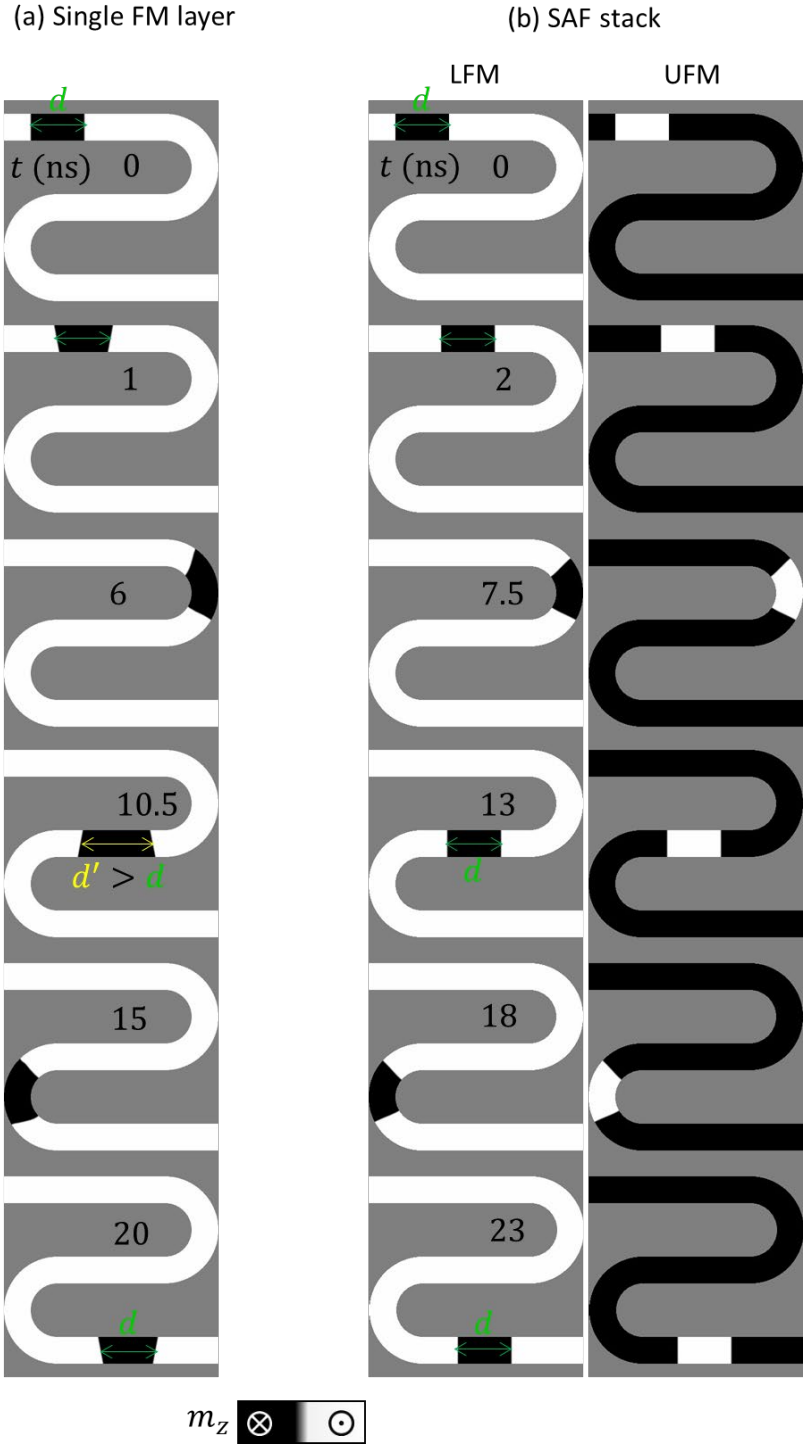

**Figure 24**. Micromagnetic snapshots showing the current-driven DW dynamics along a single FM layer stack (a), and along a SAF (b) with curved and straight parts. Here the radius of the curved parts is $r_i = w$, and $w = 256$ nm. The thickness of the FM layer in the single FM stack is $t = 0.6$ nm, and the material parameters are the same as in the LFM of the SAF (given below). For the SAF (b), the thickness of the FM layers (LFM and UFM) and the spacer (S) are $t_{LFM} = t_S = t_{UFM} = 0.6$ nm. The rest of material parameter are given in the text. The magnitude $\vec{J}_{LHM}$ in the straight parts through the HM is $J_0 = 1$ TA/m². As in Figure 18, the non-uniform spatial distribution of $\vec{J}_{LHM}$ is taken into account.

The described features of the DW dynamics along a SAF can be also discussed in terms of an extended 1DM for a straight SAF. The energy density per unit volume ($\epsilon$) of the two FM layers coupled by exchange interaction between them can be expressed as

$$\epsilon = \epsilon_1 + \epsilon_2 + \epsilon_{12} \tag{103}$$

where here $\epsilon_1$ and $\epsilon_2$ are the energy densities per unit volume of each isolated FM layer (1: LFM; 2: UFM), and the $\epsilon_{12}$ represents exchange coupling interaction between them. $\epsilon_i$ is the same as given in Equation (47), with the corresponding parameters of the layer $i$, that is,

$$\begin{aligned}\epsilon_i = A_i \frac{\sin^2 \theta_i}{\Delta^2} &+ K_{eff,i} \sin^2 \theta_i + K_{sh,i} \sin^2 \theta_i \sin^2 \phi_i + Q_i D_i \cos \phi_i \left(\frac{\sin \theta_i}{\Delta}\right) \\ &- \mu_0 M_{s,i}\left(H_x \sin \theta_i \cos \phi_i + H_y \sin \theta_i \sin \phi_i + H_z \cos \theta_i\right)\end{aligned} \tag{104}$$

where we have assumed that the DW in the two FM layers have the same DW width ($\Delta_1 = \Delta_2 = \Delta$). $Q_i$ indicates the DW configuration within each FM layer. For strong AF coupling, $Q_1 = +1$ (*up-down*) corresponds to $Q_2 = -1$ (*down-up*) and vice versa. The interaction between the layers can be expressed in terms of the corresponding polar angles ($\theta_i$, $\phi_i$) as

$$\epsilon_{12} = -B_{12}\vec{m}_1 \cdot \vec{m}_2 = -B_{12} \sin \theta_1 \sin \theta_2 \cos(\phi_1 - \phi_2) + \cos \theta_1 \cos \theta_2 \tag{105}$$

where $B_{12} = J^{ex}/t_S$ with $t_S$ being the thickness of the spacer. Also, for strong AF coupling, the positions of the DWs along the longitudinal $x$-axis are also assumed to be equal, i.e., $q_1 = q_2 = q$. Under these circumstances, the surface energy density is

$$\sigma = \sigma_1 + \sigma_2 + \sigma_{12} \tag{106}$$

with

$$\begin{aligned}\sigma_i = \frac{2A_i}{\Delta} &+ 2\Delta\left(K_{eff,i} + K_{sh,i} \sin^2 \psi\right) + \pi Q_i D_i \cos \psi_i \\ &- \mu_0 M_{s,i} \pi \Delta\left(H_x \cos \psi_i + H_y \sin \psi_i\right) - 2Q_i \mu_0 M_{s,i} q H_z\end{aligned} \tag{107}$$

and

$$\sigma_{12} = -B_{12} 2\Delta \cos(\psi_1 - \psi_2) \tag{108}$$

where we have used $\int_{-\infty}^{+\infty} \sin \theta_1 \sin \theta_2 \, dx = 2\Delta$. The rest of integrals are similar to the ones in Sec. 3.1. By a similar procedure to that in Sec. 3.1, the following 1DM Equations are obtained

$$\left(\alpha_1 \frac{M_{s,1}}{\gamma_1} + \alpha_2 \frac{M_{s,2}}{\gamma_2}\right)\frac{\dot{q}}{\Delta} + \frac{M_{s,1}}{\gamma_1} Q_1 \dot{\psi}_1 + \frac{M_{s,2}}{\gamma_2} Q_2 \dot{\psi}_2$$
$$= Q_1 M_{s,1} \left[H_z - \frac{\pi}{2} H_{SL,1} \cos\psi_1\right] + Q_2 M_{s,2} \left[H_z - \frac{\pi}{2} H_{SL,2} \cos\psi_2\right] \quad (109)$$
$$+ \frac{M_{s,1}}{\gamma_1} \beta_1 \frac{u_1}{\Delta} + \frac{M_{s,2}}{\gamma_2} \beta_2 \frac{u_2}{\Delta}$$

$$\left(-Q_1 \frac{\dot{q}}{\Delta} + \alpha_1 \dot{\psi}_1\right)$$
$$= \gamma_1 \Bigg[ -\frac{H_{k,1}}{2}\sin(2\psi_1) + \frac{\pi}{2} Q_1 H_{D,1} \sin\psi_1$$
$$+ \frac{\pi}{2}(H_y \cos\psi_1 - H_x \sin\psi_1) - \frac{\pi}{2} H_{FL,1} \cos\psi_1 \quad (110)$$
$$+ \frac{2B_{12}}{\mu_0 M_{s,i}} \sin(\psi_1 - \psi_2) \Bigg] - Q_1 \frac{u_1}{\Delta}$$

$$\left(-Q_2 \frac{\dot{q}}{\Delta} + \alpha_2 \dot{\psi}_2\right)$$
$$= \gamma_2 \Bigg[ -\frac{H_{k,2}}{2}\cos(2\psi_2) + \frac{\pi}{2} Q_2 H_{D,2} \sin\psi_2$$
$$+ \frac{\pi}{2}(H_y \cos\psi_2 - H_x \sin\psi_2) - \frac{\pi}{2} H_{FL,2} \cos\psi_2 \quad (111)$$
$$- \frac{2B_{12}}{\mu_0 M_{s,i}} \sin(\psi_1 - \psi_2) \Bigg] - Q_2 \frac{u_2}{\Delta}$$

where

$$H_{D,i} = \frac{D_i}{\mu_0 M_{s,i} \Delta} \quad (112)$$

$$H_{k,i} = \frac{2K_{sh,i}}{\mu_0 M_s} = M_{s,i}(N_y - N_x) \quad (113)$$

$$H_{SL,i} = \frac{\hbar \theta_{SH,i} J_{HM,i}}{2|e|\mu_0 M_{s,1} t_i} \quad (114)$$

$$H_{FL,i} = k_i H_{SL,i} \quad (115)$$

$$u = -\frac{g_i \mu_B P_i J_{FM,i}}{2|e| M_{s,i}} \quad (116)$$

These 1DM Equations reduce to the ones of a single-FM layer for $B_{12} = 0$ (see Sec. 3.1). Therefore, they can be solved for a single-FM layer case ($B_{12} = 0$) and for FM coupling ($Q_1 = Q_2$, $B_{12} = J^{ex}/t_S$) or AF coupling ($Q_2 = -Q_1$, $B_{12} = J^{ex}/t_S$) cases.

In Figure 25(a) we compare the micromagnetic results (dots) and the 1DM (lines) predictions for the DW velocity as a function of the current for a single-FM layer stack and for a SAF. The 1DM results were obtained by numerically solving the Equations (109)-(111). The stack is straight along the longitudinal $x$-axis. The material parameter for the SAF stack are: $M_{s,i} = 0.6 \times 10^6$ A/m, $K_{u,i} = 0.8 \times 10^6$ J/m$^3$, $A_{ex} = 16$ pJ/m, $\alpha_i = 0.1$, $P_i = 0$ and $\gamma_i = \gamma_0$ ($g_i \approx 2$) for both Lower and Upper FM layers, $i$: L, U. The spin Hall angle and the DMI parameter in the LHM are only different from zero in the LFM, $\theta_{SH,L} = 0.1$ and $D_L = 1$ mJ/m$^2$. The interlayer exchange coupling is fixed to $J^{ex} = -0.5$ mJ/m$^2$. The thicknesses of the layers are $t_{LFM} = t_S = t_{UFM} = 0.6$ nm. The strips width is $w = 256$ nm. For the single-FM stack the material parameters and the dimensions are the same as for the LFM of the stack. Whereas the DW velocity saturates as $J_{HM}$ increases for the single-FM stack, the DW velocity increases linearly with $J_{LHM}$. Both models show good quantitative agreement. Note that the DW velocity in the single-FM is larger than the one in the SAF for current below ~1 TA/m$^2$. Above this value, the DW in the SAF increases linearly, whereas the DW velocity in the single-FM saturates. As it will be shown later, the DW velocity in the low current regime scales with the inverse of the total magnetization of the system, that is, $M_S$ in the single FM layer, and $(M_{s,1} + M_{s,2})$ for the SAF. In Figure 25(b) we represent the dependence of the terminal values of the internal DW angles ($\psi$ for the single-FM stack, and $\psi_L$ and $\psi_U$ for the LFM and UFM layers of the SAF) as functions of the density current injected along the LHM (SAF) or along the HM in the single-FM stack. Again, a good agreement is achieved. Both models indicate that the internal DW angle in the LFM (the one submitted to the action of the spin current in the LHM), remains equal to its value at rest, $\psi_L \approx \pi$. The micromagnetic results show that internal DW angle in the UFM remains also close to its value at rest $\psi_U \approx 0$, but this angle deviates from zero as the current increases. Note however, that the maximum deviation is still well below than the one achieved in the single-FM stack. The 1DM also predicts a similar qualitative behavior, but the deviation from 0 is smaller than the one given by the full micromagnetic analysis. Based on these considerations we can conclude that the 1DM is still a good tool to describe the DW dynamics in these systems. In particular, we can deduce an analytical expression for the DW velocity in the SAF as a function of $J_{LHM}$. Indeed, by imposing that $\dot{\psi}_1 = \dot{\psi}_2 = 0$ ($\dot{\psi}_L = \dot{\psi}_U = 0$) and that these angles remain equal to its initial state ($\psi_L(0) = \pi, \psi_U(0) = 0$) in Equation (100), the terminal DW velocity in a SAF with strong AF coupling is

$$\dot{q}_{st} = -\Delta \frac{\gamma_0}{\alpha} Q_1 \frac{M_{s,1}}{(M_{s,1} + M_{s,2})} \frac{\pi}{2} H_{SL,1} \cos\psi_1 = \Delta \frac{\gamma_0}{\alpha} \frac{\pi}{2} \frac{\hbar \theta_{SH,L} J_{LHM}}{(M_{s,1} + M_{s,2}) 2|e|\mu_0 t_L} \quad (117)$$

which explains the linearly scaling of the coupled DW velocity $v_{DW}$ with the current in the LHM, and also points out that the DW velocity is enhanced as the $(M_{s,1} + M_{s,2})$ decreases. Therefore, this 1DM not only explain the micromagnetic results, but also suggests a manner to further optimize the DW velocity in SAF stacks. Further details of the current-driven DW dynamics in SAF can be seen in Ref. [69,70].

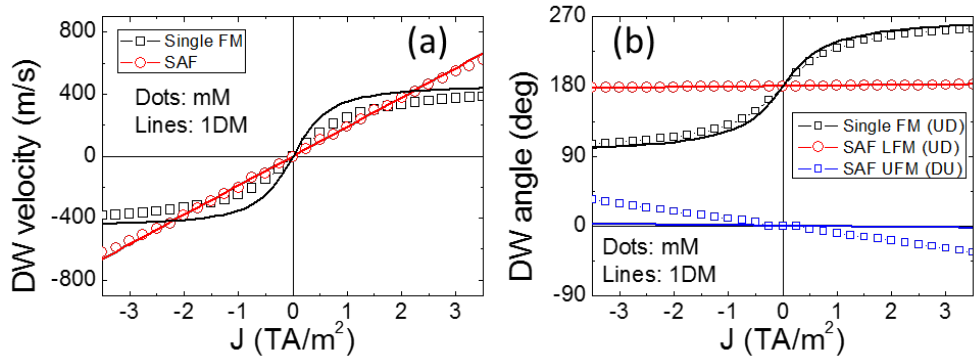

**Figure 25**. (a) Comparison of the DW velocity as a function of the current in the HM for a straight single-FM stack and for the SAF stack. Dots are micromagnetic results, whereas lines are 1DM predictions. Material parameters and dimensions are given in the main text.

### 3.8.B. Ferrimagnets (FiM)

In the SAF discussed in previous section, the magnitude of the interlayer exchange coupling depends on the thickness of the spacer ($t_S$) in a non-monotonous manner. Indeed, recent experiments [68,70] have shown that depending on such a thickness the coupling between the FM layers in the SAF can be either ferromagnetic-like ($J^{ex} > 0$, which promotes the parallel alignment of the magnetization in the FM layers), or antiferromagnetic-like ($J^{ex} < 0$, which promotes the antiparallel alignment of the magnetization in the FM layers). Only the antiferromagnetic coupling constitutes an advance over the single FM layer stack in terms of the current-driven DW motion efficiency [69,70]. Additionally, the improvement in the efficiency also depends on the saturation magnetization of the two FM layers (see [69] for further details).

A more recent and promising alternative to the SAF is the use of ferrimagnetic films (FiMs) [71-75]. These FiM films, as for example a rare earth (RE) with transitional metal (TM), are constituted by two sublattices which are antiferromagnetically coupled. Their magnetic properties, such as magnetization and coercivity, are largely influenced by the compensation temperature, which can be tuned by varying the composition or the temperature.

Here we present a micromagnetic model to describe the magnetization dynamics of a FiM formed by two components ($S_1$ specimen 1, and $S_2$ is specimen 2), which in general we name 1 and 2, and which are forming two antiferromagnetically coupled sublattices [77-82]. A schematic representation of the FiM on top of a HM is shown in Figure 26, where blue (1) and red (2) arrows represent the local magnetization of each sublattices or component of the FiM forming a ferromagnetic DW.

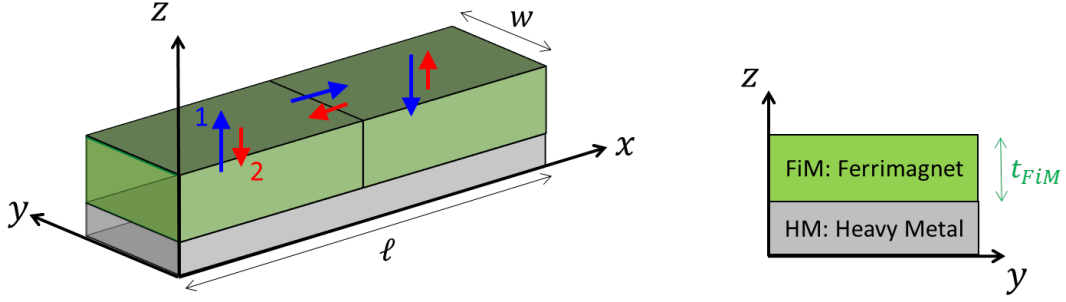

**Figure 26**. Schematic representation of a HM/FiM stack, where the FiM, consisting of two FM sublattices (1 and 2), is on top of a heavy metal (HM). The magnetization of two sublattices are represented by blue (1) and red (2) arrows, and a schematic representation of a DW in this FiM is shown. The dimensions of the system and the definition of the thickness are depicted.

As for the SAF, the temporal evolution of the magnetization of each sublattice evolves under the LLG Equation [82], which can be written as

$$\frac{d\vec{m}_i}{dt} = -\gamma_i \vec{m}_i \times \vec{H}_{eff,i} + \alpha_i \vec{m}_i \times \frac{d\vec{m}_i}{dt} + \vec{\tau}_{STT,i} + \vec{\tau}_{SOT,i} \qquad (118)$$

where here the sub-index $i$ stands for $i$: 1 and 2 sublattices respectively. $\gamma_i = g_i \mu_B/\hbar$ and $\alpha_i$ are the gyromagnetic ratios and the Gilbert damping constants, respectively. $g_i$ is the Landé factor of each layer, and $\vec{m}_i(\vec{r}, t) = \vec{M}_i/M_{s,i}$ is the normalized local magnetization to its saturation value ($M_{s,i}$), defined differently for each sublattice: $M_{s,i}$ ($i$: 1,2). In our micromagnetic model the FiM strip is formed by computational elementary cells, and within each cell we have two magnetic moments, one for each component of the FiM [82]. The respective effective field ($\vec{H}_{eff,i}$) acts on the local magnetization of each sublattice ($\vec{m}_i(\vec{r}, t)$). It is the sum of the external field, the magnetostatic field, the anisotropy field (PMA), the DMI field and the exchange field. The magnetostatic field on each local sublattice moment is numerically computed from the net magnetization of each elementary cell using similar numerical techniques as for the single FM case (see Ref. [9]). For the PMA field, the easy axis is always along the out-of-plane direction (z-axis), but we can in general consider different anisotropy constants for each sublattice $K_{u,i}$ (PMA constant). Our numerical solver also allows different DMI parameters for each lattice ($D_i$, where again, $i$: 1,2 for the S$_1$ and the S$_2$ lattices). The exchange field of each sublattice includes the interaction with itself (intralattice exchange interaction, $\vec{H}_{exch,i}$) and with the other sublattice (interlattice exchange interaction, $\vec{H}_{exch,12}$). The interlattice exchange effective field is computed as for a single FM sample, $H_{exch,i} = \frac{2A_i}{\mu_0 M_{s,i}} \nabla^2 \vec{m}_i$, (see Equation (3) and (26)), where $A_i$ is the intralattice exchange parameter. The interlattice exchange contribution $\vec{H}_{exch,12}$ to the effective field $\vec{H}_{eff,i}$ acting on each sublattice is computed from the corresponding energy density, $\omega_{exch,i} = -B_{ij} \vec{m}_i \cdot \vec{m}_j$, where $B_{ij}$ (in [J m$^{-3}$]), is a parameter describing the interlattice exchange coupling between the S₁ and S₂ sublattices (here, we used the notation $i$: 1 and $j$: 2). Finally, as for the SAF, in Equation (118), $\vec{\tau}_{STT,i}$ and $\vec{\tau}_{SOT,i}$ are the STTs and the SOTs acting on each sublattice. The STT, $\vec{\tau}_{STT,i}$ is due to the electrical current along the FiM layer ($\vec{J}_{FM}$), whereas $\vec{\tau}_{SOT,i}$ are related to the electrical current along the HM ($\vec{J}_{HM}$). Due to the different saturation magnetization ($M_{s,i}$) and the different Landé factor ($g_i$) of each sublattice, $\vec{\tau}_{STT,i}$ are also different for each sublattice. Our micromagnetic code also allows us to consider different polarization factors ($P_i$) and non-adiabatic parameters ($\beta_i$) for each sublattice ($i$: 1,2). Regarding the spin-orbit torques ($\vec{\tau}_{SOT,i}$) it is also possible to establish different spin Hall angles for each sublattice $\theta_{SH,i}$. The expressions of these torques are similar to the ones for the single-FM stack expressions (see Equations (28) and (31)), with the corresponding parameters for each FM layer.

We have also developed a 1DM for HM/FiM. The energy densities and the resulting 1DM Equations are identical to the ones deduced for the SAF. (see Equations (109)-(116)), where now the subindexes $i$: 1,2 correspond to the two sublattices of the FiM.

In order to illustrate the current-driven DW dynamics along a HM/FiM stack we have assumed a FiM with $w \times t_{FiM} = 256$ nm $\times$ 6 nm with the following common material parameters for the two sublattices $i$: 1,2: $A_i = 70$ pJ/m, $K_{u,i} = 1.4 \times 10^6$ J/m³, $\alpha_i = 0.02$, $D_i = 0.12$ J/m², $\theta_{SH} = 0.155$, $k_i = 0$ and $P_i = 0$. The antiferromagnetic coupling between the two lattices is taken into account by an interlattice exchange interaction of $B_{ij} \equiv B_{12} = -0.9 \times 10^7$ J/m³. The gyromagnetic ratios ($\gamma_i = g_i \mu_B / \hbar$) are different due to the different Landé factor: $g_1 = 2.05$ and $g_2 = 2.0$. The saturation magnetization of each sublattice $M_{s,i}$ can be tuned with the composition of the FiM and/or with the temperature of the ambient ($T$). Here, we assume the following temperature dependences for each sublattice [74,76]:

$$M_{s,1}(T) = M_{s,1}(0)\left(1 - \frac{T}{T_C}\right)^{a_1} \tag{119}$$

$$M_{s,2}(T) = M_{s,2}(0)\left(1 - \frac{T}{T_C}\right)^{a_2} \tag{120}$$

where $T_C = 450$ K is the Curie temperature of the FiM, $M_{s,1}(0) = 1.4 \times 10^6$ A/m and $M_{s,1}(0) = 1.71 \times 10^6$ A/m are the saturation magnetization at zero temperature, and $a_1 = 0.5$ and $a_2 = 0.76$ are the exponents describing the temperature dependence of the saturation magnetization of each sublattice. $M_{s,1}(T)$ and $M_{s,2}(T)$ vs $T$ are shown in Figure 27, where the vertical dashed lines indicate the temperature at which the saturation magnetization of the two sublattices vanish ($T_M = 241.5$ K, $M_{s,1}(T) = M_{s,2}(T)$), and the angular momentum compensation temperature ($T_A = 260$ K), which corresponds to $M_{s,1}(T)/g_1 = M_{s,2}(T)/g_2$.

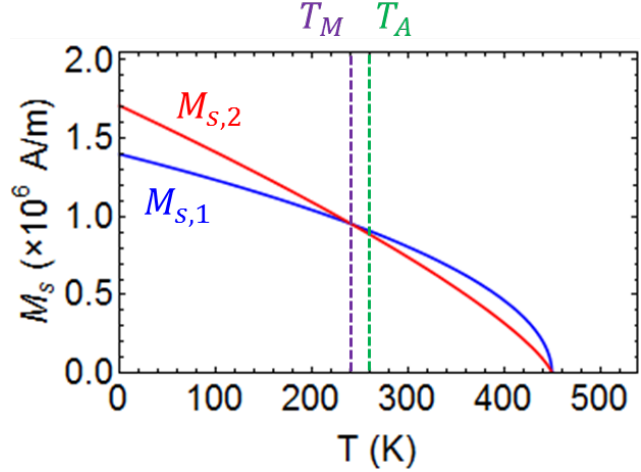

**Figure 27.** (a) Temperature dependence of the saturation magnetization of the sublattices forming the FiM. The dashed vertical lines indicate the temperature at which the saturation magnetization of the two sublattices vanish ($T_M = 241.5$ K), and the angular momentum compensation temperature ($T_A = 260$ K). Reproduced with permission [82]. Copyright 2019, Elsevier.

The micromagnetic (open symbols, mM) and the 1DM (solid lines) results of the DW velocity as functions of the current $J_{HM}$ are shown in Figure 28(a) for several representative temperatures, including $T = 240$ K $\approx T_M$ and $T = 260$ K $\approx T_A$. Both models are in remarkable agreement. Note that our model is more general than the effective 1DM used in Ref [71-74,83-85], which solve the current-driven DW dynamics for an effective ferromagnetic layer with renormalization of the damping parameter ($\alpha'$) and the gyromagnetic ratio ($\gamma'$). Moreover, our 1DM for FiMs also allows accounting for the different DW angles in the two sublattices ($\psi_i$ for $i$: 1,2). For $T < T_A$ or $T > T_A$, the velocity of the DW in the FiM saturates as the current increases. However, at $T \approx T_A$ the DW velocity increases linearly with $J_{HM} \equiv J$. The 1DM results of the internal DW angles within each sublattice are shown in Figure 28(b), which clearly shows that the linear increase of the DW velocity at $T \approx T_A$ is a direct consequence of the antiparallel alignment of the internal DW lattices moments along the longitudinal $x$-axis: $\psi_1 = \pi$ and $\psi_2 = 0$ independently on the driving current. In order words, our 1DM suggests that the linear increase of the DW velocity at the angular compensation temperature $T \approx T_A$ is due to the preservation of the antiparallel Néel DW configuration in the two sublattices. This configuration fully optimizes the spin Hall driving force.

In fact, from Equation (109), we can obtain an analytical expression which explains the current dependence of the DW velocity for different temperatures. Indeed, in the steady-state regime ($\dot{\psi}_1 = \dot{\psi}_2 = 0$), and in the absence of either STTs ($u_i = 0$) or external fields ($H_z = 0$), the DW velocity is

$$\dot{q} = v_{st} = -\Delta \frac{Q_1 M_{s,1}\left[\frac{\pi}{2} H_{SL,1} \cos \psi_{st,1}\right] + Q_2 M_{s,2}\left[\frac{\pi}{2} H_{SL,2} \cos \psi_{st,2}\right]}{\left(\alpha_1 \frac{M_{s,1}}{\gamma_1} + \alpha_2 \frac{M_{s,2}}{\gamma_2}\right)} \quad (121)$$

For $\alpha_1 = \alpha_2 = \alpha$ and $\theta_{SH,1} = \theta_{SH,2} = \theta_{SH}$, and taking into account that $Q_1 = +1$, $Q_2 = -1$, and that for $T \approx T_A$, $\frac{M_{s,1}}{\gamma_1} = \frac{M_{s,2}}{\gamma_2}$ and $\psi_{st,1} = \pi$ and $\psi_{st,2} = 0$, we have

$$\dot{q} = v_{st} = \gamma_0 \frac{\Delta}{\alpha} \frac{\pi}{2} \frac{\hbar \theta_{SH} J_{HM}}{2|e|\mu_0 t_{FiM}} \frac{g_1 g_2}{(g_2 M_{s,1} + g_1 M_{s,2})} \quad (122)$$

which scales linearly with $J_{HM}$, and therefore, explains the observed behavior at $T \approx T_A$.

Figure 28(c) plots the DW velocity as a function of the temperature $T$ as computed by the micromagnetic code and the 1DM. Both models predict that the DW velocity peaks at $T \approx T_A$ for sufficiently high current. These results are in good agreement with recent experimental observations in FiMs [73,74]. Our micromagnetic model and the extended 1DM provide a more accurate description of the current-driven DW dynamics in these FiM systems than effective 1DM models. In particular, they allow us to explain the results and suggest research lines to optimize the DW motion efficiency in future experimental setups. Furthermore, Equation (122) suggest that at $T \approx T_A$ the linear increase of the DW velocity can be further enhanced in FiMs with smaller $(g_2 M_{s,1} + g_1 M_{s,2})$ and smaller thicknesses ($t_{FiM}$).

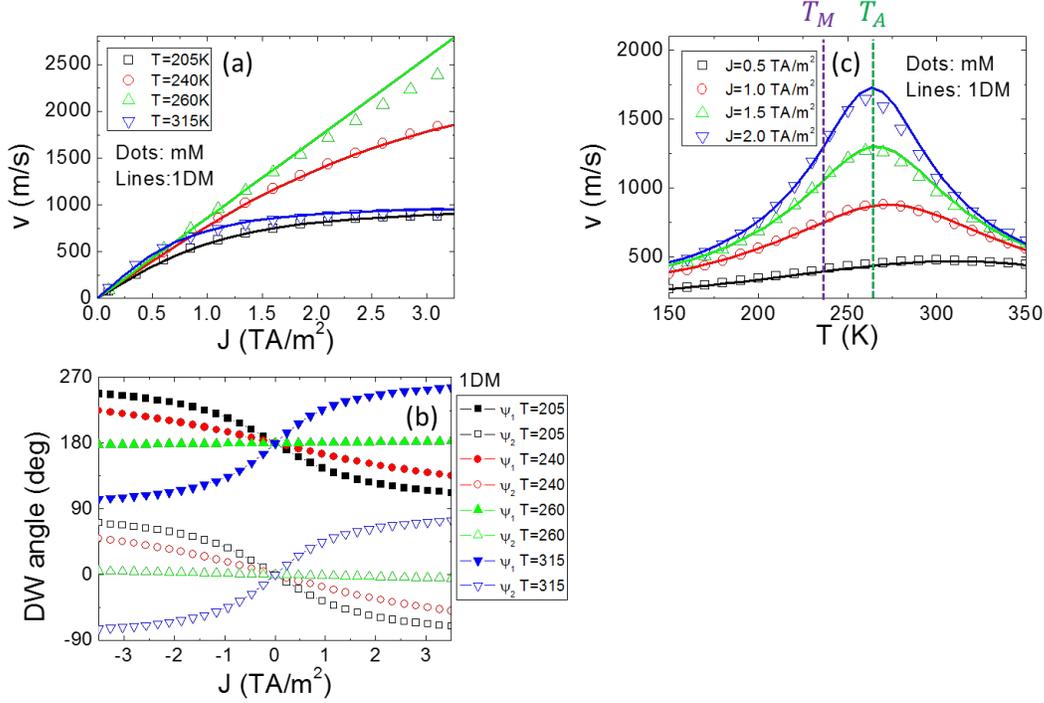

**Figure 28**. (a) Micromagnetic (open dots) and 1DM (solid lines) results of the DW velocity as a function of the density current along the HM ($J_{HM} \equiv J$) for several temperatures including $T = 240$ K $\approx T_M$ and $T_A = 260$ K. (b) 1DM results for the internal DW angles within each lattice as a function of $J_{HM} \equiv J$ for the same temperatures as in (a). (c) DW velocity as a function of $T$ for different values of $J_{HM}$. All material parameters and dimensions are given in the main text. Reproduced with permission [82]. Copyright 2019, Elsevier.

Finally, we have also micromagnetically studied the current-driven DW along a HM/FiM with straight and curved parts at the angular compensation temperature $T_A$. The transient micromagnetic snapshots are shown in Figure 29, where the same material parameters and FiM cross section as for Figure 28 were adopted. In the curved parts, the internal radius ($r_i$) is equal to the FiM strip width ($w$), i.e. $r_i = w = 256$ nm. Both in the straight and curved parts, the FiM DWs are very efficiently driven by the current injected along the HM without tilting. Similar as in the SAF, the relative distance between adjacent DWs is preserved after passing through the curved parts, which makes FiM systems promising candidates to develop fast and efficient DW devices.

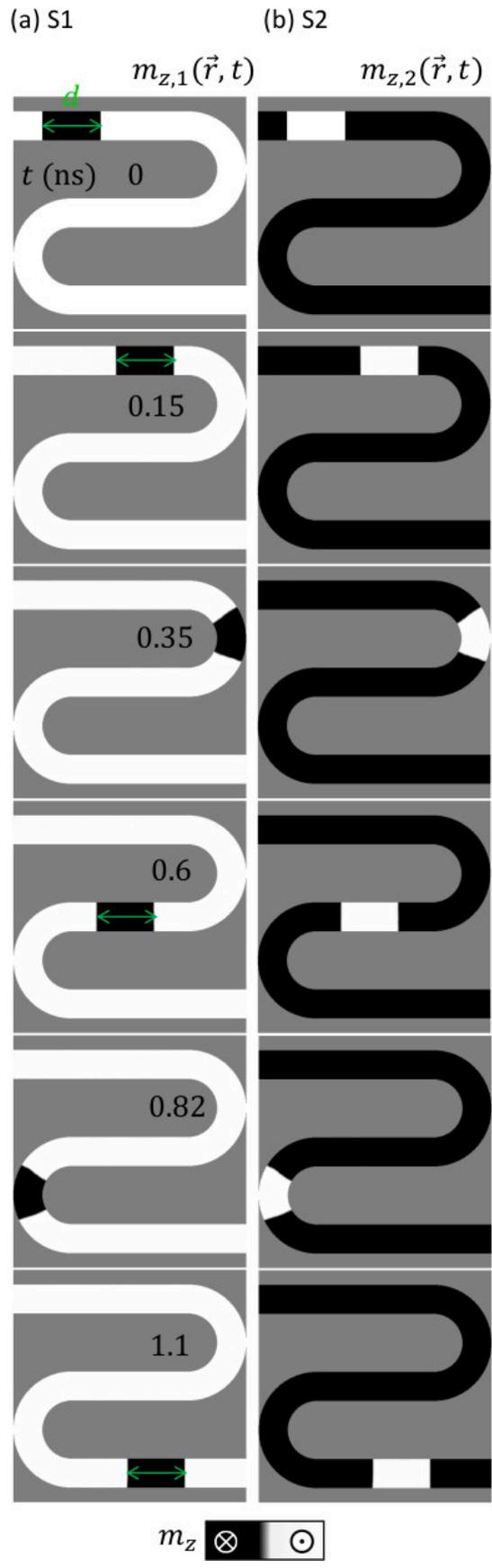

**Figure 29**. Micromagnetic snapshots showing the current-driven DW dynamics along a HM/FiM with curved and straight parts at the temperature of the angular momentum compensation ($T_A = 260$ K). Left (a) and right (b) columns show the local out-of-plane magnetization of each sublattice of the FiM. Here the radius of the curved parts is $r_i = w$, and $w = 256$ nm. The injected current in the HM is $J_{HM} = 1.5$ TA/m² in the straight parts, where it is uniform. The non-uniform distribution of this current in the curved part is taken into account. The rest of material parameters are the same as for Figure 28.

## 4. Concluding remarks and perspectives.

In this chapter we have presented a full theoretical description of the micromagnetic model and the 1D models needed to describe the DW dynamics along FM strips, HM/FM, SAF and HM/FiM stacks with high PMA. All these methods will be of great relevance for the further development of DW-based devices and to understand existing and future experimental observations. Here, we conclude this chapter by indicating some features which are present in experimental setups, and the manner they can be taken into account in the presented models.

- We have assumed perfect samples without imperfections or defects. It is known that imperfections, which are inevitable during the fabrication of the samples, introduces pinning or act against the free DW motion. In the micromagnetic formalism, it is possible to take into account the defects in different manners. For example, the sample can be divided into grains where the magnetic parameters can vary from grain to grain. The characteristic grain size is usually assumed to be random, with an average grain size typically in the order of ~10 nm. Regarding the material parameters, they are also assumed to change from grain to grain in a random manner. For instance, recent experimental results have been properly described by just considering that the easy-axis direction is mainly directed along the out-of-plane direction but with a random non-null in-plane component [86]. Other alternatives also include varying the magnitude of the anisotropy constant, the saturation magnetization and even the DMI parameter [87]. In the framework of the 1DM, the pining can be phenomenologically included by adding a periodic pinning field $H_p(x) = -H_{pin} \sin(2\pi x/L_p)$ to the external out-of-plane applied field in the 1DM equations, where $H_{pin}$ represents the magnitude of the experimentally observed pinning field (defined as the minimum field needed to overcome the pinning), and $L_p$ is related to the space periodicity of the pinning profile along the strip axis. See [63,64] for further details of the inclusion of pinning in the 1DM.
- We have also considered that the driving force on the DW, either the applied field and/or the current, are static once they are applied. However, experimental measurements of the current-driven DW dynamics are usually done by injecting current pulses with a finite duration. Different behaviors have been experimentally observed depending on the material parameters. For example, the experiments by Vogel *et al*. [88] indicated that a DW starts to move as soon as the current pulse is turned on in Pt/Co stacks, and it stops once the current pulse is switched off. On the other hand, under certain conditions of low damping and moderate DMI [89], the acceleration and deceleration times can be different, a fact which results in a dependent velocity on the duration of the current pulse, and consequently, on an inertial DW motion [89]. Of course, the micromagnetic and the 1D models developed here can be also used to study the DW dynamics under field and/or current pulses.

- Thermal effects are also other ingredients which need to be taken into account to provide a more accurate description of DW experiments. Although here we have assumed deterministic conditions, our methods can be extended to account for these thermal effects. In particular, thermal fluctuations can be directly included in the formalisms by adding a stochastic thermal field to the deterministic effective in LLG Equations. The magnetization dynamics is therefore governed by the corresponding stochastic Langevin-LLG equation. Details of this stochastic formalism and the resulting stochastic DW dynamics can be consulted in Ref. [63,90,91]. See [92] and references therein for an analysis of numerical aspects when solving micromagnetic problems with thermal noise. Additionally, when a current is flowing along the HM and/or along the conducting magnetic layers, Joule heating effects could become relevant to understand the current-driven DW dynamics. If the temperature of the system approaches Curie temperature, the LLG Equation must be replaced by the Landau-Lifshitz-Bloch Equation, which permits to evaluate the magnetization dynamics for high temperatures. We have also developed a micromagnetic formalism which solves the magnetization dynamics self-consistently as coupled to the heat transport in the system. Indeed, the formalism solves the Landau-Lisfhitz-Bloch Equation coupled to the heat transport equation, which describes the temporal evolution of the temperature. Details of this extended micromagnetic framework and the resulting current-driven DW dynamics can be consulted elsewhere [93,94,95].

Together with the preceding aspects, there exist other points which need to be further designed and optimized to develop DW based devices. In particular, adjacent DWs in a strip experience a magnetostatic repulsion [96,97] which influences their position at rest once the driving force is turned off. Besides, all DW based devices require the nucleation of DWs in the magnetic layer as a preliminary step. To conclude, we briefly comment our suggestions to address these still open questions, based on the emergent ideas from micromagnetic simulations:

- The energy landscape along a ferromagnetic strip experienced by DWs can be also spatially modulated by ion-irradiation with different doses [98]. Specifically, it has been experimentally shown that the PMA can be controlled by this irradiation, and this can be used to generate a periodic pinning profile along the strip axis where the DWs stay at rest at the minima of the PMA landscape. It has been also shown that such PMA modulation favors a unidirectional ratchet-like propagation DW motion by field pulses [98]. Micromagnetic and the 1DM can be also used to describe such a unidirectional ratchet-like field-driven DW motion [95,99]. Both models have been used to describe the experimental observation, and to propose also other energy profiles designed to promote the bi-directional DW movement of series of DWs by current pulses.

- Apart from the efficient current DW motion, DW-based devices also require to generate series of DWs in an efficient and controlled manner. Nucleation of a reversed domain in a strip initially magnetized along the out-of-plane direction is usually done by injecting a current pulse along an orthogonal conducting line to the ferromagnetic strip axis. This current pulse generates a magnetic field which can locally reverse the initial direction of the magnetization in the ferromagnetic strip under the conductive line (see, for instance, Ref. [22]). This process can be also described by means of micromagnetic simulations [100]. Moreover, micromagnetic simulations have been used to show that it is possible to nucleate and drive series of DWs along a ferromagnetic strip using two conductive orthogonal strips which serve to generate a local in-plane magnetic field [100]. The nucleation of series of DWs, and their shifting along the strip take place as due to the Oersted field generated by current pulses along the conducting strips and the effective field due to the spin Hall effect in the HM underneath [100]. If, in addition, the ferromagnetic strip presents a periodic modulation of the PMA as discussed in the previous paragraph, this method could be exploited to develop novel efficient and controlled DW-based devices.

In summary, these are just some ideas among many others which could be suggested to develop novel DW based devices. Therefore, the numerical and theoretical methods presented in this chapter will be of relevance, not only for theoreticians, but also for experimenters in the design and the interpretation of their experiments during the next years. As mentioned at the beginning, the aim of the present chapter goes in that direction. Besides of the study of DW motion, the developed micromagnetic models that we have presented here will be also useful to study the magnetization dynamics of other systems. For instance, the switching of the magnetization in a magnetic element, the analysis of the excitation and propagation of spin waves, and their interaction with other magnetic patterns, and even the dynamics of other magnetic patterns such as skyrmions can be naturally studied with the presented micromagnetic models.

## Acknowledgment

The authors are grateful to Luis Sanchez-Tejerina for valuable comments. This work was supported by projects MAT2014- 52477-C5-4-P, MAT2017-87072-C4-1-P, and MAT2017-90771-REDT from the Spanish government, and projects SA090U16 and SA299P18 from the Junta de Castilla y Leon.